\documentclass[journal ,12]{IEEEtran}
\usepackage{colortbl}
\usepackage[table]{xcolor}
\usepackage[dvipsnames]{xcolor}

\usepackage{tikz,tcolorbox}
\usepackage{lipsum}
\usepackage{algorithm}
\usepackage{algorithmic}
\usepackage{float,multirow} 

\usepackage{amsmath,amsfonts,amssymb,mathrsfs,amsthm,amsbsy,graphicx,color,bm,balance,mathtools,xfrac,nccmath,siunitx,cite,xspace}
\usepackage{multirow}

\usepackage[font=scriptsize,labelfont=bf]{caption}

\usepackage{array} 
\usepackage{makecell}
\usepackage{tabularx}
\usepackage{bm}

\usepackage{mdframed}
\usepackage{xcolor,acronym}

\newmdenv[
  backgroundcolor=gray!5,        
  linecolor=black,               
  linewidth=1pt,
  roundcorner=5pt,              
  frametitlefont=\color{white},
  frametitlealignment=\raggedright,
  frametitlebackgroundcolor=black, 
  frametitlerule=false,          
  innertopmargin=0.8\baselineskip,
  innerbottommargin=0.8\baselineskip,
  innerleftmargin=10pt,
  innerrightmargin=10pt,
  skipabove=\baselineskip,
  skipbelow=\baselineskip,
]{mybox}

\usepackage[shortlabels]{enumitem}

\usepackage{pgfplots}
\pgfplotsset{compat=newest}
\usetikzlibrary{plotmarks}
\usetikzlibrary{arrows.meta}
\usepgfplotslibrary{patchplots}
\usepackage{grffile}
\usetikzlibrary{plotmarks}
\usepackage[hidelinks]{hyperref}

 \usepackage[utf8]{inputenc}
\usepackage{booktabs, caption, makecell}

\usepackage{threeparttable}

\newcommand{\q}[1]{{\bf{#1}}}

\newcommand{\pfa}{P_{fa }}
\newcommand{\pd}{P_{d }}
\newcommand{\E}[1]{{\mathbb{E} \left\{ #1 \right\}}}

\newcommand{\thr}{\rm{th}}

\theoremstyle{definition}
\makeatletter
\newcommand{\printfnsymbol}[1]{%
  \textsuperscript{\@fnsymbol{#1}}%
}
\makeatother

\begin{document}
\bstctlcite{IEEEexample:BSTcontrol}
\title{Reconfigurable Intelligent Surfaces for Cognitive Radio Networks: Design, Optimization, and Emerging Trends\vspace{-2mm}}

\author{Diluka Galappaththige, \IEEEmembership{Member, IEEE}, Chintha Tellambura, \IEEEmembership{Fellow, IEEE}, Ranga Kulathunga, \IEEEmembership{Graduate Student Member, IEEE}, and Gayan Amarasuriya Aruma Baduge, \IEEEmembership{Senior Member, IEEE \vspace{-8mm}} 
\thanks{D. Galappaththige and C. Tellambura are with the Department of Electrical and Computer Engineering, University of Alberta, Edmonton, AB, T6G 1H9, Canada (e-mail: \{diluka.lg, ct4\}@ualberta.ca).} 
\thanks{R. Kulathunga and G. A. Aruma Baduge are with the School of Electrical, Computer, and Biomedical Engineering, Southern Illinois University, Carbondale, IL, USA (email: \{ranga.kulathunga, gayan.baduge\}@siu.edu). Their work in part was supported by the U.S. NSF under Grant CCF-2326621.} }

\maketitle  
\begin{abstract}Reconfigurable intelligent surfaces (RISs) enable programmable wireless propagation environments, offering new opportunities for cognitive radio networks (CRNs) to improve spectrum utilization, enhance spectral and energy efficiency, and operate reliably under low signal-to-noise ratio conditions. By combining the complementary strengths of RISs and CRNs, RIS-assisted CRNs (RCNs) have emerged as a promising architecture for future 6G wireless systems. Despite their growing importance, a comprehensive survey of this rapidly evolving field has been lacking. This paper fills this gap by providing a systematic and comprehensive review of RCNs. The paper first introduces the fundamentals of CRNs and RISs, including dynamic spectrum access  models, spectrum sensing techniques, RIS operating principles, and RIS architectures. It then examines the design of RCNs, covering their system architectures, deployment strategies, channel estimation, spectrum access mechanisms, communication protocols, and the joint optimization of RIS and CRN parameters. Next, the existing literature is organized into six major research directions: performance analysis, resource allocation and optimization, secure RCNs, active RISs, simultaneously transmitting and reflecting RISs, and machine learning-enabled RCNs. Finally, the paper discusses key research challenges and future directions, including scalability, practical deployment, integration with emerging 6G technologies, coexistence with evolving network architectures, standardization, and security and privacy. This survey provides a unified reference for researchers and practitioners and establishes a roadmap for the future development of RCNs.

\end{abstract}


\thispagestyle{empty}

\section*{List of Acronyms}
\addcontentsline{toc}{section}{Nomenclature}
\begin{IEEEdescription}[\IEEEusemathlabelsep\IEEEsetlabelwidth{$V_1,V_2,V_3,V_4$}]
\fontsize{0.33cm}{0.4cm}\selectfont
\item[4(5)(6)G]   Fourth (Fifth) (Sixth)  generation 
\item[BC]         Backscatter communication
\item[CDS]        Cognitive dynamic system 
\item[CF]         Cell-free 
\item[CR]         Cognitive radio 
\item[CRN]        Cognitive radio network  
\item[CSS]        Cooperative spectrum sensing 
\item[DSA]        Dynamic spectrum access 
\item[EE]         Energy efficiency 
\item[IoT]        Internet of Things  
\item[ISAC]       Integrated sensing and communication 
\item[MIMO]  	  Multiple-input multiple-output 
\item[NOMA]  	  Non-orthogonal multiple access  
\item[NTN]        Non-terrestrial network
\item[NF]         Near-field 
\item[PLS]        Physical-layer security  
\item[PT]         Primary transmitter 
\item[PU]         Primary user  
\item[RCN]        Reconfigurable intelligent surface-assisted cognitive radio network
\item[RIS]        Reconfigurable intelligent surface 
\item[SE]         Spectral efficiency 
\item[SINR]       Signal-to-interference-plus-noise ratio  
\item[SNR]        Signal-to-noise ratio  
\item[SR]         Symbiotic radio 
\item[ST]         Secondary transmitter 
\item[SU]         Secondary user  

\end{IEEEdescription}

\section{Introduction}\label{sec:introduction}
Cognitive radio (CR) has emerged as a key technology for improving spectral efficiency (SE) and energy efficiency (EE), making it an important enabler of sixth-generation (6G) wireless networks~\cite{Hilal2023,Arjoune2019,Khasawneh2023,Kusaladharma2017}. By opportunistically accessing temporarily unused licensed spectrum while protecting primary users (PUs), including television white spaces (TVWS), idle cellular spectrum, and radar bands such as the \qty{3.5}{\GHz} Citizens Broadband Radio Service (CBRS)~\cite{Ian2006,ieee802222011}, CR significantly improves spectrum utilization, network capacity, and overall wireless efficiency. The resulting dynamic spectrum access (DSA) paradigm enables secondary networks (SNs) to opportunistically reuse licensed spectrum owned by primary networks (PNs) without causing harmful interference to PUs, a process commonly represented by the CR cycle and described in detail in Section~\ref{sec_CRN_overview}


Beyond conventional spectrum sharing, CR is increasingly integrated with emerging wireless technologies, including massive multiple-input multiple-output (MIMO), cell-free (CF) networks, and backscatter communication (BC)~\cite{Hilal2023,Arjoune2019,Khasawneh2023,Alsabah2021,Ngo2017,Rezaei2023Coding,Diluka2022}. It is also expected to play a central role in beyond-fifth-generation (5G) and 6G networks by enabling intelligent spectrum management for integrated sensing and communication (ISAC), Internet of Things (IoT), ultra-dense deployments, and artificial intelligence (AI)-native wireless systems~\cite{Diluka2026WBCRN,Saad2020,Nguyen2022,Yang2020AI}.

\begin{table*}[t]\vspace{-2mm}
\centering
\renewcommand{\arraystretch}{1.0}
\caption{Key challenges in CRNs.}
\label{tab:crn_challenges}\vspace{-2mm}
\begin{tabular}{p{3cm}p{6.5cm}p{7cm}}
\hline
\textbf{Challenge} & \textbf{Description} & \textbf{Key Issues} \\ \hline \hline

Spectrum sensing & Reliable detection of PUs under dynamic conditions & Low SNR, noise uncertainty, hidden node problem, sensing delay \\ \hline

DSA & Opportunistic channel selection and access decisions & Uncertainty, fast adaptation, throughput vs. PU protection trade-off \\ \hline

Interference management & Limiting interference to PUs and among SUs & Co-channel interference, heterogeneous environments \\ \hline

Spectrum mobility & Channel switching when PU activity is detected & Handover delay, packet loss, QoS degradation \\ \hline

Channel estimation & Estimating channel coefficients accurately & Intermittent access, estimation overhead \\ \hline

Scalability & Supporting dense and large-scale CRNs & Coordination overhead, distributed control complexity \\ \hline

EE & Reducing energy consumption for sensing and adaptation & Continuous sensing overhead, IoT constraints \\ \hline

Security & Protecting CRNs from malicious attacks & PU emulation (PUE) attacks, data falsification, eavesdropping \\ \hline

Learning and adaptation & Use of AI/ML for intelligent decision-making & Training data, generalization, computational complexity \\ \hline

Hardware constraints & Practical limitations of RF front-ends and sensing systems & Wideband sensing complexity, analog-to-digital converter (ADC) limitations \\ \hline

Regulation & Compliance with spectrum policies & Lack of unified standards, regional variations \\ \hline

QoS and latency & Maintaining reliable communication performance & Variable latency, unreliable connectivity \\ \hline

\end{tabular}\vspace{-5mm}
\end{table*}


\subsection{Challenges in CR networks (CRNs)}
Dynamic and opportunistic spectrum access gives rise to a well-documented set of challenges~\cite{Hilal2023, Arjoune2019, Khasawneh2023, Kusaladharma2017}, summarized in Table~\ref{tab:crn_challenges}: unreliable spectrum sensing under low signal-to-noise ratio (SNR) and hidden-node conditions; timely and accurate DSA decisions under uncertain network states; interference management between PNs and SNs; spectrum mobility and handover overhead; imperfect channel estimation under intermittent access; high-dimensional, non-convex resource allocation problems; scalability and coordination overhead in dense deployments; energy and hardware constraints, particularly for IoT devices; security threats such as PU emulation and sensing-data falsification; and the practical limitations (training data, generalization, complexity) of AI/machine learning (ML)-based solutions. These challenges are interrelated; for instance, improving sensing reliability typically requires longer observation windows, which reduces the time available for data transmission, while stricter interference protection for PUs comes at the cost of secondary throughput. A large body of research has addressed these issues, particularly those related to spectrum-sensing reliability, interference management, and channel estimation \cite{Hilal2023, Arjoune2019, Khasawneh2023, Kusaladharma2017}.

\subsection{The Promise of Reconfigurable Intelligent Surfaces (RISs)}
RISs enable programmable wireless environments by transforming the propagation channel from an uncontrollable medium into a design variable~\cite{Yuan2021,Ahmed2024RISAdvances,Chen2022}. By intelligently controlling the amplitude, phase, and polarization of reflected electromagnetic (EM) waves, RISs improve coverage, particularly in non-line-of-sight scenarios. Conversely, their largely passive, thin, lightweight, and low-cost architecture provides a hardware- and energy-efficient means of realizing beamforming, interference management, and channel reconfiguration. These properties make RISs well suited for large-scale deployment and facilitate emerging technologies such as ISAC, millimeter-wave (mmWave)/terahertz (THz) communications, CF architectures, and enhanced physical-layer security (PLS), establishing RISs as a key enabler for future programmable, reliable, and energy-efficient wireless networks.

Consequently, RIS-assisted CRNs (RCNs) have attracted significant attention as a promising means of improving spectrum utilization and communication performance~\cite{Yuan2021,Ahmed2024RISAdvances,Chen2022,Diluka2020,Diluka2021CFIRS}. By intelligently reconfiguring the propagation environment, RISs can enhance the SE and/or EE while improving spectrum sensing, secondary transmission, primary transmission, or their joint operation. For example, an RIS can simultaneously strengthen secondary communication links while inducing destructive combining at PUs to suppress interference~\cite{Nasser2022}. However, the coexistence of PUs and SUs, together with the tight coupling between spectrum sensing, communication, and RIS control, makes resource allocation, interference management, and beamforming design significantly more challenging than in conventional CRNs, motivating the cross-layer treatment developed throughout this survey~\cite{Yuan2021,Ahmed2024RISAdvances,Chen2022,Diluka2020,Diluka2021CFIRS}.

\subsection{Existing Surveys}
To the best of our knowledge, there are no surveys on RCNs. To provide context, we briefly describe several recent surveys on CRNs, IoT-CRN convergence, and 6G-enabled CRNs. Table~\ref{tab:existing_surveys} summarizes their scope, key topics, and the extent to which RIS/RCN aspects are addressed.

Reference \cite{Hilal2023} surveys cognitive dynamic systems (CDSs), which emulate human cognitive functions, i.e., perception-action, memory, attention, intelligence, and language, for autonomous adaptation, with applications in CR and cognitive radar such as adaptive waveform design, DSA, and enhanced sensing. Although RCNs are not explicitly considered, their memory and learning components could conceptually support ML/reinforcement-learning-based RIS optimization and environment-aware adaptation, offering a conceptual foundation for autonomous RCN architectures \cite{Zhang2020, He2020, Zhou2023}.

\begin{table*}[t]\vspace{-2mm}
\centering
\renewcommand{\arraystretch}{1.0}
\caption{Summary of existing surveys related to CRNs and their relation to RCNs.}
\label{tab:existing_surveys}\vspace{-2mm}
\begin{tabular}{p{0.8cm}p{2.0cm}p{6.5cm}p{6.5cm}}
\hline
\textbf{Ref.} & \textbf{Focus} & \textbf{Key Topics Covered} & \textbf{RIS/RCN Consideration} \\ \hline \hline

\cite{Hilal2023} & CDS 
& Perception-action cycle, memory, attention, intelligence; adaptive waveform design; DSA; cognitive radar 
& Not addressed; CDS framework could conceptually support RIS-based adaptive sensing/access, but not explored \\ \hline

\cite{Arjoune2019} & Spectrum sensing techniques 
& Energy detection, cyclostationary detection, matched filtering, compressive sensing, ML/DL-based sensing, cooperative sensing 
& None; assumes fixed/passive channel; no RIS-aided sensing enhancement \\ \hline

\cite{Khasawneh2023} & IoT-CRN convergence 
& Cross-layer taxonomy (sensing, communication, intelligence, system, application, security layers); ML, edge computing, blockchain 
& None; resource-constrained IoT focus; RIS-enabled reconfiguration not considered \\ \hline

\cite{Aslam2021} & CRNs in 6G 
& AI, THz, mMIMO, satellite-aerial-terrestrial integration; XR/smart city applications; 6G security challenges 
& RIS briefly mentioned as 6G enabler; no RIS-CRN joint optimization or design framework \\ \hline

\textbf{This survey} & RCNs 
& RIS-aided sensing, channel estimation, spectrum access, resource allocation, secure/active/STAR-RIS, ML for RCNs, joint optimization 
& Comprehensive and dedicated treatment \\ \hline

\end{tabular}\vspace{-5mm}
\end{table*}


Reference \cite{Arjoune2019} surveys spectrum sensing techniques for DSA, classifying narrowband and wideband approaches (energy detection, matched filtering, cyclostationary detection, compressive sensing, and ML-based methods; see Section~\ref{sec_basics_RIS_CRN}), analyzing trade-offs among detection performance, complexity, and prior knowledge, and identifying robust, low-complexity, and data-driven sensing as future directions. The survey assumes a fixed, passive channel throughout and does not consider RIS-aided sensing enhancement.


Reference \cite{Khasawneh2023} addresses the convergence of IoT and CRNs, motivated by IoT growth and spectrum scarcity, reviewing cognitive IoT architectures and techniques, cooperative/distributed sensing, energy-efficient protocols, and ML/AI-based spectrum prediction, tailored to resource-constrained devices, and identifying ML, edge computing, blockchain, and 6G as future directions. RIS-enabled reconfiguration is not considered.


Reference \cite{Aslam2021} surveys CRNs for 6G, covering enabling technologies (AI, THz, massive MIMO (mMIMO), satellite-aerial-terrestrial integration) and applications (XR, smart cities, industrial automation, healthcare) that demand intelligent spectrum management, along with security challenges arising from AI integration and heterogeneous architectures. RIS is briefly mentioned as a 6G enabler, but no RIS-CRN joint optimization or design framework is provided.


\subsection{Gaps in the Literature}
Although existing surveys provide valuable overviews of CRNs, CDS, and IoT-enabled CRNs~\cite{Hilal2023,Arjoune2019,Khasawneh2023}, a comprehensive survey dedicated to RCNs remains unavailable. These surveys primarily address spectrum sensing, DSA, resource allocation, and learning-based adaptation under the conventional assumption of a passive wireless propagation environment. Conversely, RIS surveys focus on channel modeling, beamforming, and optimization for conventional wireless systems without considering the unique challenges of spectrum sharing between PN and SN~\cite{Yuan2021,Ahmed2024RISAdvances,Chen2022}. Consequently, neither body of literature provides a unified treatment of RCNs, leaving several gaps.

First, the impact of RIS on \emph{spectrum sensing} remains largely unexplored. Existing sensing techniques, from energy detection to ML-based methods, are developed under fixed-channel assumptions and target robustness against low SNR and noise uncertainty~\cite{Saber2022,Nasser2022,Lin2022,Wu2021}, whereas RIS can actively manipulate the propagation environment to improve SNR and spatial diversity~\cite{Yuan2021,Ahmed2024RISAdvances,Chen2022}, fundamentally altering sensing metrics such as probability of detection $\pd$, false-alarm probability $\pfa$, and sensing latency. The joint design of RIS configuration and spectrum sensing, however, has received little attention.

Second, \emph{RIS-aware DSA, interference management, and cross-layer design} remain underdeveloped. Conventional interweave, underlay, and overlay paradigms regulate interference through access policies, without exploiting programmable propagation~\cite{Liu2021RIS,Gong2020,Chen2022}. Since RIS tightly couples sensing, communication, and interference shaping through its influence on channel propagation, existing studies that optimize these functions independently leave SE and coexistence gains unrealized, motivating holistic frameworks that jointly optimize RIS control, sensing, access, and resource allocation.

Third, several \emph{application-specific and methodological gaps} persist, i.e., RIS integration with IoT-enabled CRNs remains largely unexamined despite IoT being a frequently cited CRN use case~\cite{Khasawneh2023}; AI/ML frameworks for CRNs do not yet account for RIS-specific control variables such as reflection coefficients and phase-shifts~\cite{Arjoune2019}; and RIS's potential to enhance PLS, by shaping propagation to suppress information leakage and generate artificial interference, is largely overlooked in existing security-focused CRN surveys~\cite{Jacek2022}.

Finally, \emph{practical and analytical characterizations} are lacking. RIS introduces additional impairments beyond conventional channel state information (CSI) and hardware limitations, including double-fading channels, phase quantization, inter-element coupling, and channel-estimation overhead~\cite{Yuan2021,Ahmed2024RISAdvances,Chen2022}, which are rarely incorporated into RCN design. Likewise, little is known about how RIS parameters, such as the number of reflecting elements, scale with sensing accuracy, interference management, and computational complexity, and RIS has yet to be integrated into a unified 6G vision encompassing CRNs, IoT, ISAC, and edge AI~\cite{Ahmed2024RISAdvances}.

Addressing these gaps requires a unified RCN framework that considers spectrum awareness, programmable channel control, intelligent resource allocation, and PU protection under realistic propagation, the objective of this survey.

\subsection{Our Contribution}
A high-quality survey should do more than summarize and classify the existing literature; it should synthesize prior work into a coherent framework, identify fundamental challenges and research gaps, and establish a roadmap for future research. Guided by this objective, this survey provides the first comprehensive and systematic review of RCNs, addressing the gaps identified above while highlighting emerging trends, open problems, and promising research directions. Beyond consolidating existing knowledge, the survey grounds its technical discussion in six worked case studies with numerical results, illustrating how RIS reshapes spectrum sensing, resource allocation, security, and learning-based design in CRNs, and offers a concrete roadmap toward integrating RCNs within the broader 6G ecosystem.

\begin{figure*}[!t]\vspace{-2mm}
\centering
    \def\svgwidth{435pt} 
    \fontsize{8}{8}\selectfont 
    \graphicspath{{Figures/}}
    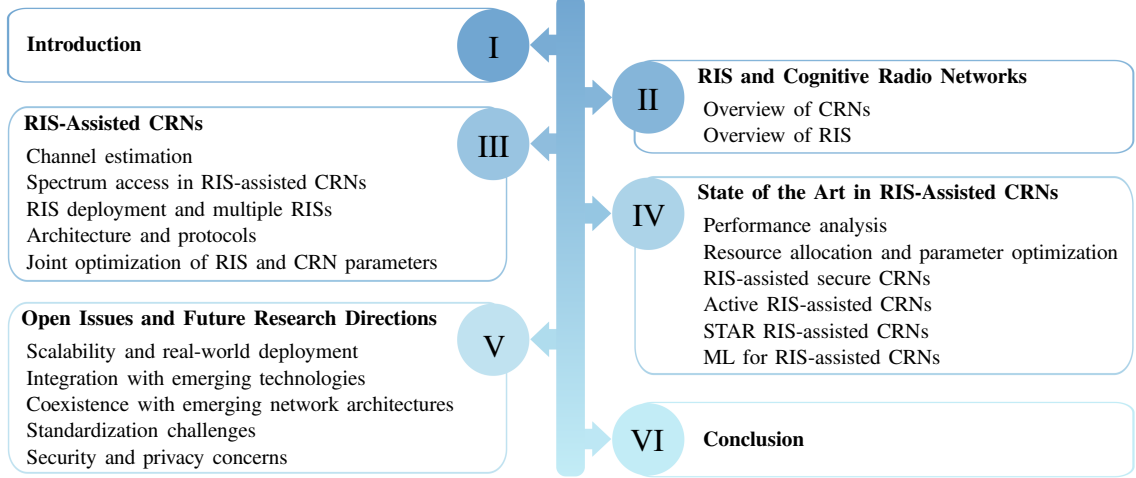 \vspace{-2mm}
    \caption{Outline of the main contributions of this paper.}\vspace{-5mm} \label{fig_Outline}
\end{figure*}

The contributions of this survey paper are summarized as follows (Fig.~\ref{fig_Outline}):
\begin{enumerate}
    \item We review the fundamentals of CRNs and RIS, including DSA models, spectrum-sensing techniques, and RIS principles and types, establishing how RIS reshapes traditional CRN assumptions on sensing and propagation.

    \item We examine RCNs and their unique characteristics, analyzing the benefits and challenges of integrating RIS into CRNs. Key aspects include channel estimation, RIS-assisted spectrum access, deployment and multi-RIS configurations, architectures and protocols, and joint optimization of RIS and CRN parameters.

    \item We classify recent RCN advances into six categories: performance analysis, resource allocation and optimization, security, active RIS, simultaneously transmitting and reflecting (STAR)-RIS, and ML-enabled designs. For each category, we review methodologies, key results, and insights, and illustrate representative designs through worked case studies with numerical results.

    \item We explore the remaining challenges, open issues, and emerging trends, including scalability and deployment, integration with emerging technologies such as the IoT, unmanned aerial vehicles (UAVs), mobile edge computing (MEC), ISAC, near-field (NF) communication, next-generation multiple access technologies (e.g., rate-splitting multiple access (RSMA)),  fluid antenna systems (FAS), coexistence with emerging architectures, including CF mMIMO, open-radio access networks (O-RANs), non-terrestrial networks (NTNs), and agentic AI, standardization hurdles, and security and privacy concerns.
\end{enumerate}

The remainder of this paper is structured as follows: Section~\ref{sec_basics_RIS_CRN} reviews  CRN and RIS fundamentals. Section~\ref{sec_ris_crn} explores RCNs, emphasizing the impact of RIS integration, key features, benefits, and challenges. Section~\ref{sec_state_art} reviews state-of-the-art technical contributions under six main categories. Finally, Section~\ref{sec_future} examines future research directions, opportunities, and challenges.

\section{RIS and Cognitive Radio Networks}\label{sec_basics_RIS_CRN}
This section briefly reviews the fundamental concepts of CRNs and RISs.

\subsection{Key features  of CRNs}\label{sec_CRN_overview}
The opportunities and challenges of CRNs have been discussed in the Introduction (Section \ref{sec:introduction}). Here we briefly review the underlying access and sensing models.

\begin{figure}[!t]\vspace{-2mm}
    \centering 
    \def\svgwidth{220pt} 
    \fontsize{8}{8}\selectfont 
    \graphicspath{{Figures/}}
    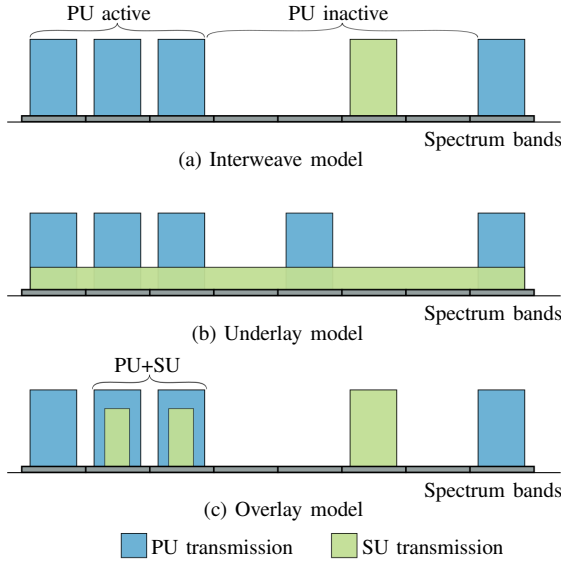 \vspace{-2mm}
    \caption{DSA models.}\label{fig_SpectrumAccessMode}\vspace{-5mm} 
\end{figure}

\subsubsection{Dynamic Spectrum Access}\label{sec_DSA}
DSA refers to the capability of unlicensed SUs to adaptively and opportunistically access spectrum that is temporarily unused by licensed PUs, through either sensing-based approaches or database-driven access, where SUs obtain spectrum availability from centralized geolocation databases or spectrum access systems~\cite{Nasser2021, Sangeeta2022}, an approach widely adopted for TVWS networks and the \qty{3.5}{\GHz} CBRS band~\cite{Ian2006, ieee802222011} that reduces sensing overhead and mitigates hidden-node problems relative to purely sensing-based access. DSA involves three key functions: (i) spectrum sensing, to detect spectrum holes by monitoring PU activity; (ii) spectrum decision, to select the most suitable band based on channel conditions, quality-of-service (QoS), and interference constraints; and (iii) spectrum access and mobility, to access the selected band while vacating or switching when the PU becomes active.

DSA can be categorized into three primary models (Fig.~\ref{fig_SpectrumAccessMode})~\cite{Nasser2021, Sangeeta2022, Kusaladharma2017}.
\begin{enumerate}[label=\roman*)]
    \item \textit{Interweave model:} SUs access the spectrum only when PUs are inactive, as PUs have absolute priority~\cite{Nasser2021}. SUs must vacate the band immediately upon PU activity, requiring reliable spectrum sensing to detect idle channels while addressing hidden nodes and multipath fading.

    \item \textit{Underlay model:} SUs transmit concurrently with PUs provided their aggregate interference remains below a tolerable threshold, either by spreading power over a broad spectrum (e.g., ultra-wideband systems) or via the \textit{interference temperature model}, where transmit power is controlled such that total interference stays below a regulated threshold~\cite{Sangeeta2022,6308772}.

    \item \textit{Overlay model:} SUs share the spectrum concurrently with PUs~\cite{Nasser2021}, but rather than merely constraining interference, they leverage cooperative techniques such as coding and relaying to preserve or enhance PU performance, e.g., by assisting PU transmission to improve the SINR at the primary receiver (PR), yielding a mutually beneficial operation at the cost of requiring detailed knowledge of PU signals and tight coordination.
\end{enumerate}

\subsubsection{Spectrum Sensing}\label{sec_spectrum_sensing}
This is necessary for SUs to continuously monitor bands for PU presence, ensuring interference-free operation~\cite{Arjoune2019}. Common techniques offer different trade-offs between complexity and detection performance~\cite{khattab2012cognitivebook, Akyildiz2008, Hilal2023, Diluka2021, Diluka2020NOMAUnderlay, Diluka2019, Kusaladharma2017}: \textit{energy detection} compares received signal energy against a threshold-low complexity, but degrades under low SNR and noise uncertainty \cite{5429879,6987540}; \textit{matched filter detection} coherently correlates the received signal with a known PU template, achieving optimal performance at the cost of requiring prior signal knowledge, synchronization, and higher complexity; \textit{cyclostationary feature detection} exploits periodic PU signal features for robust, noise-tolerant detection at the expense of processing power; \textit{cooperative spectrum sensing (CSS)} aggregates sensing information across multiple SUs to mitigate the hidden-node problem and improve accuracy; and \textit{ML-based sensing} uses deep learning (DL) models to learn spectrum occupancy patterns, improving adaptability and precision over model-based methods~\cite{Arjoune2019}.

\subsubsection{Variants of CRNs}
Beyond conventional CRNs, symbiotic radio (SR) and ambient BC have emerged as promising technologies for improving SE and EE~\cite{Rezaei2023Coding, Galappaththige2023, Diluka2022, Rezaei2024, Long2020, Galappaththige2023SR}. Unlike traditional CRNs, where SUs opportunistically access spectrum while minimizing interference to PUs, secondary devices in SR systems modulate and reflect primary signals to convey their own information, establishing a symbiotic relationship between PNs and SNs. Depending on the degree of cooperation, SR is classified into parasitic (SN benefits, slightly degrading PN performance), commensal (SN transmits without affecting PN performance), mutualistic (both PN and SN benefit, e.g., through cooperative reflection and relaying), and competitive (systems compete for resources, degrading each other's performance) modes, with mutualistic and commensal SR being particularly attractive since they enable spectrum sharing with limited or beneficial impact on primary transmissions~\cite{Rezaei2023Coding, Galappaththige2023, Diluka2022, Rezaei2024, Long2020, Galappaththige2023SR}.

Ambient BC is a key implementation of SR, in which low-power tags communicate by modulating and reflecting ambient RF signals from cellular, WiFi, or TV transmissions~\cite{Rezaei2023Coding, Galappaththige2023, Diluka2022}, achieving extremely low power consumption suited for IoT connectivity. These ultra-low-power devices may operate with energy harvesting (EH) \cite{9187234}. While SR and BC share the spectrum-sharing objective of CRNs, they tightly couple primary and secondary transmissions through cooperative interactions rather than relying on opportunistic or interference-constrained access, positioning them as an evolution of CR toward more collaborative frameworks. Although their integration with RCNs remains largely unexplored, these paradigms offer promising opportunities for enhancing SE and EE in 6G.

\subsection{Overview of RISs}
RIS surveys include \cite{Yuan2021, Ahmed2024RISAdvances, Chen2022}. An RIS is a two-dimensional array of reconfigurable elements, made out of ``meta-materials'', that can be programmed to dynamically control the direction, phase, amplitude, and polarization of incident EM waves, thereby transforming the wireless channel into a controllable propagation environment~\cite{Liu2021RIS, Gong2020, Chen2022, Diluka2020, Galappaththige2023, Diluka2022RISSWIPT, Diluka2022CFRIS}. Owing to their low profile, lightweight, and low-cost form factor, RISs can be seamlessly deployed on building walls, street lamps, billboards, vehicles, UAVs, and satellites, integrating into existing networks with only protocol-level adjustments rather than hardware modifications~\cite{Liu2021RIS, Gong2020, Chen2022}. Their passive scattering mechanism further supports full-duplex operation without the noise and interference typical of active relays (except in active RIS), enhancing SE while ensuring backward compatibility~\cite{Liu2021RIS, Gong2020, Chen2022}.

A typical RIS comprises meta-materials made up of numerous meta-atoms, which are engineered elements, typically conductive or semiconductor-based, that govern how incident EM waves interact with the surface. These are organized into reconfigurable unit cells, each of which adjusts its impedance or capacitance to control the phase (and, in active RIS, amplitude) of the reflected wave, collectively enabling beamforming, interference reduction, and coverage enhancement. A smart controller coordinates these adjustments across all elements, operating in either a centralized (base station (BS)-managed) or decentralized fashion, and in active RIS configurations, a small external power supply drives the amplification components to enhance reconfigurability~\cite{Liu2021RIS, Gong2020, Chen2022, Ahmed2024}.



\subsubsection{Types of RIS}
RIS architectures are classified by their reconfigurability, signal processing capability, and energy requirements~\cite{Liu2021RIS, Gong2020, Chen2022, Long2021, Ahmed2024, Liu2021}, spanning passive RIS, active RIS, STAR-RIS, stacked RIS (SRIS), and beyond-diagonal RIS (BD-RIS),
with different trade-offs in hardware complexity, energy consumption, spatial degrees of freedom (DoF), and signal manipulation capabilities.
\begin{itemize}
    \item \textit{Passive RIS:} Reflects incident EM waves via controllable phase shifts without amplification, offering high EE and low hardware complexity, but suffering from double fading across the cascaded transmitter-RIS-receiver link~\cite{Liu2021RIS, Gong2020, Chen2022,Gunasinghe2024a}. Passive RISs are well suited to EE-critical scenarios such as urban coverage enhancement and rural connectivity extension.

    \item \textit{Active RIS:} Incorporates amplifiers powered by external energy sources to jointly adjust phase and amplify incident signals, compensating for double-fading loss at the cost of higher power consumption and amplification noise~\cite{Long2021, Ahmed2024}. Active RIS suits high-capacity, low-latency systems and scenarios with severe fading or long transmission distances.

    \item \textit{STAR-RIS:} Unlike conventional RIS, which serves only one side (half-space) of the propagation environment, each STAR-RIS element can simultaneously reflect and transmit the incident signal, enabling full-space coverage and additional spatial DoF~\cite{Liu2021,Gunasinghe2024b}. This makes STAR-RIS attractive for dynamic, high-mobility scenarios such as dense urban and smart-city deployments.

    \item \textit{SRIS:} A SRIS employs multiple sequential layers, exploiting inter-layer coupling to manipulate EM waves multiple times before reception, achieving higher passive beamforming gains and finer wavefront control than conventional RIS~\cite{Liu2025SRIS, Abbas2026SRIS}. Although SRIS has not yet been investigated in the context of CRNs, its enhanced beamforming gains and spatial DoF could improve spectrum-sensing reliability, secondary transmission performance, and interference suppression toward PUs, particularly in low-SNR regimes where conventional RIS-assisted sensing suffers from severe path loss and double fading. Realizing these gains requires addressing increased hardware complexity, channel-estimation overhead, and joint optimization across coupled layers, making SRIS-assisted CRNs an important future research direction~\cite{Liu2025SRIS, Abbas2026SRIS}.

    \item \textit{BD-RIS:} Whereas conventional RISs use a diagonal scattering matrix with independently adjusted elements, BD-RIS introduces controllable coupling among elements, yielding a non-diagonal scattering matrix that enables signals to be jointly routed and processed for substantially greater spatial DoF~\cite{Li2026BDRIS}. Although unexplored in CRNs, this finer control over the propagation environment could enable more precise interference nulling toward PUs while enhancing spectrum sensing and secondary communications, particularly valuable in dense spectrum-sharing scenarios, at the cost of increased hardware complexity, control overhead, and optimization sophistication, positioning BD-RCN design as an important future research direction~\cite{Li2026BDRIS}.
\end{itemize}

\section{RIS-Assisted CRNs}\label{sec_ris_crn}

\begin{figure}[!t]\vspace{-2mm}
\centering
    \def\svgwidth{220pt} 
    \fontsize{8}{8}\selectfont 
    \graphicspath{{Figures/}}
    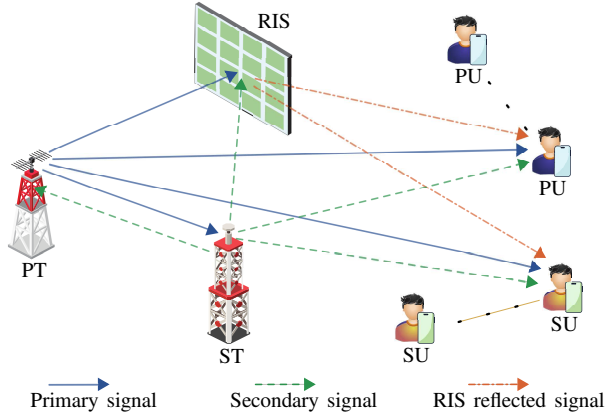 \vspace{-1mm}
    \caption{System model of an RCN. The PT serves multiple PUs, while the ST communicates with SUs. The RIS reflections enhance desired links and improve the performance of spectrum sensing and sharing.}\label{fig_RCN_SystemModel}\vspace{-5mm} 
\end{figure}

The RIS can enhance the performance of CRNs in several ways (Fig.~\ref{fig_RCN_SystemModel}), including \cite{Wu2021, Lin2022, Nasser2022, Saber2022, Ge2022, Zhang2020, He2020, Zhou2023, Wu2023, Srivastava2024}: (i) \textit{spectrum sensing enhancement}, by strengthening received PU signals at SUs and mitigating fading and hidden-node effects; (ii) \textit{interference management}, by shaping reflected signals to suppress interference toward PUs; (iii) \textit{coverage extension}, by establishing favorable propagation paths around blockages; (iv) \textit{SE improvement}, through spatial signal control and passive beamforming gains; (v) \textit{EE enhancement}, by reducing transmit power requirements via passive array gain; (vi) \textit{PLS}, by degrading unintended signal reception while strengthening desired links; and (vii) \textit{adaptive environment reconfiguration}, by dynamically adjusting the propagation environment in response to channel variations and PU activity. These gains stem from the additional spatial DoF that RIS provides: by adjusting the phase shifts of its reflecting elements, RIS coherently combines signals at the receiver without generating additional RF signals, improving sensing accuracy and communication reliability with lower complexity than the active sensing and complex cooperative strategies required in conventional CRNs.

\begin{figure*}[!t]\vspace{-2mm}
    \centering 
    \def\svgwidth{445pt} 
    \fontsize{8}{8}\selectfont 
    \graphicspath{{Figures/}}
    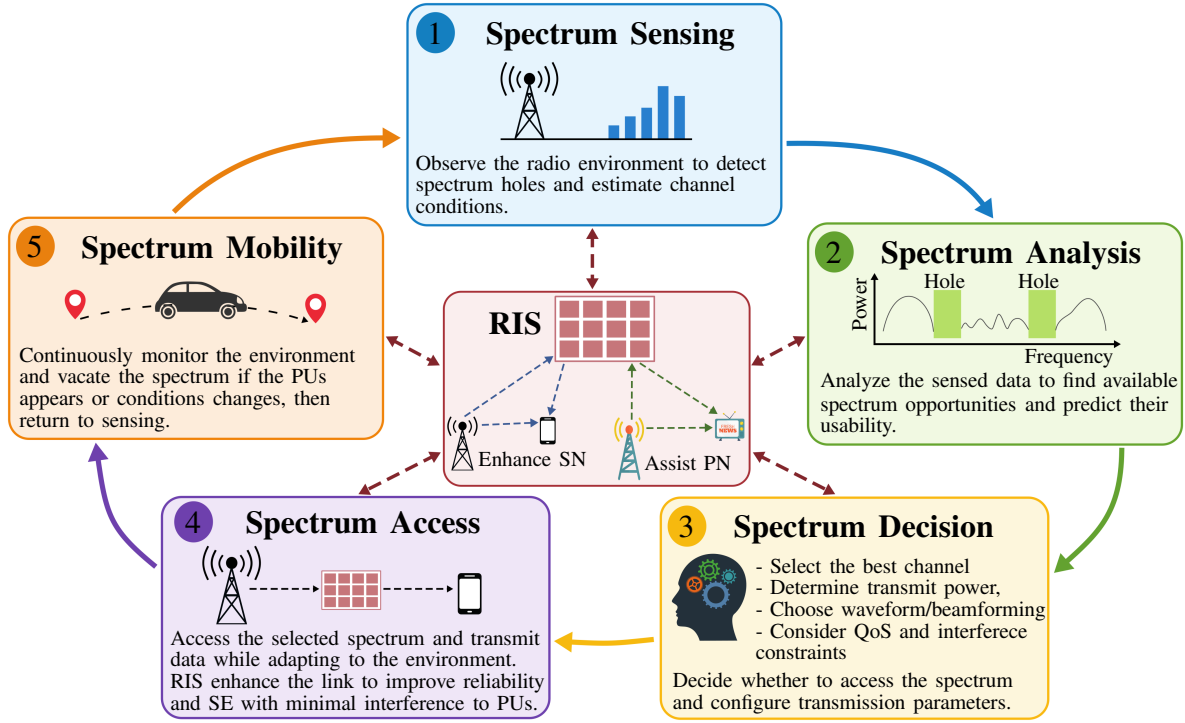 \vspace{-2mm}
    \caption{RCN concept.} \label{fig_RCNCycle} \vspace{-3mm}
\end{figure*}

These gains are especially valuable against the hidden-node (or primary transmitter (PT)) problem, where deep fading, shadowing, and blockage prevent SUs from reliably detecting PU signals, leading to harmful interference~\cite{Wu2021, Lin2022, Nasser2022, Saber2022, Ge2022, Zhang2020, He2020, Zhou2023, Wu2023, Srivastava2024}. By redirecting reflected signals toward SUs, strategically deployed RISs enhance signal detectability, reduce false alarms and missed detections, shorten the required sensing duration, and enable more efficient DSA (Fig.~\ref{fig_RCNCycle}). These benefits come at the cost of additional challenges, including complex channel estimation and the need for joint optimization of RIS and CRN parameters, discussed in the remainder of this section. Table~\ref{tab:RIS_CRN_metrics} summarizes the performance metrics most directly affected by RIS configuration; among these, detection probability, false-alarm probability, achievable rate/SE, and interference-related metrics are quantitatively evaluated through analytical and case-study results in Section~\ref{sec_state_art}, while channel estimation and RIS-configuration overheads are discussed further in Section~\ref{sec_channel_est} and Section~\ref{sec_future}.

\begin{table*}[t]
\centering
\caption{Critical performance metrics for RCNs.}
\label{tab:RIS_CRN_metrics}\vspace{-2mm}
\renewcommand{\arraystretch}{1.0}
\begin{tabular}{p{3cm}p{4.5cm}p{9cm}}
\hline
\textbf{Category} & \textbf{Metric} & \textbf{Description} \\ \hline \hline

\multirow{3}{*}{Spectrum sensing} 
& $P_d$ 
& Enhanced detection of PU signals via RIS-assisted signal reflection and constructive combining. \\ \cline{2-3}

& $P_{fa}$ 
& Influenced by RIS configuration, improper phase design may increase false alarms. \\ \cline{2-3}

& $P_m$ 
& Reduced through RIS-enhanced received signal strength and spatial diversity. \\ \hline

\multirow{2}{*}{SE} 
& Throughput/achievable rate 
& Improved via RIS-enabled passive beamforming and interference shaping. \\ \cline{2-3}

& Spectrum utilization efficiency 
& Higher utilization due to more reliable sensing and reduced unnecessary channel vacating. \\ \hline

\multirow{2}{*}{Interference management} 
& Interference to PUs 
& Controlled via RIS-based interference nulling or redirection. \\ \cline{2-3}

& Collision probability 
& Reduced due to improved sensing and predictive environment awareness. \\ \hline

\multirow{2}{*}{QoS} 
& Outage probability 
& Reduced via RIS-enhanced link reliability and diversity. \\ \cline{2-3}

& Reliability/SINR 
& Improved effective SINR through passive beamforming. \\ \hline

EE 
& EE (bits/Joule) 
& Improved due to the passive nature of RIS (no RF chains). \\ \hline

\multirow{2}{*}{\shortstack[l]{Channel estimation\\\& configuration}}
& CSI acquisition overhead 
& Increased due to cascaded BS-RIS-user channel estimation (Section~\ref{sec_channel_est}). \\ \cline{2-3}

& Reconfiguration latency/quantization error 
& Overhead from updating RIS phase shifts and discrete phase resolution. \\ \hline

\end{tabular}\vspace{-5mm}
\end{table*}

\subsection{Channel Estimation}\label{sec_channel_est}
Accurate channel estimation is essential to fully harness the benefits of RIS, as it enables precise beamforming and phase adaptation \cite{Yu2023, Swindlehurst2022}. In RCNs, multiple channels must be estimated, including the PT/secondary transmitter (ST)-to-RIS channels, the RIS-to-PR/SR  channels, and the direct PT/ST-to-PR/SR channels. While multiple works treat channel estimation in RIS networks (i.e., with a single network or PN), channel estimation in CRNs poses additional challenges due to the coexistence of two networks (primary and secondary) sharing the same spectrum resources \cite{Tarek2017}. Conversely, the passive nature of RIS transmission also complicates individual link/channel estimation. A potential solution is to adopt cascade channel estimation methods, in which the combined effect of the transmitter-RIS and RIS-receiver links is estimated jointly rather than separately \cite{Yu2023, Swindlehurst2022}. However, this method incurs significant training overhead, particularly for large-scale RIS arrays with numerous reflecting elements \cite{Yu2023, Swindlehurst2022}.

In addition, hardware limitations and limited cooperation between primary and secondary systems pose major challenges for channel estimation in RCNs \cite{Yu2023, Swindlehurst2022, Tarek2017}. One viable alternative is to maximize the average SNR using statistical CSI, with RIS configurations adjusted based on long-term fading statistics rather than real-time estimation \cite{Gunasinghe2024a, Gunasinghe2024b}. Additionally, ML-based approaches, such as DL and reinforcement learning, can be leveraged to improve estimation accuracy while reducing training overhead \cite{Yu2023}.

\subsection{Spectrum Access in RCNs}
In conventional CRNs (i.e., without RISs), spectrum access is typically classified into interweave, underlay, and overlay paradigms (Section~\ref{sec_DSA}). The integration of RIS into CRNs can significantly enhance these spectrum access models. In the interweave mode, RIS can improve the reliability of spectrum sensing by redirecting primary signals toward SRs (users), thereby eliminating the hidden node problem and enhancing detection accuracy \cite{Zhang2020, He2020, Zhou2023, Wu2023, Srivastava2024}. Furthermore, RIS can enable intelligent spectrum-aware beamforming to ensure SUs opportunistically access the spectrum without interfering with primary transmission \cite{Zhang2020, He2020, Zhou2023, Wu2023, Srivastava2024}. In underlay mode, RIS can support interference-aware beamforming, i.e., it can steer secondary transmitted signals away from PRs, enabling coexistence under stringent interference constraints \cite{Zhang2020, He2020, Zhou2023, Wu2023, Srivastava2024}. This extends the feasibility of underlay CRNs to denser environments. In the overlay mode, RIS can intelligently relay primary signals toward SRs/users, enabling efficient overlay communication without requiring complex cooperative relays \cite{Zhang2020, He2020, Zhou2023, Wu2023, Srivastava2024}.

\subsection{RIS Deployment and Multiple RISs}
The deployment of RIS in CRNs is crucial for enhancing SE, coverage, and interference management. In particular, the RIS should be strategically placed with high visibility to the PT to ensure optimal reflection and redirection of signals for spectrum sensing \cite{Yuan2021, Asiedu2023, Ge2024}. Also, it must cover as many SUs as possible. Thus, the optimal placement of RIS depends on several factors, including the operating environment, the position of the PT, and the spatial distribution of SUs.

On the other hand, in practice, a single RIS may not provide adequate coverage, especially in dense urban environments with severe multipath fading and obstructions \cite{Yuan2021, Asiedu2023, Ge2024}. Deploying multiple RISs is thus essential to extend coverage, enhance spectrum sensing and communication reliability, and support a larger number of users. However, multiple RISs operating in the vicinity introduce several technical challenges, including inter-RIS interference and synchronization among RISs to avoid phase misalignment. Moreover, joint optimization of RIS parameters is required to enable cooperative transmission, thereby minimizing destructive interference and maximizing overall system performance \cite{Yuan2021, Asiedu2023, Ge2024}. This requires control channels to enable information exchange between the RISs and the CR systems \cite{Yuan2021, Asiedu2023, Ge2024}.

\subsection{Architecture and Protocols}
A key advantage of RIS is its ability to dynamically reconfigure its reflection properties based on network requirements \cite{Gong2020, Chen2022}. Integrating RIS into communication systems, including CRNs, requires dedicated control channels between the network entities and the RIS, as well as protocol adaptations. However, this integration does not require significant hardware modifications, making RIS a cost-effective solution for improving SE and interference management \cite{Gong2020, Chen2022}.

In conventional systems (a single network), integrating a RIS typically requires two dedicated control channels \cite{Liu2021RIS, Gong2020, Chen2022}, i.e., one for configuring the RIS parameters, such as reflection coefficients and phase shifts, to optimize signal reception at the receiver, and the other for adapting the transmission features of the BS based on the target data rate and QoS requirements of the receiver \cite{Liu2021RIS, Gong2020, Chen2022}. However, in RCNs, additional control channels may be required to accommodate both primary and secondary systems. If the RIS assists only the SN, two control links are generally sufficient, as in conventional networks. However, when the RIS interacts with both primary and secondary systems, more complex control signaling is required to dynamically adjust RIS operation in the presence of both networks \cite{Zhang2020, He2020, Zhou2023, Wu2023}.

Several studies investigate transmission and routing protocols to integrate RIS into existing wireless networks. For instance, reference \cite{Yang2020} proposes a transmission protocol for RIS-assisted orthogonal frequency division multiplexing (OFDM) systems to reduce the overhead associated with channel training. The approach groups adjacent RIS elements and estimates the combined channels for each group, thereby reducing training overhead. In \cite{Wang2020}, a three-phase channel estimation protocol is introduced to minimize channel estimation time in RIS-assisted systems. For mobile ad hoc networks, \cite{Sadreddini2021} presents an RIS-assisted routing protocol in which RIS operates as an intelligent node, improving routing efficiency, minimizing delay, and mitigating interference.

Despite these advancements in RIS-enabled communication protocols, RCNs introduce unique challenges that require further investigation. Protocols designed for RCNs must account for the dynamic nature of spectrum access in CRNs and the coexistence of PUs and SUs. Specifically, RIS must be aware of the SN's sensing period to determine when to assist spectrum sensing and when to enhance secondary transmissions \cite{Zhang2020, He2020, Zhou2023, Wu2023, Srivastava2024}. This necessitates the development of adaptive exchange protocols that dynamically adjust RIS functions based on real-time spectrum occupancy \cite{Zhang2020, He2020, Zhou2023, Wu2023, Srivastava2024}. Moreover, the network architecture must be modified to accommodate RIS control signaling, particularly when multiple RISs operate cooperatively.

\subsection{Joint Optimization of RIS and CRN Parameters}
To fully leverage RIS in CRNs, joint optimization of both primary and secondary system parameters, such as transmit beamforming and power control, as well as RIS parameters, such as phase shifts (or beamforming), is essential \cite{Zhang2020, He2020, Zhou2023, Wu2023, Srivastava2024, Yuan2021, Asiedu2023, Ge2024}.  Optimizing the RIS phase shifts in CRNs is particularly challenging, as it requires a careful balance with transmission power levels in both PN and SN to enhance throughput while adhering to interference constraints \cite{Zhang2020, He2020, Zhou2023, Wu2023, Srivastava2024, Yuan2021, Asiedu2023, Ge2024}. Furthermore, the presence of multiple PUs and SUs introduces additional complexity, requiring efficient resource allocation strategies to maintain the required QoS for all users. For example, when an RIS assists multiple SUs with spectrum sensing and data transmission with different specifications, it must allocate resources to each user by considering several objectives, such as minimizing sensing time, improving SE, and reducing interference with primary transmission \cite{Zhang2020, He2020, Zhou2023, Wu2023, Srivastava2024, Yuan2021, Asiedu2023, Ge2024}. This necessitates sophisticated optimization techniques that can dynamically adapt to real-time network conditions and user demands.

To address these challenges, advanced algorithms, such as convex optimization and deep reinforcement learning (DRL), can be used to obtain optimal configurations \cite{Khalek2024, Kayraklik2024}. These optimization frameworks must account for several factors, including path loss, multipath effects, and user mobility. Additionally, adaptive reconfiguration of RIS parameters based on real-time traffic demands and interference conditions can further enhance CRN efficiency \cite{Khalek2024, Kayraklik2024}.

\section{State of the Art in RIS-Assisted CRNs}\label{sec_state_art}
This section reviews the existing RCN literature under several categories, including performance analysis, resource allocation and parameter optimization, security, active and STAR RISs, and ML techniques. 

\subsection{Performance Analysis} Performance analysis helps evaluate the impact of RIS on spectrum sensing, detection accuracy, and overall network efficiency. Key metrics, such as $\pd$ and $\pfa$, quantify performance gains and provide insights for optimizing system design.

Recent works have established the significant potential of RIS to enhance spectrum sensing in CRNs by enabling programmable control of the wireless propagation environment \cite{Wu2021, Lin2022, Nasser2022, Saber2022, Ge2022}. In particular,  energy detection schemes have been developed for single-user and CSS, both with and without a direct link between PUs and SUs \cite{Wu2021}. This work shows that optimized phase shifts can coherently enhance the received PU signal and substantially improve $\pd$ at low SNR. Closed-form expressions for $\pd$ are derived using the Gamma distribution approximation and the central limit theorem (CLT). Additionally, energy-detection-based CSS with multiple SUs and multiple RIS-enhanced square-law selection diversity reception is also proposed. The average probability of detection for these schemes is derived using the $K$-rank fusion and square-law selection criteria, respectively.

A comprehensive framework for RIS-assisted spectrum sensing is developed in \cite{Lin2022}, where system-level design and RIS phase optimization are jointly considered to enhance sensing performance across CRN scenarios. Focusing on a single-PU, single-SU setup, the RIS is leveraged to improve codebook-based spectrum sensing by exploiting its large aperture and passive beamforming gains to strengthen the PU signal at the SU. To minimize the missed-detection probability under a fixed false-alarm constraint, a weighted energy-detection scheme is proposed that combines the received signal power across multiple RIS reflection patterns. The results demonstrate that optimized RIS configurations significantly improve detection accuracy while reducing false-alarm rates.

In \cite{Nasser2022}, the impact of RIS deployment on $\pd$ is further investigated for a CRN with a single PU and a single SU. Two cases are analyzed: RIS assisting the SU or the PU. Closed-form expressions for $\pd$ are derived for both cases, showing substantially enhanced sensing performance, even when the RIS is configured to assist the PU.

Reference \cite{Saber2022} investigates an RCN with a single SU and derives analytical expressions for the false alarm probability $P_{fa}$, detection probability $P_d$, and achievable rate, thereby quantifying the impact of RIS on SE.  The key idea is that the RIS intelligently adjusts its reflection coefficients to enhance the received PU signal strength at the SU, thereby improving sensing reliability under weak-signal or fading conditions. The paper analyzes the received sensing signal in RIS-assisted environments and evaluates the impact of RIS on classical energy-detection-based spectrum sensing.

The paper shows through simulations that RIS significantly improves $\pd$ while reducing the probability of missed detection, particularly at low SNR. They also show that increasing the number of RIS elements improves sensing performance by enhancing passive beamforming gain. In addition, the paper highlights that RIS-assisted sensing can mitigate hidden-node problems and extend sensing coverage without requiring additional active transmission power. Overall, the paper positions RIS as a promising low-power technology for improving the accuracy and reliability of spectrum sensing in future CRNs.

The work in \cite{Ge2022} provides a fundamental theoretical analysis of RIS-assisted spectrum sensing by determining the number of reflecting elements required for near-perfect detection probability in a CRN with a single-antenna PU and a multi-antenna SU. Specifically, it characterizes the scaling behavior of RIS-assisted sensing and establishes the relationship between RIS size and sensing reliability, offering key design insights. In the asymptotic regime, the detection probability is derived using a maximum-eigenvalue-based detector and random matrix theory. Furthermore, the RIS configuration relies on statistical CSI, eliminating the need for real-time channel estimation.

Table~\ref{tab_performance_summary} summarizes several works on performance analysis in RCNS. Overall, these studies consistently show that RIS transforms spectrum sensing from a passive detection task into an actively controllable process. By enabling adaptive propagation shaping, RCNs substantially improve detection accuracy, robustness, and SE, particularly in low-SNR and propagation-impaired environments.

\begin{table*}[htbp]\vspace{-2mm}
\centering
\begin{threeparttable}
\renewcommand{\arraystretch}{1.0}
\caption{Summary of performance analysis literature in RCNs.\vspace{-2mm}}
\label{tab_performance_summary}
\begin{tabular}{llllp{2.5cm}p{10cm}}
\hline
\textbf{Ref.} & \textbf{PUs} & \textbf{SUs} & \textbf{RISs} & \textbf{Performance Metric} & \textbf{Key Contribution} \\  \hline \hline

\cite{Wu2021}  
& 1 & 1, M & 1, M 
& $P_d$ 
& Energy detection schemes for single-SU and multi-SU CSS \\  \hline

\cite{Lin2022} 
& 1 & 1 & 1 
& $P_m$ and $P_{fa}$ 
& RIS-enhanced codebook-based spectrum sensing using weighted energy detection \\  \hline

\cite{Nasser2022} 
& 1 & 1 & 1 
& $P_d$ 
& Impact of RIS on spectrum sensing when assisting SU versus PU \\ \hline

\cite{Saber2022} 
& 1 & 1 & 1 
& $P_{fa}$, $P_d$, and rate 
& The SE improvement through RIS in CRNs \\  \hline

\cite{Ge2022} 
& 1 & 1 (MA) & 1 
& $P_d$  
& Theoretical analysis of RIS  for near-optimal detection with eigenvalue detection  \\  \hline

\end{tabular}
\begin{tablenotes}
      \scriptsize{
      \item M - Multiple, \quad  MA - Multi-antenna.}
    \end{tablenotes}
\end{threeparttable}\vspace{-5mm}
\end{table*}

\textit{Case Study 1:} 
A brief performance analysis of RCNs is provided. 

\begin{figure}[!t]\vspace{-2mm}
\centering
    \def\svgwidth{220pt} 
    \fontsize{8}{8}\selectfont 
    \graphicspath{{Figures/}}
    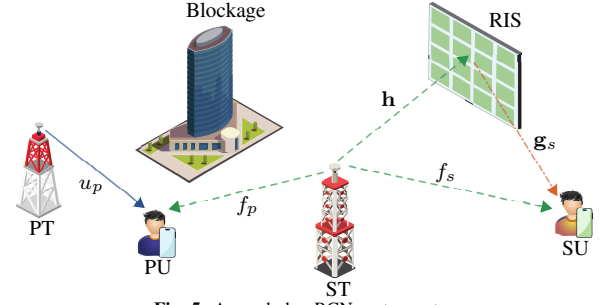 \vspace{-2mm}
    \caption{An underlay RCN system setup.}\vspace{-5mm} \label{fig_PA_SystemModel}
\end{figure}

The underlay RCN comprises a single-antenna PT, a single-antenna ST, a single-antenna PU, a single-antenna SU, and an RIS with $L$ reflecting elements (Fig.~\ref{fig_PA_SystemModel}). The channels between the RIS and the PU, and between SU and PT, are assumed to be unavailable due to severe blockage. The channel from ST to the $n$-th reflective element is denoted by $h_n$, while the channel from the $n$-th reflective element to SU is given by $g_{s,n}$. Moreover, the direct channels from ST-to-SU, ST-and-PU, and PT-to-PU are denoted as $f_s$, $f_p$, and $u_p$, respectively. All these channels are assumed to be independent Rayleigh-distributed, and hence they can be modeled as $v = \sqrt{\zeta_v} \tilde{v}$, where $v \in \{h_n, g_{s, n}, f_s, f_p, u_p\}$, $\tilde{v}$ follows a complex Gaussian distribution with zero mean and unit variance, i.e., $\mathcal{CN}(0, 1)$, and $\zeta_v$ captures the path-loss of the channel $v$. Moreover, the channel $v$ can also be represented as $v = \alpha_{v} e^{j \theta_{v}}$, where $\alpha_{v} = |v|$ is the channel amplitude with Rayleigh distribution and $\theta_{v}$ is the corresponding channel phase, which are uniformly distributed between $(-\pi, \pi]$.

We let the channel from ST-to-RIS, and RIS-to-SU as $\q{h} = [h_1, \cdots, h_L]^{\rm T}$ and $\q{g}_s = [g_{s,1}, \cdots, g_{s,L}]$, respectively. Thus, the effective channel from ST to SU is given by $f_s + \q{g}_s \boldsymbol{\Phi} \q{h}$, where $\boldsymbol{\Phi} = \mathrm{diag}([\eta_1 e^{j \phi_1}, \cdots, \eta_n e^{j \phi_n}, \cdots, \eta_L e^{j \phi_L}])$ is the reflection matrix of RIS, which captures the reflection properties of the $L$ reflective elements of the RIS. Here, $\eta_n$ and $\phi_n$ are the reflection coefficient and the phase shift introduced by the $n$-th reflective element of the RIS, respectively. This formulation is general and applies to both passive and active RIS architectures; for passive RIS, $0 \leq \eta_n \leq 1$, whereas for active RIS, $\eta_n > 1$ due to signal amplification \cite{Nguyen2025}.

The signal transmitted at ST is reflected by the RIS towards SU. The received signal at SU through $L$ reflective elements can be written as $y_s = \sqrt{P_{s}} (f_s + \q{g}_s \boldsymbol{\Phi} \q{h}) x_s + n_s = \sqrt{P_{s}} (f_s + \sum_{n = 1}^{L} \eta_n e^{j \phi_n} g_{s, n} h_n) x_s + n_s$, where $P_s$ is the maximum transmit power at ST, $x_s$ is the transmitted symbol by ST satisfying $\E{|x_s|^2} = 1$, and $n_s \sim \mathcal{CN}(0, \sigma^2)$. Then, the SNR at SU is given as 
\begin{eqnarray} \label{eqn_SNR_s}
     \Gamma_s &=& \bar{\gamma}_s | f_s + \q{g}_s \boldsymbol{\Phi} \q{h} |^2, \nonumber\\
      &=& \bar{\gamma}_s \left| \alpha_{f_s} e^{j \theta_{f_s}} + \sum\nolimits_{n = 1}^{L} \eta_n \alpha_{h_n} \alpha_{g_{s, n}} e^{j (\phi_n + \theta_{h_n} + \theta_{g_{s, n}})} \right|^2\!\!, \quad
\end{eqnarray}
where $\bar{\gamma}_s = P_s / \sigma^2$ is the average transmit SNR at ST. The optimal choice of $\phi_n$ can be given by $\underset{-\pi < \phi_n \le \pi}{\arg\max}~\Gamma_s = \phi_n^\star  = \theta_{f_s} - (\theta_{h_n} + \theta_{g_{s, n}})$ \cite{Qingqing2019, Diluka2020dRIS}, and thereby, the optimal received SNR at SU can be written as
\begin{eqnarray} \label{eqn_SNR_s_final}
     \Gamma_s^\star = \bar{\gamma}_s \left| \alpha_{f_s} + \sum\nolimits_{n = 1}^{L} \eta_n \alpha_{h_n} \alpha_{g_{s, n}} \right|^2.
\end{eqnarray}

On the other hand, the received signal at PU can be written as $y_p = \sqrt{P_{p}} u_p x_p + \sqrt{P_{s}} f_p x_s + n_p$, where $P_p$ is the maximum transmit power at PT, $x_p$ is the transmitted symbol by PT satisfying $\E{|x_p|^2} = 1$, and $n_p \sim \mathcal{CN}(0, \sigma^2)$. The SINR at PU is thus given by
\begin{eqnarray} \label{eqn_SINR_p}
     \Gamma_p = \frac{\bar{\gamma}_p |u_p|^2}{\bar{\gamma}_s |f_p|^2 + 1} = \frac{\bar{\gamma}_p |\alpha_{u_p}|^2}{\bar{\gamma}_s |\alpha_{f_p}|^2 + 1},
\end{eqnarray}
where $\bar{\gamma}_p = P_p / \sigma^2$.

Since all $\alpha_v$ are independently Rayleigh distributed RVs, by using the CLT, $X = \sum_{n = 1}^{L} \eta_n \alpha_{h_n} \alpha_{g_{s, n}}$ converges to a Gaussian distribution for a sufficiently large number of passive elements in the RIS \cite{papoulis2002probability, Diluka2020dRIS}. Thereby, the distribution of $|Y|^2 = \big| \sum_{n = 1}^{L} \eta_n \alpha_{h_n} \alpha_{g_{s, n}} \big|^2$ can be tightly approximated by a non-central chi-squared distribution with a single DoF \cite{papoulis2002probability, Diluka2020dRIS}, i.e., $|Y|^2 \sim \mathcal{X}_1^2(u)$, where $u$ is the non-centrality parameter. Then, the average communication rate at SU is defined as
\begin{eqnarray} \label{eqn_SU_rate}
     \mathcal{R}_s = \E{\log_2(1 + \Gamma_s^*} \approx \log_2(1 + \E{\Gamma_s^*}),
\end{eqnarray}
where $\E{\Gamma_s^*}$ is given in \eqref{eqn_E_SNR} \cite{Diluka2020dRIS},
\begin{figure*}[t!]
    \begin{eqnarray} \label{eqn_E_SNR}
     \E{\Gamma_s^\star} &=& \bar{\gamma}_s \left(\E{| \alpha_{f_s} |^2} +  \E{\left|\sum\nolimits_{n = 1}^{L} \eta_n \alpha_{h_n} \alpha_{g_{s, n}} \right|^2} + 2 \E{\alpha_{f_s}} \sum\nolimits_{n = 1}^{L} \eta_n \E{\alpha_{h_n}} \E{\alpha_{g_{s, n}}} \right) \nonumber\\
     &=& \bar{\gamma}_s \left(\zeta_{f_s} + \frac{(16 - \pi^2) (1 + \kappa)}{4} \sum\nolimits_{n = 1}^{L} \lambda_n^2 + \sqrt{\pi}^3 \zeta_{f_s} \sum\nolimits_{n = 1}^{L} \frac{\lambda_n^2}{\eta_n} \right)
\end{eqnarray}

\vspace{-2mm}

\hrulefill

\vspace{-5mm}

\end{figure*}
where $\lambda_n^2 = \zeta_{h_n} \zeta_{g_{s,n}} \eta_n^2 / 4$, $\kappa = \mu_Y^2 / \sigma_Y^2$, $\mu_Y = \sum_{n = 1}^{L} \pi \lambda_n / 2$, and $\sigma_Y^2 = \sum_{n = 1}^{L} \lambda_n^2 (16 - \pi^2) / 4$. Similarly, the communication rate at PU is given by
\begin{eqnarray} \label{eqn_PU_rate}
     \mathcal{R}_p = \E{\log_2(1 + \Gamma_p} \approx \log_2(1 + \E{\Gamma_p}),
\end{eqnarray}
where $\E{\Gamma_p}$ is given as
\begin{eqnarray} \label{eqn_E_SINR}
     \E{\Gamma_p} &\approx& \frac{\bar{\gamma}_p \E{|\alpha_{u_p}|^2}}{\bar{\gamma}_s \E{|\alpha_{f_p}|^2} + 1} = \frac{\bar{\gamma}_p \zeta_{u_p}}{\bar{\gamma}_s \zeta_{f_p} + 1}.
\end{eqnarray} 

\begin{figure}[!t]\vspace{-2mm}	
\centering
    \includegraphics[width=0.40\textwidth]{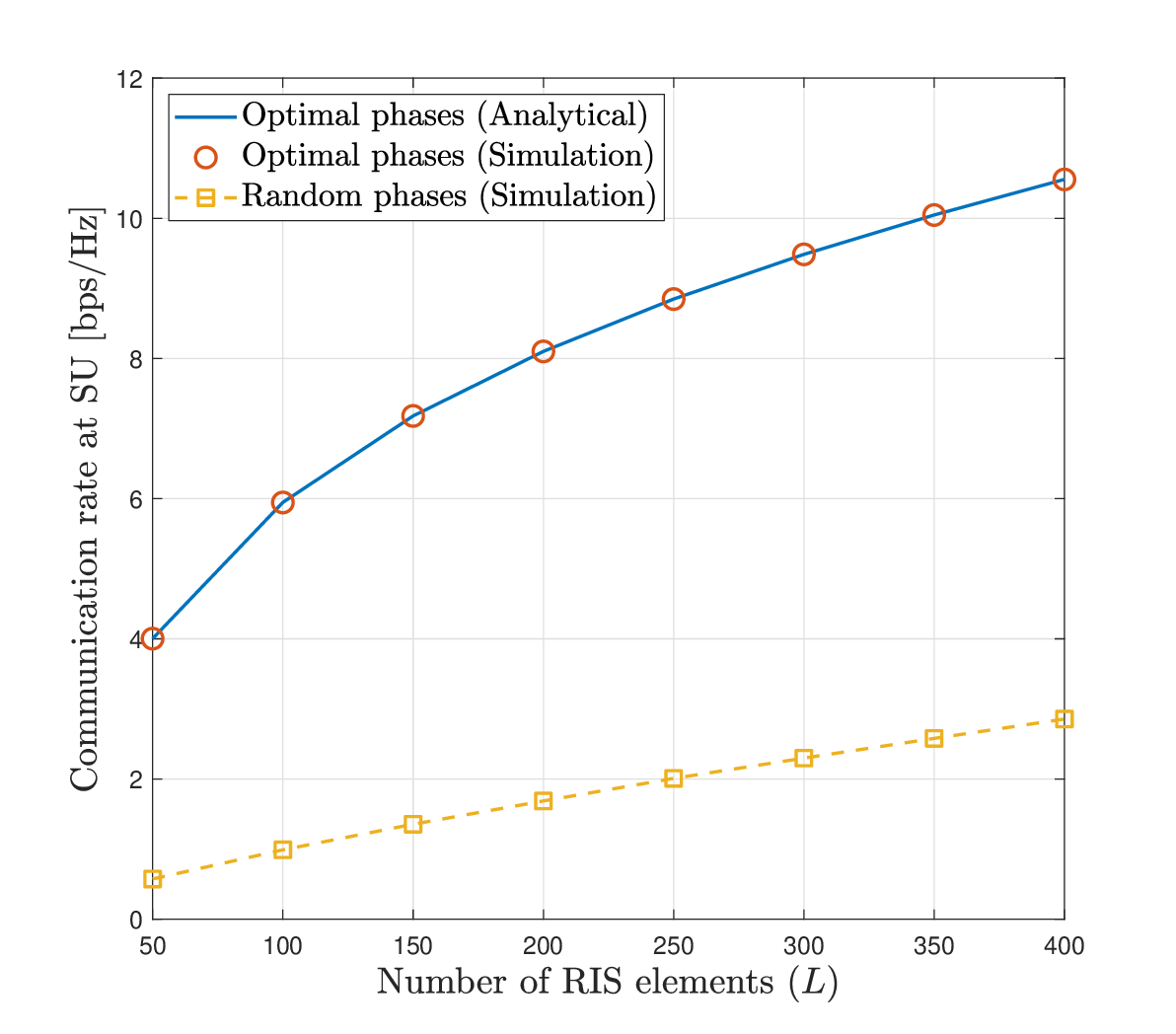}\vspace{-3mm}
\caption{The achievable rate at the SU as a function of the RIS size.} \label{fig_SU_rate_RIS_elements}\vspace{-5mm}
\end{figure}

For the numerical evaluation, all wireless links are modeled as independent Rayleigh fading channels. The large-scale fading is generated according to the 3GPP Urban Micro (UMi) model at a carrier frequency of $f_c=\qty{3}{\GHz}$ \cite[Table B.1.2.1]{3GPP2010}. The noise power is computed as $\sigma^2=10\log_{10}(N_0BN_f)\,\qty{}{\dB m}$, where $N_0=\qty{-174}{\dB m/\Hz}$ is the thermal noise power spectral density, $B=\qty{10}{\MHz}$ is the system bandwidth, and $N_f=\qty{10}{\dB}$ is the receiver noise figure. Furthermore,  a passive RIS ($\eta_n = 1$) is utilized.

Fig.~\ref{fig_SU_rate_RIS_elements} depicts the SU data rate as a function of the number of RIS elements, $L$, under both random and optimal phases, \eqref{eqn_SNR_s}. The data rate increases with $L$ in both cases, since a larger RIS provides additional spatial DoF; under optimal phase control, increasing $L$ from $\num{50}$ to $\num{400}$ yields a \qty{163.66}{\percent} improvement in the SU rate. A significant gap also exists between the two configurations: at $L=\num{200}$, the SU rate with optimal phases is \qty{380}{\percent} higher than with random phases, underscoring the importance of proper RIS phase design.


\begin{figure}[!t]\vspace{-0mm}	
\centering
    \includegraphics[width=0.40\textwidth]{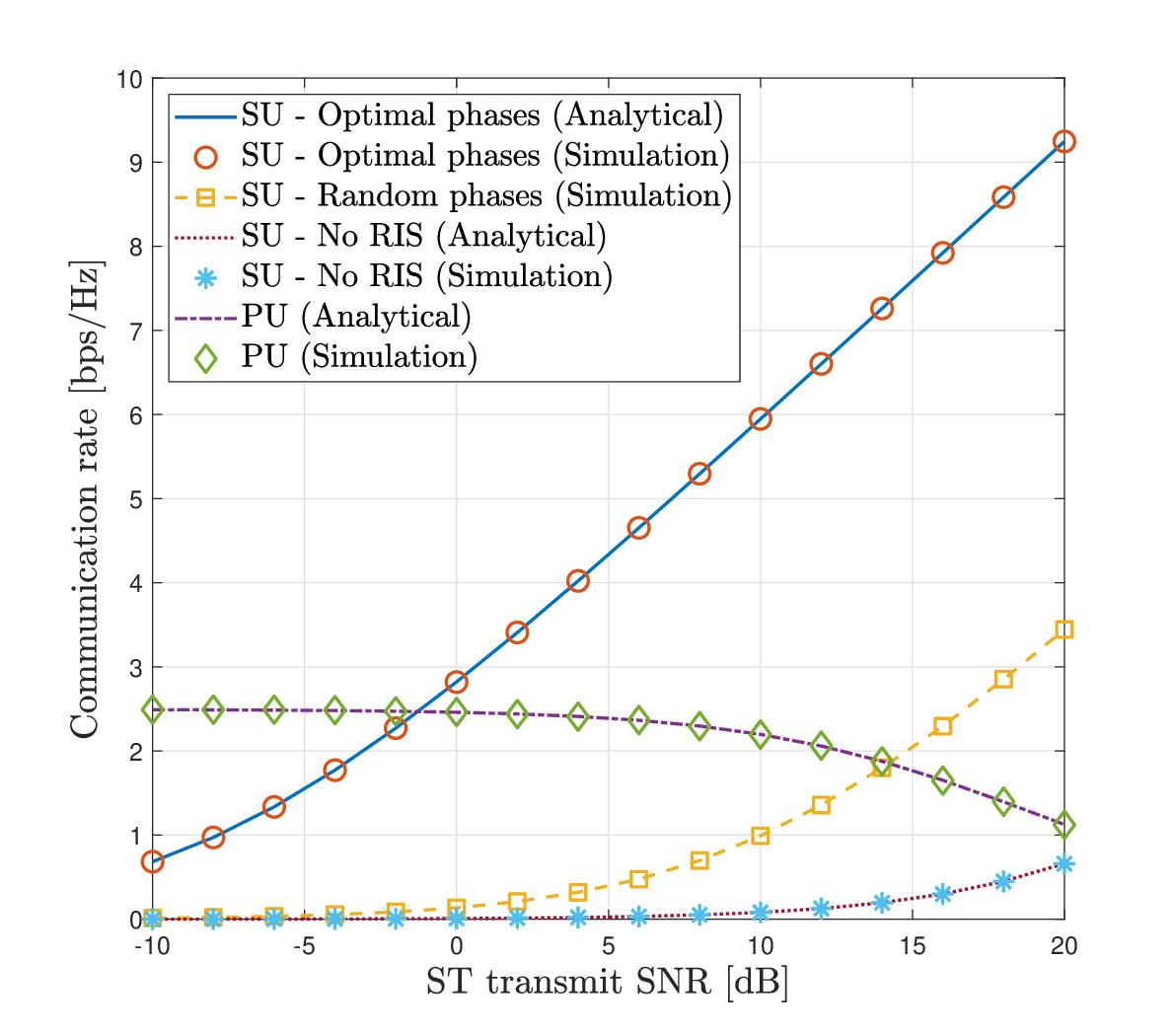}\vspace{-3mm}
\caption{The achievable rate at the SU and PU as a function of the ST's average transmit SNR.} \label{fig_rate_ST_transmit_SNR}\vspace{-5mm}
\end{figure}

Fig.~\ref{fig_rate_ST_transmit_SNR} shows the PU and SU rates versus the ST's average transmit SNR. Increasing the ST transmit SNR raises the desired signal power at the SU but also increases interference at the PU, so the secondary rate improves at the expense of the primary rate. For example, raising the ST transmit SNR from \qty{-10}{\dB} to \qty{20}{\dB} increases the SU rate by \qty{1251.19}{\percent}, while the PU rate drops by only \qty{54.93}{\percent}. At $\gamma_S =  \qty{10}{\dB}$, the SU rate with optimal phases exceeds the random-phase and no-RIS cases by \qty{499.74}{\percent} and \qty{7219.93}{\percent}, respectively.


\begin{figure}[!t]\vspace{-2mm}	
\centering
    \includegraphics[width=0.40\textwidth]{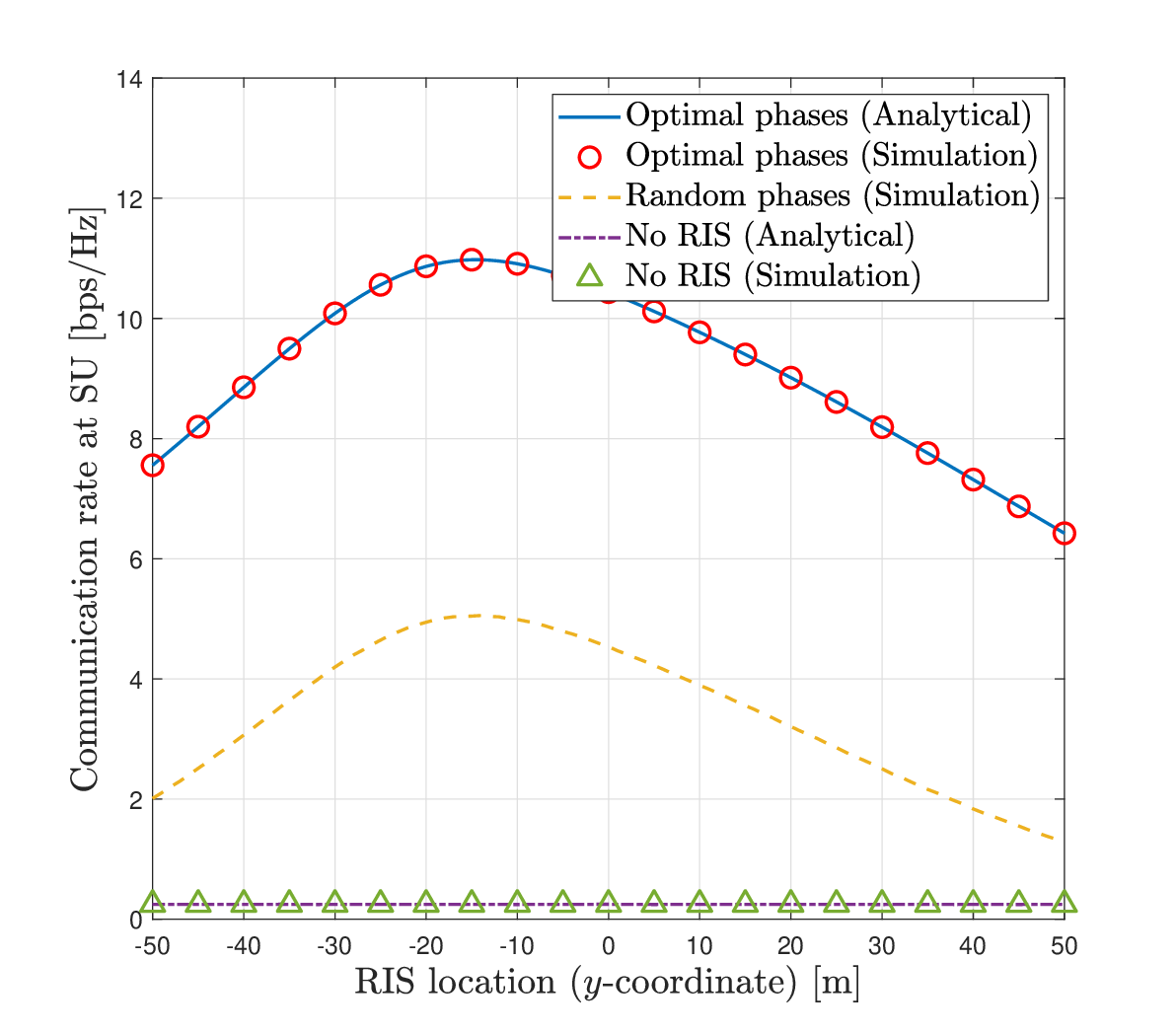}\vspace{-3mm}
\caption{The achievable rate at the SU as a function of the RIS location along the $ y$-coordinate.} \label{fig_case_1_secondary_rate_vs_RIS_location}\vspace{-5mm}
\end{figure}

Fig.~\ref{fig_case_1_secondary_rate_vs_RIS_location} evaluates the impact of RIS location by varying its position along the $y$-axis (with $x=0$), with the ST and SU fixed at $(50, 20)$ and $(-20, -20)$, respectively. Optimal phases consistently outperform random phases across all locations (e.g., a \qty{227.01}{\percent} gain at $y=\qty{30}{\m}$), and the SU rate peaks at \qty{10.97}{bps/\Hz} when the RIS is close to the SU (around $(0,-15)$), where reduced RIS-SU path loss allows the reflected and direct signals to combine constructively. As the RIS moves away from this region, path loss and phase misalignment gradually reduce the achievable rate. In contrast, the no-RIS case is unaffected by RIS placement, since it relies solely on the direct ST-SU link; at $y=30$, RIS deployment still improves the SU rate by roughly \qty{3216}{\percent} (optimal phases) or \qty{914}{\percent} (random phases) over the no-RIS case. These results confirm that RIS placement, not just phase configuration, is critical to realizing performance gains, motivating joint optimization of location, phase shifts, and transmission parameters in practical deployments.

Overall, RISs are particularly effective at enhancing secondary system performance under limited-power, underlay spectrum-sharing conditions, without imposing a significant performance penalty on the primary system, offering a robust solution to large-scale fading, path loss, blockage, and power limitations in CRNs.

\begin{mybox}[frametitle={Key Insights}]
\begin{enumerate}[leftmargin=*]

\item \textit{RIS size and phase design drive SU gains:} The SU rate increases monotonically with the number of RIS elements $L$ and depends critically on accurate phase configuration, with the gap between optimal and random phases widening as $L$ and SNR increase (Figs.~\ref{fig_SU_rate_RIS_elements}--\ref{fig_rate_ST_transmit_SNR}).

\item \textit{RIS balances SU performance and PU protection:} Increasing ST transmit SNR substantially enhances SU performance while only moderately degrading PU performance, showing that RCNs can balance secondary throughput against primary protection under interference constraints (Fig.~\ref{fig_rate_ST_transmit_SNR}).

\item \textit{RIS deployment location critically impacts performance:} The SU rate is highly sensitive to RIS placement, peaking when the RIS balances the cascaded ST-RIS and RIS-SU channel gains (typically near the SU); suboptimal placement significantly erodes gains, so location must be jointly optimized with phase configuration (Fig.~\ref{fig_case_1_secondary_rate_vs_RIS_location}).

\end{enumerate}
\end{mybox}

\subsection{Resource Allocation and Parameter Optimization}\label{sec_RA_pssive}
These are crucial in optimizing SE, power distribution, and signal enhancement while ensuring seamless coexistence between PN and SN. RCNs must dynamically adjust their resources to accommodate spectrum availability, interference constraints, and the  RIS elements. The unique combination of cognitive-spectrum access and RIS beamforming introduces new challenges in balancing signal enhancement for SUs while preserving PUs' performance. Standard resource allocation methods may not fully exploit the benefits of RIS, necessitating specialized optimization strategies that account for the interplay between reconfigurable reflections and cognitive network dynamics. To address these challenges, various resource allocation and parameter optimization schemes have been developed for both single-RIS scenarios \cite{Zhang2020, He2020, Zhou2023, Wu2023, Srivastava2024} and multiple-RIS scenarios \cite{Yuan2021, Asiedu2023, Ge2024}.

\subsubsection{Single RIS}
Reference \cite{Zhang2020} maximizes the weighted sum rate of SUs in an RCN with a single PU, multiple SUs, and a multi-antenna ST. Here, the RIS assists in the secondary transmission. This is achieved by jointly optimizing the transmit beamforming at the ST and the RIS phase shifts, subject to the ST's power constraint, the PU's interference temperature limit, and unit-modulus constraints on the RIS phase shifts. To solve this optimization problem, a block coordinate descent (BCD) algorithm is proposed that leverages Lagrange duality and an inner approximation method. Reference \cite{He2020} investigates a transmit power minimization problem in an RCN with multiple PUs, multiple SUs, and a multi-antenna ST. The RIS enhances secondary transmission in the absence of direct links between the ST and SUs. It jointly optimizes transmit beamforming at the ST and the RIS phase-shift matrix, while satisfying the SINR requirements for each SU, the PU's interference temperature constraint, and the unit-modulus constraints on the RIS phase shifts. To address this optimization problem, an alternating optimization (AO) algorithm is proposed that incorporates the difference-of-convex (DC) method and the semidefinite relaxation (SDR) technique. 

In \cite{Zhou2023}, a UAV serves as a mobile aerial BS to communicate with a single SU while mitigating potential spectrum interference for multiple PUs. A RIS is deployed to assist UAV-SU transmission in the absence of a direct link. To maximize its rate, the SU optimizes the UAV trajectory, transmit power, and RIS phase shifts subject to a given interference temperature threshold at the PUs. A BCD algorithm incorporating the successive convex approximation (SCA) technique is presented. The study \cite{Wu2023} employs an RIS to enhance both spectrum-sensing accuracy and secondary transmission in a CRN with multiple PUs, multiple SUs, a single-antenna PT, and a multi-antenna ST. For spectrum sensing, a novel detection threshold based on the false-alarm probability is derived. To maximize the SU sum rate, the transmit beamforming at the ST, time allocation for spectrum sensing, and RIS phase shifts are jointly optimized. The optimization is constrained by the maximum interference temperature at the PUs, the secondary transmit power limit, the required minimum probability of detection, the maximum tolerable probability of false alarm, and the RIS phase-shift constraints. To address the non-convex optimization problem, a computationally efficient BCD-based algorithm is proposed that utilizes SCA and SDR techniques. In \cite{Srivastava2024}, a RIS is deployed to enhance spectrum sensing, EH at the single-antenna ST, and data transmission in a non-orthogonal multiple access (NOMA)-enabled SN with two SUs, in the presence of a single PU. To maximize spectrum-sensing performance, the study optimizes RIS phase shifts to maximize the received primary-signal SNR at the ST, subject to RIS phase-shift constraints. A Grey Wolf Optimization (GWO) algorithm is proposed to achieve this objective.

\subsubsection{Multiple RISs}
Reference \cite{Yuan2021} investigates a CRN where multiple RISs assist downlink transmission from a multi-antenna ST to a single-antenna SU, coexisting with a PN of multiple PUs. The objective is to maximize the SU rate while satisfying the total transmit power constraint at the ST and interference temperature constraints at the PUs. This is achieved by jointly optimizing the ST's beamforming and the RIS reflecting coefficients. A BCD algorithm leveraging second-order cone programming (SOCP) and SDR methods is proposed for both perfect and imperfect CSI scenarios. In \cite{Asiedu2023}, a CRN in which NOMA-based PUs coexist with multiple RISs with backscattering capabilities, acting as SUs to share the same spectrum, is investigated. The RISs not only transmit their own data to the SR but also enhance the primary transmission, thereby emphasizing the symbiotic relationship between the PN and the SN. The sum rate of PUs and SUs is maximized by jointly optimizing the PT's power allocation for NOMA PUs and the RIS phase shifts, subject to RIS phase shift constraints, the primary transmit power constraint, and minimum-rate requirements for both PUs and SUs. To solve the formulated problem, an iterative algorithm is proposed that transforms it into a convex weighted MMSE problem. 

Reference \cite{Ge2024} studies a multiple RIS CRN  with a single PU and multiple SUs, where multiple  RISs enhance CSS within a limited sensing time at the SUs. The optimization maximizes cooperative $\pd$ at the SUs while ensuring the maximum tolerable $\pfa$ by optimizing the phase shifts at each RIS under two CSS schemes: data fusion and decision fusion. More practical scenarios without instantaneous CSI are also considered.  In both cases, the optimal RIS phase shifts can be obtained numerically.

Table~\ref{tab_resource_allocation} gives a summary of the state of the art. 

\begin{table*}[htbp]\vspace{-2mm}
\centering
\begin{threeparttable}
\renewcommand{\arraystretch}{1.0}
\caption{Summary of resource allocation and parameter optimization in RCNs.\vspace{-2mm}}
\label{tab_resource_allocation}
\begin{tabular}{llllp{2cm}p{1.6cm}p{3cm}p{2cm}p{3.8cm}}
\hline
\textbf{Ref.} & \textbf{PUs} & \textbf{SUs} & \textbf{RIS} & \textbf{RIS Role} & \textbf{Objective} & \textbf{Constraints} & \textbf{Method} & \textbf{Key Contribution} \\  \hline \hline

\cite{Zhang2020}  
& 1  & M  & 1          
& Assist secondary transmission 
& WSR maximization     
& Power constraints, PU IT, RIS PS modulus 
& BCD, Lagrange duality,  inner approximation  
& Maximizes SU SR by optimizing BF and RIS PS    \\ \hline

\cite{He2020}     
& M  & M  & 1          
& Assist secondary transmission   
& Transmit power minimization    
& SUs's SINR requirements, PU IT, RIS PS modulus 
& AO, DC, SDR 
& Minimizes transmit power by optimizing transmit BF and RIS PS \\  \hline

\cite{Zhou2023}   
& M  & 1  & 1  
& Assist secondary transmission 
& SU's rate maximization  
& PU IT 
& BCD, SCA 
& Maximizes SU rate by optimizing the UAV trajectory, transmit power, and RIS PS    \\ \hline

\cite{Wu2023}  
& M  & M  & 1 
& Assist secondary spectrum sensing and transmission 
& SR maximization  
& IT, transmit power budget, RIS PS constraint, minimum detection probability, maximum $P_{fa}$ 
& BCD, SCA, SDR 
& Maximizes SUs' SR by optimizing BF, sensing time allocation,  and RIS PS \\  \hline

\cite{Srivastava2024} 
& 1  & 2 & 1          
& Assist spectrum sensing and EH 
& Received primary SNR maximization 
& RIS PS constraints 
& GWO   
& Optimizes RIS phase shifts to enhance spectrum sensing and EH in a NOMA-enabled CRN  \\  \hline

\cite{Yuan2021}  
& M   & 1  & M   
& Assist secondary transmission    
& SU rate maximization     
& Transmit power budget, PU IT, RIS PS constraints 
& BCD, SOCP, SDR 
& Maximizes SU rate by optimizing BF and each RIS PS  \\ \hline

\cite{Asiedu2023} 
& M  & M  & M   
& Assist both PU and SU 
& PUs and SUs SR maximization    
& RIS PS constraints, transmit power budget, minimum PU/SU rates 
& Iterative algorithm 
& Maximizes PUs and SUs SR by optimizing NOMA power allocation and each RIS PS \\  \hline

\cite{Ge2024}  
& 1   & M   & M   
& Assist SUs CSS 
& Detection probability maximization 
& Maximum $P_{fa}$, RIS PS constraints 
& Numerical solution 
& Enhances cooperative detection probability by optimizing each RIS PS \\  \hline

\end{tabular}
\begin{tablenotes}
    \scriptsize{
    \item M - Multiple, \quad WSR - Weighted sum rate, \quad BF - Beamforming, \quad PS - Phase shift, \quad SR - Sum rate, \quad IT - interference temperature.}
\end{tablenotes}
\end{threeparttable}\vspace{-5mm}
\end{table*}

\textit{Case Study 2:}
A representative case study illustrates joint resource allocation and parameter optimization. 

\begin{figure}[!t]\vspace{-2mm}
\centering
    \def\svgwidth{220pt} 
    \fontsize{8}{8}\selectfont 
    \graphicspath{{Figures/}}
    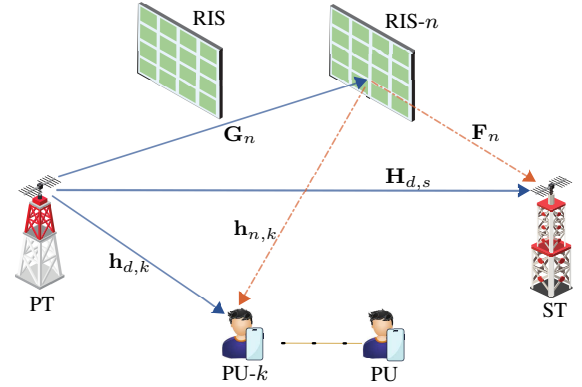 \vspace{-2mm}
    \caption{An interweave RCN system model.}\vspace{-5mm} \label{fig_RA_SystemModel}
\end{figure}

We consider an interweave RCN consisting of a PT equipped with $M_t$ antennas, $K$ single-antenna PUs, an ST equipped with $M_s$ antennas for spectrum sensing, and $N$ distributed RISs (Fig.~\ref{fig_RA_SystemModel}). Each RIS comprises $L_n$ passive reflecting elements. The PT transmits $K$ independent data streams using linear beamforming. The transmitted signal is given by $ \mathbf{x} = \sum_{k=1}^{K} \mathbf{w}_k s_k$, where $\mathbf{w}_k \in \mathbb{C}^{M_t \times 1}$ is the beamforming vector for PU $k$, and $s_k$ is the corresponding unit-power data symbol. The total transmit power is constrained as $\sum_{k=1}^{K} \|\mathbf{w}_k\|^2 \le P_{\max}$. Each RIS applies a diagonal phase-shift matrix $\mathbf{\Phi}_n = \operatorname{diag}(e^{j\theta_{n,1}}, \ldots, e^{j\theta_{n,L_n}})$, where each element satisfies the unit-modulus constraint $|e^{j\theta_{n,l}}|=1$.

Let $\mathbf{h}_{d,k}$ denote the direct PT-to-the $k$-th PU channel, $\mathbf{G}_n$ the PT-to-the $n$-th RIS channel, and $\mathbf{h}_{n,k}$ the $n$-th RIS-to-the $k$-th PU channel. Then, the effective channel from the PT to the $k$-th PU is expressed as $\mathbf{h}_k^{\mathrm{H}}(\boldsymbol{\Phi}) = \mathbf{h}_{d,k}^{\mathrm{H}} + \sum_{n=1}^{N} \mathbf{h}_{n,k}^{\mathrm{H}} \mathbf{\Phi}_n \mathbf{G}_n$. Accordingly, the received signal at the $k$-th PU is $y_k = \mathbf{h}_k^{\mathrm{H}}(\boldsymbol{\Phi}) \sum_{i=1}^{K} \mathbf{w}_i s_i + n_k$, and the corresponding SINR is given by
\begin{align}
    \Gamma_k(\mathbf{W},\boldsymbol{\Phi}) = \frac{|\mathbf{h}_k^{\mathrm{H}}(\boldsymbol{\Phi})\mathbf{w}_k|^2}{\sum_{i\neq k}^{K}|\mathbf{h}_k^{\mathrm{H}}(\boldsymbol{\Phi})\mathbf{w}_i|^2 + \sigma^2}.
\end{align}

During the sensing phase, the ST receives the primary signal through both direct and RIS-reflected links. Let $\mathbf{H}_{d,s}$ denote the direct PT-to-ST channel and $\mathbf{F}_n$ the $n$-th RIS-to-ST channel. The effective channel is $\mathbf{H}_s(\boldsymbol{\Phi}) = \mathbf{H}_{d,s} + \sum_{n=1}^{N} \mathbf{F}_n \mathbf{\Phi}_n \mathbf{G}_n$. Thus, the received signal at the ST is $\mathbf{y}_s = \mathbf{H}_s(\boldsymbol{\Phi}) \sum_{k=1}^{K} \mathbf{w}_k s_k$. The sensing performance is characterized by the received signal power, i.e.,
\begin{align}
    P_{\mathrm{sen}}(\boldsymbol{\Phi}) = \sum\nolimits_{k=1}^{K} \|\mathbf{H}_s(\boldsymbol{\Phi})\mathbf{w}_k\|^2.
\end{align}

The objective is to jointly optimize the PT beamforming vectors and RIS phase shifts to maximize the total PU sum rate while satisfying the PT transmit power constraint, the RIS phase-shift constraint, and the ST sensing requirement. However,  in interweaving CRNs,  there is no direct cooperation between the PN and SN. Hence, the PT optimizes its beamforming solely for primary transmission, while the RISs are configured to simultaneously assist both primary communication and spectrum sensing at the ST.

Accordingly, the optimization is decoupled into two subproblems. The beamforming design at the PT is formulated as
\begin{subequations}
\begin{align} \label{eqn_P_w}
    \mathbf{P}_w:~& \max_{\mathbf{W}} \quad
     \sum\nolimits_{k=1}^{K} \log_2 \left(1+ \Gamma_k(\mathbf{W},\boldsymbol{\Phi})\right), \\
    \text{s.t.} \quad & \sum\nolimits_{k=1}^{K} \|\mathbf{w}_k\|^2 \le P_{\max} \label{eqn_PT_power_const}.
\end{align}
\end{subequations}
For a given RIS configuration, problem $\mathbf{P}_w$ remains non-convex due to the coupled interference terms in the SINR expressions. Efficient solutions can be obtained using fractional programming (FP) and manifold optimization (MO) techniques, which enable direct handling of the power constraint while iteratively improving the beamforming vectors \cite{zargari2025riemannian, zargari2024CFISAC, liu2020simple}.

The RIS optimization problem is formulated as
\begin{subequations}
\begin{align} \label{eqn_P_phi}
    \!\!\! \mathbf{P}_{\phi}:~& \max_{\boldsymbol{\Phi}} \quad \sum\nolimits_{k=1}^{K} \log_2 \left(1+ \Gamma_k(\mathbf{W},\boldsymbol{\Phi})\right) + P_{\mathrm{sen}}(\boldsymbol{\Phi}),\!\! \\
    \!\!\! \text{s.t.} \quad & |\phi_{n, l}| = 1, \quad \forall n, l \label{eqn_RIS_phi_const}.
\end{align}
\end{subequations}
Problem $\mathbf{P}_{\phi}$ is also non-convex due to the unit-modulus constraints and the coupled structure of the objective function. To address this, MO can be employed by treating the feasible set of RIS phase shifts as a complex circle manifold \cite{lee2003introduction, Boumal2023book}. This allows direct optimization over the feasible set without requiring relaxation or approximation of the unit-modulus constraints. Iterative algorithms such as Riemannian gradient or conjugate gradient methods can be used to update the RIS phase shifts until convergence.

By alternately optimizing  $\mathbf{P}_w$ and $\mathbf{P}_{\phi}$, a locally optimal solution can be obtained. This AO framework effectively captures the interplay between active beamforming at the PT and passive beamforming via the RISs.

\begin{figure}[!t]\vspace{-2mm}	
\centering
    \includegraphics[width=0.40\textwidth]{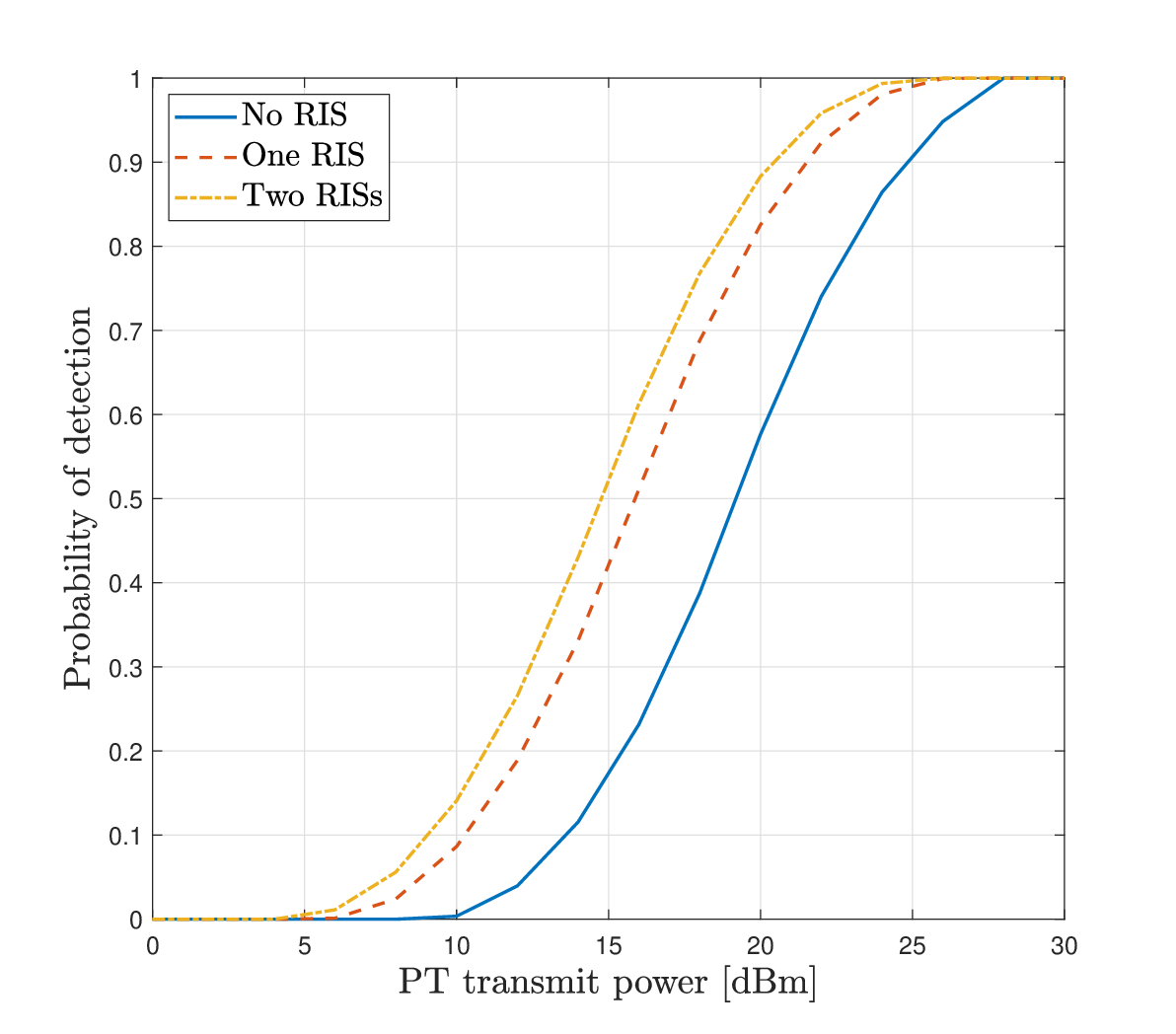}\vspace{-3mm}
\caption{The spectrum detection probability at the ST as a function of the PT's transmit power.} \label{fig_RA_P_detection}\vspace{-5mm}
\end{figure}

All channels are modeled as independent Rayleigh fading channels with the 3GPP UMi large-scale pathloss model at a carrier frequency of $f_c=\qty{3}{\GHz}$. Unless otherwise stated, the PN comprises $K=\num{4}$ PUs, and both the PT and ST are equipped with $M_t=M_s=\num{8}$ antennas. In each case, it is assumed that $N\times L_n=64$. 
The PT is located at $(0,0)$ and the ST at $(80,0)$. The PUs are randomly distributed within a circular region centered at $(40,0)$ with a radius of \qty{10}{\m}. The RIS are uniformly placed between $(30,10)$ and $(60,10)$.

To evaluate the spectrum sensing ability of the ST, the probability of detection at the ST is a good proxy, which is defined as $P_d = \mathbb{P}\!\left(P_{\mathrm{sen}}(\boldsymbol{\Phi}) \geq P_{\mathrm{th}}\right),$ where $P_{\mathrm{sen}}(\boldsymbol{\Phi})$ denotes the received primary signal power at the ST and $P_{\mathrm{th}}=10^{-4}$ is the detection threshold.

Fig.~\ref{fig_RA_P_detection} compares three deployment scenarios, no RIS, one RIS, and two distributed RISs, as $\pd$ at the ST varies with the PT power, $P_{\max}$. While $\pd$ increases monotonically with $P_{\max}$ in all cases, RIS-assisted configurations accelerate this improvement by strengthening the cascaded PT-RIS-ST channel through constructive signal combining, and consistently outperform the no-RIS case across the entire power range. The gain from one to two distributed RISs is comparatively modest, since the total number of reflecting elements is fixed at $N \times L_n = 64$; distributing them across multiple RISs mainly adds spatial diversity rather than additional passive beamforming gain. Notably, the leftward shift of the detection curves with RIS deployment shows that RCNs can achieve a target $\pd$ at significantly lower PT transmit power, indicating improved sensing sensitivity in low-SNR regimes.

\begin{figure}[!t]\vspace{-2mm}	
\centering
    \includegraphics[width=0.40\textwidth]{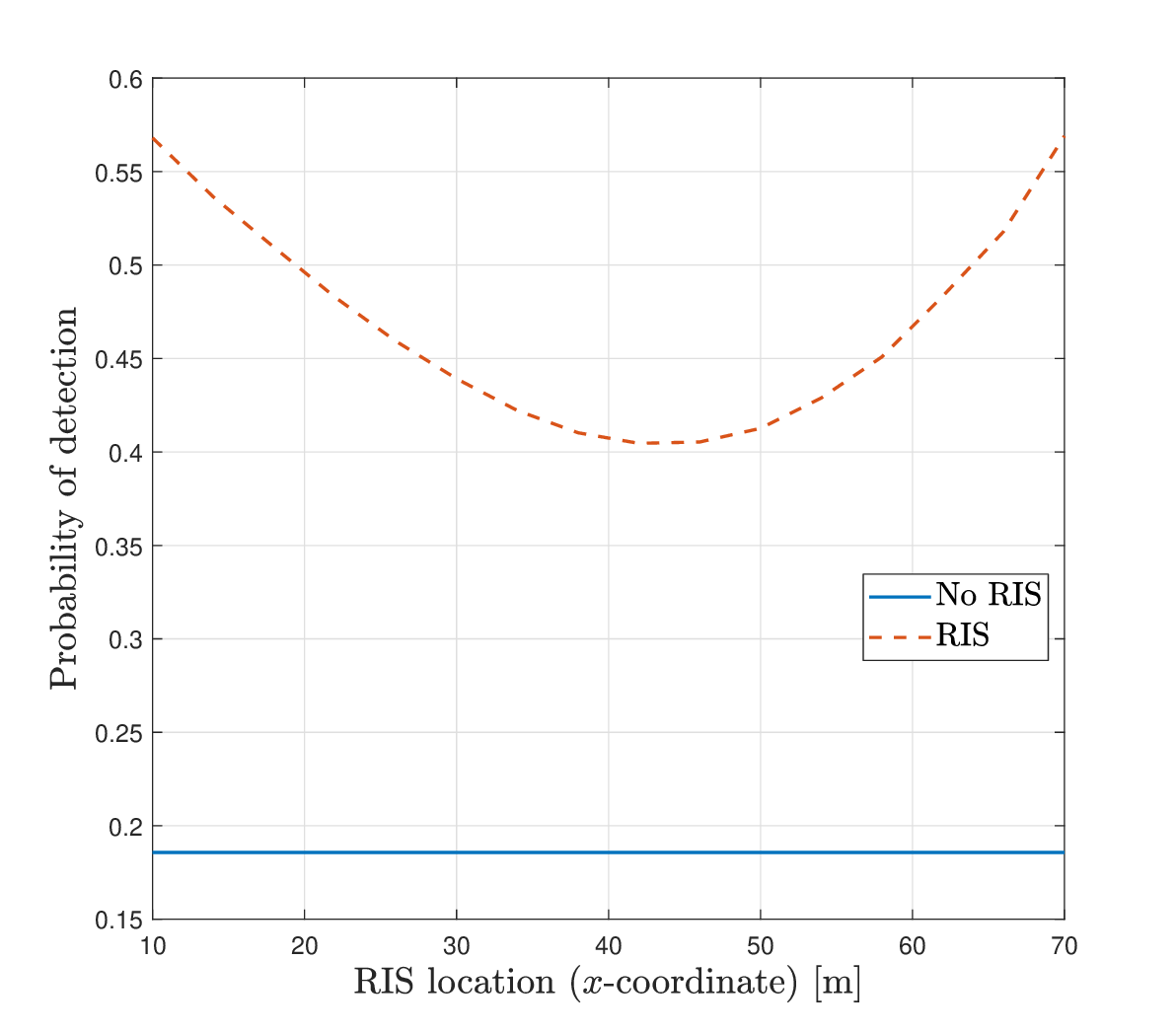}\vspace{-3mm}
\caption{The spectrum detection probability at the ST as a function of the RIS location.} \label{fig_RA_RIS_location}\vspace{-5mm}
\end{figure}

Fig.~\ref{fig_RA_RIS_location} examines the effect of RIS location for $P_{\max}=\qty{15}{\dB m}$, varying the RIS position along $10 \leq x\leq 70$ m with $y=10$ m fixed. The detection probability with RIS is non-monotonic, maximized when the RIS is placed close to either the PT or the ST, where the corresponding link experiences lower path loss and dominates the cascaded gain, and minimized at the midpoint, where both links experience moderate path loss. The no-RIS case remains constant across all locations, since it relies solely on the direct PT-ST link; the RIS-assisted configuration outperforms it even under this worst-case midpoint placement, demonstrating the robustness of RIS-enhanced sensing.



These results indicate that RIS deployment should not be treated independently of sensing performance: location, transmit power, and sensing thresholds must be jointly optimized, with placement near the transmitter or receiver preferred over uniform or midpoint placement, particularly in low-SNR regimes~\cite{Galappaththige2022Distributedt, Diluka2020}.


\begin{mybox}[frametitle={Key Insights}]
\begin{enumerate}[leftmargin=*]

\item \textit{RIS deployment enhances spectrum sensing and reduces the power needed for it:} RIS assistance consistently outperforms the no-RIS case by strengthening the cascaded PT-RIS-ST channel, increasing $\pd$ across all transmit power levels while lowering the PT power required to reach a target $\pd$ (Fig.~\ref{fig_RA_P_detection}).

\item \textit{Distributed RISs add modest diversity gains:} Deploying multiple RISs improves sensing further through spatial diversity, though gains beyond a single RIS are limited once the dominant PT-RIS-ST link is established (Fig.~\ref{fig_RA_P_detection}).

\item \textit{RIS placement critically impacts sensing performance:} $\pd$ is highly sensitive to RIS location, peaking near the PT or ST and dipping at the midpoint, underscoring the need for deployment-aware resource allocation that jointly optimizes RIS location, transmit power, and sensing thresholds (Fig.~\ref{fig_RA_RIS_location}).

\end{enumerate}
\end{mybox}

\subsection{Secure RCNs}
Unauthorized users may opportunistically access licensed spectrum, thereby increasing security risks, including PU emulation attacks, interference, and eavesdropping \cite{Pooja2023}. These risks are more severe than in conventional wireless networks because CRNs rely on DSA, distributed spectrum sensing, and adaptive decision-making, which introduce vulnerabilities in spectrum sensing, channel access, and network coordination. Moreover, the open and opportunistic nature of spectrum access enables malicious users to exploit spectrum-sharing mechanisms, manipulate sensing information, impersonate PUs, and intercept secondary transmissions. These challenges are summarized in Table~\ref{tab_crn_security}. 

\begin{table*}[t]\vspace{-2mm}
\centering
\caption{Major Security Approaches in CRNs.}
\label{tab_crn_security}\vspace{-2mm}
\renewcommand{\arraystretch}{1.0}
\begin{tabular}{p{3.4cm}p{5.8cm}p{5.8cm}}
\hline
\textbf{Security Approach} & \textbf{Main Idea} & \textbf{Typical Techniques/Applications} \\
\hline \hline

Secure spectrum sensing &
Improve the reliability and robustness of spectrum sensing against malicious users and sensing errors &
CSS, trust/reputation-based sensing, anomaly detection, ML-based attack detection \\
\hline

PLS &
Exploit wireless channel characteristics to enhance confidentiality and suppress eavesdropping &
AN, secure beamforming, cooperative jamming, RSMA-based secure transmission \\
\hline

RIS-assisted security &
Use RIS to intelligently manipulate wireless propagation for secure transmission and interference suppression &
Secure beam steering, eavesdropper nulling, RIS-assisted anti-jamming, STAR-RIS security \\
\hline

Authentication and access control &
Prevent unauthorized spectrum access and PU emulation attacks &
RF fingerprinting, location verification, lightweight authentication, cryptographic techniques \\
\hline

Anti-jamming techniques &
Mitigate intentional interference and denial-of-service attacks &
Frequency hopping, adaptive spectrum access, game-theoretic anti-jamming strategies \\
\hline

Blockchain-based security &
Provide decentralized and tamper-resistant trust management and spectrum sharing &
Secure spectrum trading, distributed trust management, blockchain-assisted CSS \\
\hline

AI-driven security &
Use AI to detect, predict, and mitigate attacks in dynamic environments &
Reinforcement learning, federated learning, DL-based intrusion detection \\
\hline

Cross-layer security &
Jointly optimize security mechanisms across multiple network layers &
Secure routing, joint sensing-access-security optimization, QoS-aware secure spectrum allocation \\
\hline

Emerging 6G security techniques &
Address new security challenges introduced by future 6G-enabled CRNs &
NF secure communications, ISAC security, NTN-integrated CRNs, semantic-aware security \\
\hline

\end{tabular}\vspace{-5mm}
\end{table*}

As this table shows, several approaches exist to improve security. Since this survey focuses on the physical layer of wireless networks, we expand on PLS only.   PLS exploits the inherent randomness and spatial characteristics of wireless channels, such as fading, noise, and spatial selectivity, to ensure security without relying solely on higher-layer cryptographic mechanisms \cite{Wang2019PLS, Chen2017, Angueira2022, Liu2017}. They assume computationally bounded adversaries, whereas PLS provides security by leveraging differences between the legitimate and eavesdropping channels. The central objective is to establish a channel advantage for the legitimate receiver, such that the eavesdropper's channel is fundamentally degraded \cite{Wang2019PLS, Chen2017, Angueira2022, Liu2017}.

To achieve this, several transmission and signal processing techniques have been developed. First, artificial noise (AN) injection introduces structured interference that lies in the null space of the legitimate receiver while disrupting the eavesdropper \cite{Wang2019PLS, Chen2017, Angueira2022, Liu2017}. Second, beamforming and precoding enable spatial focusing of signal energy toward intended users while minimizing leakage. Third, cooperative jamming further enhances secrecy by allowing auxiliary nodes to transmit interference that selectively degrades the eavesdropper's reception \cite{Wang2019PLS, Chen2017, Angueira2022, Liu2017}. In addition, channel-based secret key generation exploits channel reciprocity and randomness to establish shared keys between legitimate users. More recently, RIS-assisted techniques have emerged as a powerful tool, enabling programmable control of the propagation environment to simultaneously enhance legitimate links and suppress eavesdropping channels \cite{Wang2019PLS, Chen2017, Angueira2022, Liu2017}.

Several fundamental insights underpin PLS. First, security is intrinsically a channel advantage problem: secure communication is achievable when the legitimate channel outperforms the eavesdropper's channel \cite{Wang2019PLS, Chen2017, Angueira2022, Liu2017}. Second, channel randomness, traditionally viewed as a challenge, becomes a valuable resource that can be harnessed for secrecy. Third, spatial DoF, enabled by technologies such as MIMO and RIS, plays a critical role in shaping signal propagation for secure transmission \cite{Wang2019PLS, Chen2017, Angueira2022, Liu2017}. Finally, PLS introduces inherent trade-offs among secrecy, rate, reliability, and power, necessitating joint optimization in system design \cite{Wang2019PLS, Chen2017, Angueira2022, Liu2017}.

The performance of PLS schemes is commonly evaluated using information-theoretic metrics. The secrecy rate, defined as the difference between the capacities of the legitimate and eavesdropping channels, serves as the primary measure of secure throughput \cite{Wang2019PLS, Chen2017, Angueira2022, Liu2017}. Secrecy capacity quantifies the maximum achievable secure rate under ideal conditions, whereas secrecy outage probability quantifies the probability that a target secrecy rate cannot be maintained due to channel fluctuations. Additional metrics, such as the eavesdropper's bit error rate and secrecy EE, further quantify system robustness and efficiency \cite{Wang2019PLS, Chen2017, Angueira2022, Liu2017}.

Overall, PLS transforms security from a purely algorithmic problem into a fundamentally physical one, where confidentiality is achieved through the intelligent exploitation and control of the wireless medium \cite{Pooja2023}.

RIS-assisted PLS leverages the programmable propagation control enabled by RISs to enhance confidentiality in wireless networks \cite{Khoshafa2025}. The basic insight is simple.  By intelligently adjusting the phase shifts (and possibly amplitudes) of reflected signals, RISs can strengthen the legitimate channel while simultaneously degrading the eavesdropper's channel, thereby creating a favorable channel advantage for secure communication. This capability is particularly effective in scenarios with limited transmitter-side DoF or strong line-of-sight (LoS) conditions, where conventional beamforming alone may be insufficient \cite{Khoshafa2025}. RISs can also facilitate AN shaping, interference nulling, and spatial signal alignment, further improving secrecy performance. Compared with traditional approaches, RIS-assisted PLS offers a low-cost and energy-efficient means of enhancing security, as it does not require active RF chains and provides additional spatial DoF for fine-grained control of the wireless environment \cite{Khoshafa2025}.

Conversely, RISs can further enhance the PLS CRNs by intelligently reconfiguring wireless propagation \cite{Wu2022, Wu2022Secure, Zhangyu2023, Khoshafa2023, Shixiong2024}. For example, by leveraging passive beamforming, RISs can steer signals toward legitimate users while minimizing leakage to unintended receivers, thereby improving secrecy performance. Additionally, RISs can create secure communication zones, generate interference or AN to mask confidential signals, and even act as passive jammers to disrupt eavesdroppers. Moreover, by exploiting CSI, RISs can detect potential eavesdroppers and dynamically adjust phase shifts to degrade the quality of their received signals, thereby reinforcing security \cite{Wu2022, Wu2022Secure, Zhangyu2023, Khoshafa2023, Shixiong2024}.

The security issues in RCNs have been studied recently~\cite{Wu2022, Wu2022Secure, Zhangyu2023, Khoshafa2023, Shixiong2024}. In~\cite{Wu2022}, the PLS of an RCN with multiple PUs, a single SU, and multiple eavesdroppers is investigated. The RIS supports secondary transmission while leveraging secondary interference to enhance the PLS of the PN. The objective is to minimize the total transmit power while satisfying the SU's rate requirement, the PUs' secrecy rate constraint, and the maximum interference temperature at the PUs. To achieve this, an AO algorithm is proposed that incorporates a penalty-based rank-1 relaxation and SOCP methods to jointly optimize the primary and secondary transmit beamforming, along with RIS reflect beamforming (i.e., phase shifts). To balance secrecy rate and energy consumption, \cite{Wu2022Secure} investigates the secrecy EE of the secondary system in an RIS-assisted underlay CRN with a single PU, a single SU, and multiple eavesdroppers. This is achieved by jointly optimizing the secondary transmit beamforming and the RIS phase shifts while ensuring a minimum secrecy-rate threshold and adhering to the maximum interference temperature at the PU. To tackle the non-convex problem with coupled variables, an AO algorithm is proposed that utilizes the iterative penalty-based method, the DC method, and SOCP approximation, demonstrating the effects of RIS on SE and PLS in CRNs.

In \cite{Zhangyu2023}, the secrecy of secondary transmission in a RIS-assisted single PU, single SU CRN is examined in the presence of an eavesdropper. To counter eavesdropping, AN is introduced at the secondary BS. The objective is to minimize the transmit power of the secondary BS while satisfying the PU's interference temperature constraint and confidentiality rate requirement by jointly optimizing the secondary BS beamforming, AN at the secondary BS, and the RIS phase shift matrix. Reference \cite{Khoshafa2023} utilizes RIS to enhance the PLS of the PN and the data transmission of the SN in a CRN with a single PU, single SU, and a passive eavesdropper attempting to intercept PU data. The study considers scenarios in which the eavesdropper employs either selection or maximal-ratio combining to process signals from the PN. Analytical expressions for the secrecy outage probability and the probability of nonzero secrecy capacity of the PN are derived. Additionally, an expression for the outage probability of the SN is presented.

Practical RIS implementations are subject to hardware constraints, most notably the quantization of phase shifts, as continuous phase control is not feasible. This raises the fundamental question of how phase quantization impacts secrecy performance. To address this, \cite{Shixiong2024} investigates the PLS of an RCN  under discrete phase control. The RIS phase shifts are designed to enhance secondary transmission by maximizing the SNR at the SU, without requiring eavesdropper CSI. For a passive eavesdropper, with and without a direct link between the secondary BS and the SU, SOP expressions are derived.

An asymptotic analysis reveals the impact of the number of RIS elements and phase quantization levels. The study provides three key insights. First, discrete phase control still enables tractable secrecy analysis, allowing a rigorous performance evaluation. Second, even without eavesdropper CSI, enhancing the legitimate channel via RIS optimization is sufficient to improve secrecy performance. Third, increasing the number of RIS elements significantly reduces SOP, indicating that the performance loss due to phase quantization can be effectively mitigated through RIS scaling. Overall, the results highlight a fundamental trade-off between hardware simplicity and secrecy performance, which can be alleviated by deploying large-scale RISs.

Secure RCN works are summarized in Table~\ref{tab_security}.

\begin{table*}[htbp]\vspace{-2mm}
\centering
\begin{threeparttable}
\renewcommand{\arraystretch}{1.0}
\caption{Summary of secure RCN  literature.\vspace{-2mm}}
\label{tab_security}
\begin{tabular}{llllp{1.8cm}p{3.3cm}p{7.5cm}}
\hline
\textbf{Ref.} & \textbf{PUs} & \textbf{SUs} & \textbf{Eaves} & \textbf{Eavesdropping} & \textbf{RIS Role} & \textbf{Objective} \\ \hline \hline

\cite{Wu2022} 
& M & 1 & M 
& Primary data 
& Assist SU \& enhance PPLS 
& Minimize total transmit power while satisfying SU rate requirement, PU secrecy rate, PU IT \\ \hline

\cite{Wu2022Secure} 
& 1 & 1 & M 
& Secondary data 
& Assist SU \& enhance SPLS 
& Maximize secrecy EE of secondary system adhering to minimum secrecy rate and PU IT \\ \hline

\cite{Zhangyu2023} 
& 1 & 1 & 1 
& Secondary data 
& Assist SU \& enhance SPLS 
& Minimize secondary transmit power subject to PU IT and confidentiality rate requirement with AN at secondary BS \\ \hline

\cite{Khoshafa2023} 
& 1 & 1 & 1 
& Primary data 
& Assist SU \& enhance PPLS 
& Analytical expressions for primary secrecy outage probability and  probability of nonzero secrecy capacity as well as SN outage probability \\ \hline

\cite{Shixiong2024} 
& 1 & 1 & 1 
& Secondary data 
& Assist SU \& enhance SPLS 
& Analytical expressions for secrecy outage probability as well as asymptotic analysis investigating impacts of the number of RIS elements and quantization bits \\ \hline
 
\end{tabular}
\begin{tablenotes}
      \scriptsize{
      \item M - Multiple, \quad  IT - interference temperature, \quad PPLS - primary PLS, \quad SPLS - secondary PLS.}
    \end{tablenotes}
\end{threeparttable}\vspace{-5mm}
\end{table*}

\textit{Case Study 3:}
 PLS in RCNs is illustrated here. 

We consider an underlay RCN comprising an $M$-antenna ST, a single-antenna SU, a single-antenna PU, an RIS with $L$ passive reflecting elements, and a passive eavesdropper (Eve). The RIS is deployed to enhance the secondary legitimate transmission while suppressing information leakage toward Eve and limiting interference to the PU. The ST transmitted signal is given by $\mathbf{x} = \mathbf{w}s$, where $\mathbf{w} \in \mathbb{C}^{M \times 1}$ is the beamforming vector and $s$ is a unit-power information symbol. The ST transmit power is constrained as $\|\mathbf{w}\|^2 \le P_{\max}$. The RIS applies a diagonal phase-shift matrix $\mathbf{\Phi} = \operatorname{diag}(e^{j\theta_1}, \ldots, e^{j\theta_L})$, where each element satisfies the unit-modulus constraint $|e^{j\theta_l}|=1$.

Let $\mathbf{h}_{d,a} \in \mathbb{C}^{M \times 1}$ denote the direct ST-to-node $a$ channel, $\mathbf{G} \in \mathbb{C}^{L \times M}$ the ST-to-RIS channel, and $\mathbf{f}_a \in \mathbb{C}^{L \times 1}$ the RIS-to-node $a$ channel, where $a \in \{s,p,e\}$ correspond to the SU, PU, and Eve, respectively. The effective channel between the ST and the node $a$ is expressed as $\mathbf{h}_a^{\mathrm{H}}(\boldsymbol{\Phi}) = \mathbf{h}_{d,a}^{\mathrm{H}} + \mathbf{f}_a^{\mathrm{H}} \mathbf{\Phi} \mathbf{G}$. Accordingly, the received signals at the SU and Eve are given by $y_a = \mathbf{h}_a^{\mathrm{H}}(\boldsymbol{\Phi}) \mathbf{w} s + n_a$ for $a \in \{s,e\}$, and the corresponding SNRs are given as
\begin{align}
    \Gamma_a(\mathbf{w},\boldsymbol{\Phi}) = \frac{|\mathbf{h}_a^{\mathrm{H}}(\boldsymbol{\Phi})\mathbf{w}|^2}{\sigma^2}, \quad a \in \{s,e\}.
\end{align}
The secrecy rate is then defined as
\begin{equation}
    \mathcal{R}^{\rm Sec} = \left[\log_2(1+\Gamma_s(\mathbf{w},\boldsymbol{\Phi})) - \log_2(1+\Gamma_e(\mathbf{w},\boldsymbol{\Phi}))\right]^+,
\end{equation}
where $[x]^+ = \max(0, x)$. Meanwhile, the interference imposed on the PU can be expressed as
\begin{equation}
    P_{\mathrm{Int}}(\mathbf{w},\boldsymbol{\Phi}) = |\mathbf{h}_p^{\mathrm{H}}(\boldsymbol{\Phi})\mathbf{w}|^2.
\end{equation}

The objective is to maximize the secrecy rate while satisfying the ST transmit power constraint, the RIS phase-shift constraint, and the PU interference temperature constraint. The resulting problem is formulated as
\begin{subequations}
\begin{align}
    \mathbf{S}:~& \max_{\mathbf{w},\boldsymbol{\Phi}} \quad \mathcal{R}^{\rm Sec}, \\
    \text{s.t.} \quad & P_{\mathrm{Int}}(\mathbf{w},\boldsymbol{\Phi}) \leq P_{\thr}, \\
    & \|\mathbf{w}\|^2 \le P_{\max}, \\
    & |\phi_l| = 1, \quad \forall l,
\end{align}
\end{subequations}
where $P_{\thr}$ is the maximum allowable interference temperature at the PU. Due to the coupled variables and the non-concave secrecy-rate objective, problem $\mathbf{S}$ is non-convex. An AO framework can be adopted to decouple the problem into two subproblems.

For a given RIS configuration, the beamforming design is formulated as
\begin{subequations}
    \begin{align}
        \mathbf{S}_w:~& \max_{\mathbf{w}} \quad \mathcal{R}^{\rm Sec}, \\
        \text{s.t.} \quad & P_{\mathrm{Int}}(\mathbf{w},\boldsymbol{\Phi}) \leq P_{\thr}, \\
        & \|\mathbf{w}\|^2 \le P_{\max}.
    \end{align}
\end{subequations}
This subproblem can be efficiently addressed using MO combined with the augmented Lagrangian method (ALM), which enables direct handling of the power constraint while incorporating the interference constraint \cite{zargari2025riemannian, zargari2024CFISAC, liu2020simple}.

For a given beamforming vector, the RIS phase-shift optimization problem is given by
\begin{subequations}
    \begin{align}
        \mathbf{S}_{\phi}:~& \max_{\boldsymbol{\Phi}} \quad \mathcal{R}^{\rm Sec}, \\
        \text{s.t.} \quad & P_{\mathrm{Int}}(\mathbf{w},\boldsymbol{\Phi}) \leq P_{\thr}, \\
        & |\phi_l| = 1, \quad \forall l.
    \end{align}
\end{subequations}
By expressing the effective channels as linear functions of the RIS phase vector, this problem is reformulated over a complex circle manifold. The interference constraint is incorporated using the ALM framework, and the resulting problem is efficiently solved via MO techniques \cite{zargari2025riemannian, zargari2024CFISAC, liu2020simple}.
By alternately solving $\mathbf{S}_w$ and $\mathbf{S}_{\phi}$, a locally optimal solution is obtained. This framework highlights the role of RIS in improving the PLS of  CRNs, where the propagation environment is jointly optimized to enhance the legitimate link while suppressing the eavesdropping channel under interference constraints.

\begin{figure}[!t]\vspace{-2mm}	
\centering
    \includegraphics[width=0.40\textwidth]{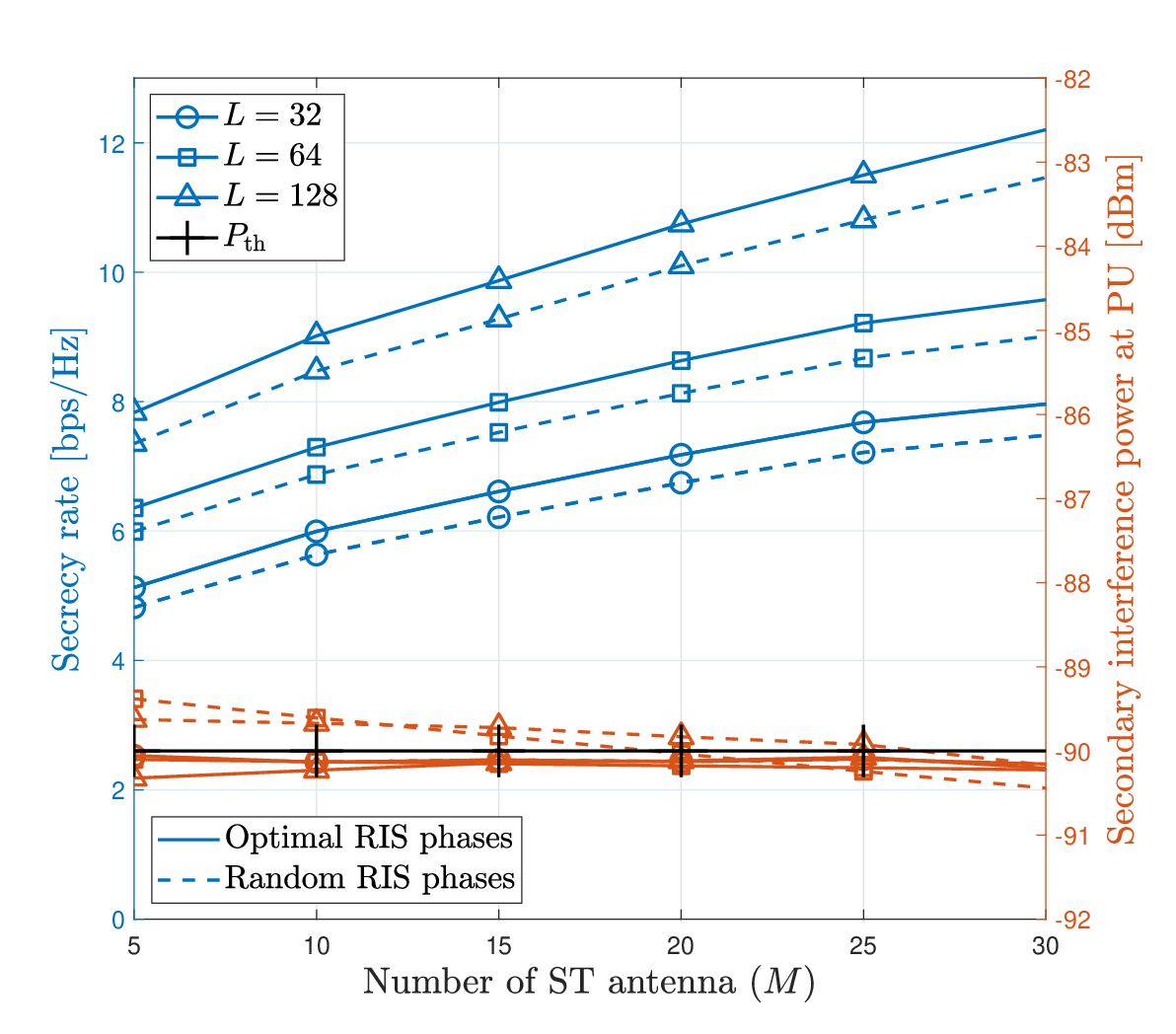}\vspace{-3mm}
\caption{The secrecy rate (left $y$-axis) and the secondary interference power (right $y$-axis) at the PU as a function of the number of ST transmit antennas.} \label{fig_SecrecyRate_Interf_M}\vspace{-5mm}
\end{figure}

Fig.~\ref{fig_SecrecyRate_Interf_M} illustrates the secrecy rate (left $y$-axis) and the interference power at the PU (right $y$-axis) as a function of the number of ST transmit antennas, $M$, for different RIS sizes $L =\{32, 64, 128\}$. The large-scale fading follows the 3GPP UMi model. The interference threshold is set to $P_{\mathrm{th}}=\qty{-90}{\dB m}$, close to the noise floor, reflecting a stringent PU protection requirement. The ST and RIS are located at $(0,0)$ and $(0,40)$, respectively, while the SU, PU, and Eve are positioned at $(40,20)$, $(-40,20)$, and $(40,40)$, respectively.

The secrecy rate increases monotonically with $M$ for all configurations, as additional ST antennas provide spatial DoF for beamforming toward the SU while suppressing leakage toward the eavesdropper, and this gain is amplified by RIS size: at $M=\num{30}$, increasing $L$ from $\num{32}$ to $\num{128}$ improves the secrecy rate from approximately \qty{8}{bps/\Hz} to over \qty{12}{bps/\Hz} (a \qty{50}{\percent} gain), reflecting the complementary roles of active (ST) and passive (RIS) beamforming. Optimal RIS phase configuration also matters increasingly at larger scales: at $M=\num{20}$, the secrecy-rate gain of optimal over random phases grows from about \qty{9}{\percent} at $L=\num{64}$ to \qty{12}{\percent} at $L=\num{128}$. Throughout, the interference power at the PU remains tightly controlled around $P_{\mathrm{th}}$ across all $M$ and $L$, showing that joint beamforming and RIS optimization can enhance secrecy without compromising PU protection, even under a near-noise-floor constraint.

\begin{mybox}[frametitle={Key Insights}]
\begin{enumerate}
\item \textit{Secrecy scales with spatial DoF and RIS size:} The secrecy rate grows with both the number of ST antennas and RIS size, reflecting the complementary gains of active and passive beamforming (e.g., $\approx\qty{50}{\percent}$ gain from $L=32$ to $L=128$ at $M=30$).

\item \textit{RIS phase optimization matters more at larger scale:} The gap between optimal and random RIS phases widens as $L$ increases, making accurate phase design essential for large-scale RIS deployments.

\item \textit{Strict interference constraints are met without sacrificing secrecy:} PU interference stays tightly controlled around $P_{\mathrm{th}}$ across all configurations, confirming that joint beamforming and RIS optimization can deliver strong secrecy while ensuring robust PU protection, even under near-noise-floor constraints.
\end{enumerate}
\end{mybox}





\subsection{Active RCNs}
\begin{figure}[!t]\vspace{-2mm}	
\centering
    \includegraphics[width=0.40\textwidth]{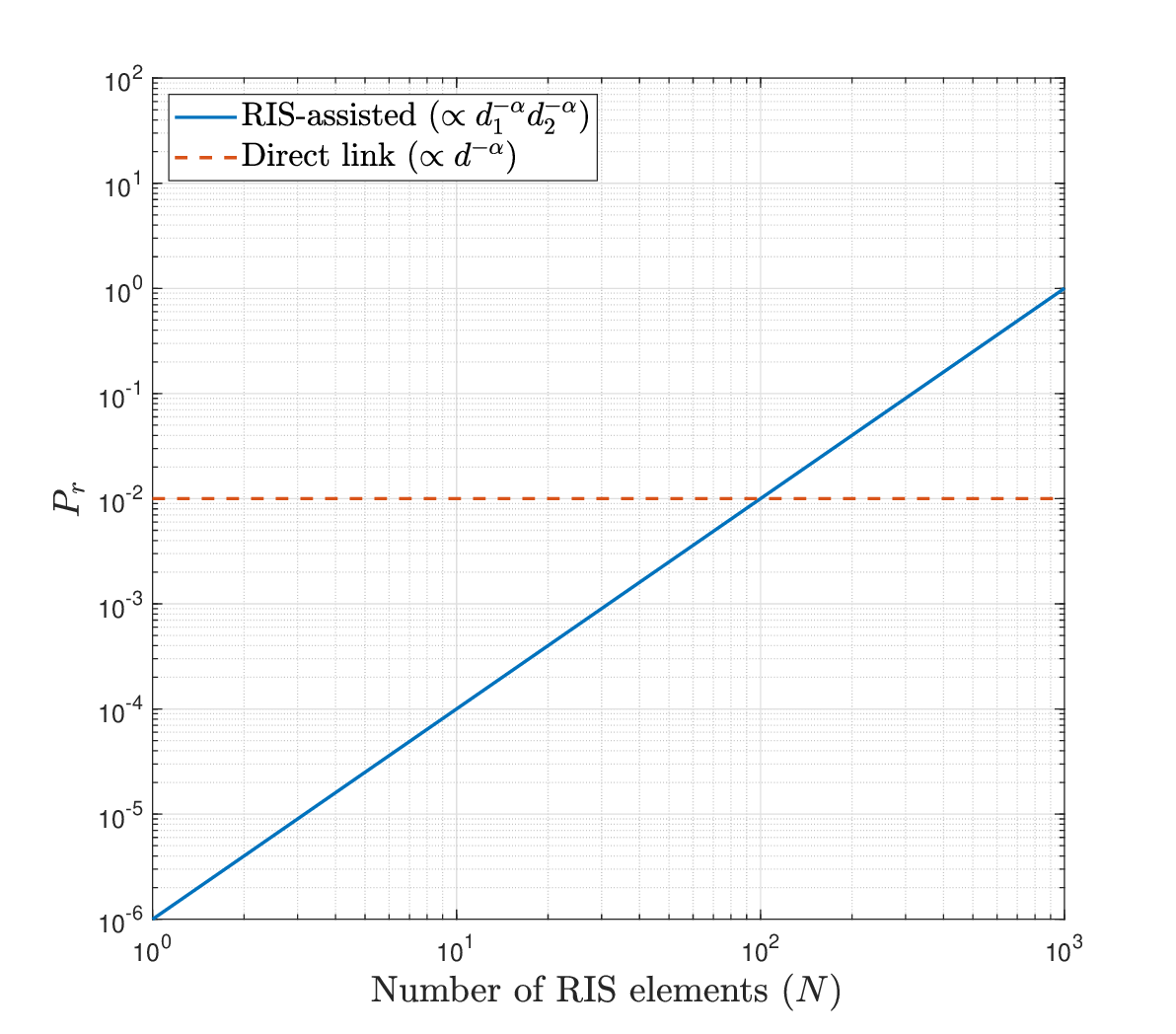}\vspace{-3mm}
\caption{Scaling of normalized received power with the number of RIS elements $N$. While RIS provides a quadratic array gain ($\mathcal{O}(N^2)$), the double path-loss $d_1^{-\alpha} d_2^{-\alpha}$ significantly attenuates the signal, requiring large $N$ to match direct-link performance.} \label{fig:RIS_scaling}\vspace{-5mm}
\end{figure}

The \emph{double-fading effect} is a fundamental limitation in passive RIS-assisted systems. It arises from cascaded transmitter-RIS and RIS-receiver links, leading to multiplicative attenuation of the channel gains \cite{Liu2021RIS, Gong2020, Chen2022, Diluka2020, Galappaththige2023, Diluka2022RISSWIPT, Diluka2022CFRIS}. Specifically, the effective channel is given by $h_{\mathrm{eff}} = h_1 h_2$, where $h_1$ and $h_2$ denote the transmitter-RIS and RIS-receiver channel coefficients (per element), respectively. Consequently, the received signal experiences \emph{double path-loss}, typically scaling as $d_1^{-\alpha} d_2^{-\alpha}$, where $d_1$ and $d_2$ are the link distances and $\alpha$ is the path-loss exponent. Although coherent phase alignment across $N$ RIS elements yields an array gain that scales as $\mathcal{O}(N^2)$, this gain is often insufficient to fully compensate for the compounded attenuation, particularly in large-scale or obstructed deployments (Fig. \ref{fig:RIS_scaling}). This cascaded structure is analogous to dyadic backscatter channels, in which the end-to-end link is governed by a product of forward- and backscatter components \cite{Diluka2022, Rezaei2023Coding}.

In CRNs, double fading adversely affects both \emph{spectrum sensing} and \emph{communication}. For sensing, the reduction in the received SNR degrades the probability of detection $\pd$ and the false alarm $\pfa$, and increases the minimum sensing time required to satisfy a target $\pd$ under a fixed $\pfa$. For communication, it lowers the achievable rate and link reliability. A common mitigation strategy is to increase the number of RIS elements to exploit the $\mathcal{O}(N^2)$ beamforming gain \cite{Long2021}. Despite the low power consumption of individual elements, large-scale RIS deployments incur non-negligible control overhead, channel estimation complexity, and hardware constraints, which limit the practically achievable aperture and, consequently, the achievable gains in both sensing and communication performance.

Consequently, active RIS has emerged as a promising alternative. Unlike passive RIS, each element of an active RIS is equipped with a low-power amplifier, enabling it to both reflect and amplify the incident signal. This effectively transforms the cascaded channel into a partially regenerative link, alleviating the severe attenuation due to double fading and achieving a higher received SNR with fewer elements. From a signal-model perspective, an active RIS introduces an additional gain matrix into the RIS response, which can be jointly optimized with the phase shifts to balance amplification and noise. However, this amplification also introduces thermal noise and potential distortion, rendering the system noise-limited unless carefully designed. Consequently, noise-aware beamforming, power allocation, and stability constraints are essential to ensure performance gains. Moreover, active RIS entails higher power consumption and hardware complexity due to RF chains and amplifiers, necessitating efficient power control and circuit design. Despite these challenges, active RIS offers a powerful means to enhance spectrum sensing and data transmission in CRNs, particularly in low-SNR and coverage-limited scenarios \cite{Long2021, Jungang2023, Jungang2024, Xie2024}.

Nevertheless, active RISs in CRNs have been investigated only to a limited extent, with the exception of \cite{Jungang2023, Jungang2024, Xie2024}. In particular, reference \cite{Jungang2023} employs an active RIS to enhance spectrum sensing by amplifying the received PU signal at the SU, reducing the required sensing time in a single-antenna PU, a multi-antenna SU CRN. A detection-probability maximization problem is formulated under a given maximum false-alarm probability constraint by optimizing the RIS reflecting-coefficient matrix. Using random matrix theory, the authors transform the problem into an equivalent one of maximizing the largest eigenvalue of the covariance matrix of the sensing signal. They propose a weighted MMSE algorithm to obtain the optimal RIS reflecting-coefficient matrix. Additionally, the study investigates the minimum power budget required for the active RIS to achieve $\pd \approx 1 $ in a simplified scenario without the PU-SU direct link and under LoS channel conditions. The study \cite{Jungang2024} extends the system model in \cite{Jungang2023} to evaluate the secondary spectrum-sensing performance in the presence of multiple interference sources.

Reference \cite{Xie2024} investigates how active or passive RISs can be leveraged to enhance spectrum sensing in CRNs, with a particular focus on two fundamental design questions: whether RIS should operate in a passive or active sensing mode, and how many reflecting elements are required to achieve reliable detection. The authors consider a standard CR setup in which an SU performs energy detection to sense the presence of a PU, assisted by an RIS that modifies the propagation environment.

Within this framework, the paper contrasts passive and active RIS architectures. As mentioned earlier, passive RIS enables coherent signal combining at the receiver, thereby improving the received SNR without introducing additional noise. In contrast, an active RIS incorporates amplification, which can significantly boost the received signal strength, particularly in low-SNR regimes, but at the cost of injecting amplification noise. The analysis reveals a key trade-off: active RIS can outperform passive RIS when the system is noise-limited, whereas passive RIS becomes more advantageous at moderate-to-high SNR due to its noise-free operation and scalability.

To quantify these effects, the authors derive analytical expressions for $\pd$ and $\pfa$ within an energy-detection framework. Their results show that RIS-assisted sensing performance improves substantially with the number of reflecting elements, owing to the coherent-combining gain provided by the RIS. In particular, the detection performance exhibits a sharp transition: below a certain number of elements, sensing is unreliable, whereas beyond this threshold, performance rapidly approaches near-optimal levels. This leads to an important design insight: determining the minimum number of RIS elements required to meet a target sensing performance is critical for practical deployment.

Overall, the paper demonstrates that RIS can effectively enhance spectrum sensing by improving received signal quality and mitigating challenges such as the hidden-node problem. It provides clear design guidelines, indicating that passive RIS is generally preferable for large-scale, energy-efficient deployments, while active RIS is beneficial in severely noise-limited scenarios despite its higher complexity. The work highlights the importance of RIS configuration and size as key factors in unlocking reliable and efficient spectrum sensing in future CR systems.

Table~\ref{tab_active} presents a summary of these active RCN works. It is clear that active RIS in CRNs remains largely unexplored, with several fundamental research challenges still open. A key gap lies in the joint design of spectrum sensing, communication, and RIS amplification, since current work typically optimizes these components in isolation. In particular, noise-aware modeling and optimization are still immature, as amplification introduces additional thermal noise whose propagation through cascaded channels is not yet fully characterized. This directly impacts both sensing reliability and data transmission, necessitating new frameworks that jointly optimize amplification gains, phase shifts, and transmission strategies under realistic noise constraints. Furthermore, hardware impairments, such as nonlinear amplifiers, quantized phase/gain control, and stability issues, remain insufficiently studied, especially in large-scale deployments. These challenges are further compounded by the need for energy-efficient designs, as active RIS consumes non-negligible power, requiring joint power allocation and energy-aware optimization in CRNs.

Beyond these foundational issues, several system-level and emerging directions remain open. These include multi-RIS and distributed active RIS architectures, where coordination, scalability, and signaling overhead become critical, as well as security vulnerabilities introduced by amplification, such as RIS-assisted jamming or spoofing attacks. In addition, learning-based control of active RIS remains in its infancy, particularly for real-time adaptation in dynamic-spectrum environments. There is also a lack of fundamental performance limits, including capacity and detection bounds under active RIS, as well as scaling laws with respect to the number of elements and power consumption. Finally, important practical scenarios, such as wideband operation, NF propagation, and user mobility, remain largely unaddressed, along with the stability and control-theoretic aspects of active RIS systems. Together, these gaps highlight significant opportunities for advancing both the theory and practice of active RCNs.

\begin{table*}[htbp]\vspace{-2mm}
\centering
\begin{threeparttable}
\renewcommand{\arraystretch}{1.0}
\caption{Summary of active RCNs literature.\vspace{-2mm}}
\label{tab_active}
\begin{tabular}{lllp{6cm}p{8cm}}
\hline
\textbf{Ref.} & \textbf{PUs} & \textbf{SUs} & \textbf{RIS Role} & \textbf{Objective} \\ \hline \hline

\cite{Jungang2023} 
& 1 & 1 (MA) 
& Amplifies  PU signals at SU 
& Maximize detection probability for a given $P_{fa}$.  \\ \hline

\cite{Jungang2024} 
& 1 & 1 (MA) 
& Amplifies  PU signals at SU to increase $\pd$ 
& Extend  \cite{Jungang2023} to evaluate the impacts of additional interference  \\ \hline

\cite{Xie2024} 
& 1 & 1 (MA) 
& Amplifies  PU signals at SU 
& Compare $\pd$ of spectrum sensing between the active and passive RISs \\ \hline

\end{tabular}
\begin{tablenotes}
      \scriptsize{
      \item MA - Multiple antennas.}
    \end{tablenotes}
\end{threeparttable}\vspace{-5mm}
\end{table*}

\textit{Case Study 4:}
Here, we extend the previous case study in Section~\ref{sec_RA_pssive} to an active RCN, where each RIS element can adjust both the amplitude and phase of the reflected signal. Thus, the reflection coefficient of the $l$-th element of the $n$-th RIS is expressed as $v_{n,l}=a_{n,l}e^{j\theta_{n,l}}$, where $a_{n,l}\geq 0$ denotes the amplification factor. Accordingly, the reflection matrix of the $n$-th RIS is given by $\mathbf{\Psi}_n = \operatorname{diag}(v_{n,1}, \ldots, v_{n,L_n})$. Let $\mathbf{v}_n \in \mathbb{C}^{L_n}$ denote the coefficient vector of the $n$-th RIS, and $\mathbf{v}= [\mathbf{v}_1^{\mathrm{T}}, \ldots, \mathbf{v}_N^{\mathrm{T}}]^{\mathrm{T}} \in \mathbb{C}^{L \times 1}$ denote the stacked vector across all RISs with $L = \sum_{n=1}^{N} L_n$.

Following the same system and channel model as in Section~\ref{sec_RA_pssive}, the received signals at the PUs and ST retain the same structure. However, the effective channels now depend on $\mathbf{v}$, yielding $\mathbf{h}_k^{\mathrm{H}}(\mathbf{\Psi}) \mathbf{w}_i = b_{k,i} + \mathbf{q}_{k,i}^{\mathrm{T}} \mathbf{v}$, where $b_{k,i} = \mathbf{h}_{d,k}^{\mathrm{H}} \mathbf{w}_i$ and $\mathbf{q}_{k,i}$ captures the cascaded PT-RIS-PU channels. Accordingly, the SINR of PU $k$ is expressed as
\begin{equation}
    \Gamma_k(\mathbf{v}) = \frac{|x_{k,k}(\mathbf{v})|^2}{\sum_{i \neq k}^{K} |x_{k,i}(\mathbf{v})|^2 + \sigma^2},
\end{equation}
where $x_{k,i}(\mathbf{v}) = b_{k,i} + \mathbf{q}_{k,i}^{\mathrm{T}} \mathbf{v}$. Similarly, the sensing power at the ST is given by
\begin{equation}
    P_{\mathrm{sen}}(\mathbf{v}) = \sum\nolimits_{k=1}^{K} \| \mathbf{d}_k + \mathbf{B}_k \mathbf{v} \|^2,
\end{equation}
where $\mathbf{d}_k = \mathbf{H}_{d,s} \mathbf{w}_k$ and $\mathbf{B}_k \in \mathbb{C}^{M_s \times L}$ denotes the equivalent cascaded channel matrix.

Unlike the passive RIS case, active RIS introduces an additional power constraint due to signal amplification. Specifically, the transmit power at the $n$-th RIS must satisfy
\begin{equation}
    \!\!\!\sum\nolimits_{k=1}^{K} \| \operatorname{diag}(\mathbf{v}_n)\mathbf{G}_n \mathbf{w}_k \|^2 + \sigma_{r,n}^2 \|\mathbf{v}_n\|^2 \le P_{{\rm RIS},n}^{\max}, ~ \forall n, \!\!
\end{equation}
where $\sigma_{r,n}^2$ denotes the amplifier noise power. This constraint reflects the trade-off between signal amplification and noise enhancement.

Accordingly, for a given beamforming matrix $\mathbf{W}$ (optimized as in the passive RIS case), the active RIS design problem is formulated as
\begin{align}
    \mathbf{P}_{\phi a}:~& \max_{\mathbf{v}} \quad 
    \sum\nolimits_{k=1}^{K}\log_2\bigl(1+\Gamma_k(\mathbf{v})\bigr) + P_{\mathrm{sen}}(\mathbf{v}), \\
    \text{s.t.} \quad 
    & \mathbf{v}_n^{\mathrm{H}} \mathbf{C}_n \mathbf{v}_n \le P_{{\rm RIS},n}^{\max}, \quad \forall n,
\end{align}
where $\mathbf{C}_n = \sum_{k=1}^{K} \operatorname{diag}(\mathbf{G}_n \mathbf{w}_k) \operatorname{diag}(\mathbf{G}_n \mathbf{w}_k)^{\mathrm{H}} + \sigma_{r,n}^2 \mathbf{I}$ is a positive semidefinite matrix capturing both the signal amplification and noise contributions at the $n$-th RIS.

Problem $\mathbf{P}_{\phi a}$ is non-convex due to the coupled SINR expressions and the quadratic constraints. Compared to the passive RIS case, the optimization is more involved because both amplitude and phase must be jointly optimized under power constraints. To address this, SCA can be employed \cite{boyd2004convex}. In each iteration, the non-convex objective is approximated by a concave surrogate, and the resulting convex problem is solved to update $\mathbf{v}$. This iterative procedure converges to a stationary solution \cite{bezdek2003convergence}.

\begin{figure}[!t]\vspace{-2mm}	
\centering
    \includegraphics[width=0.40\textwidth]{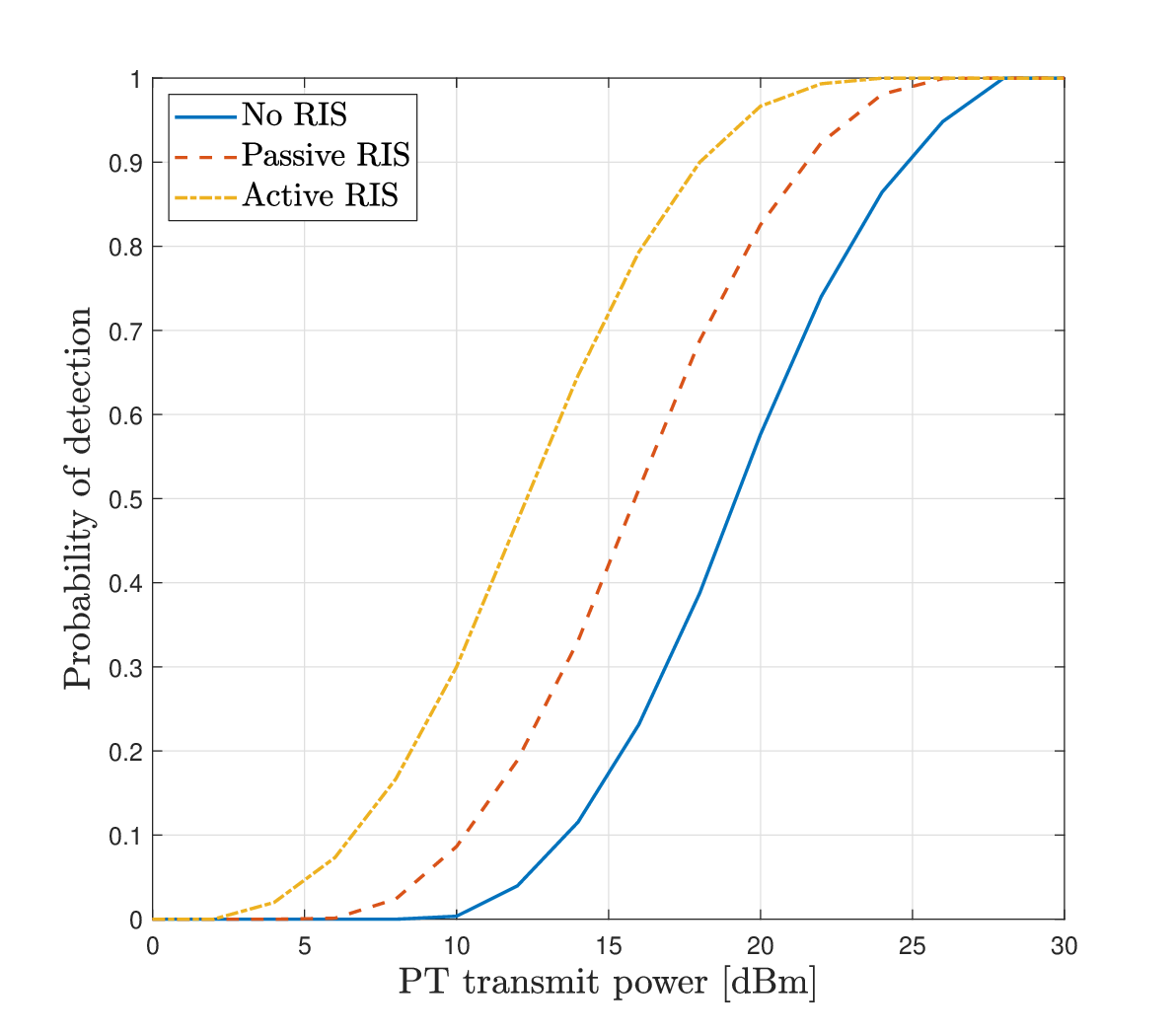}\vspace{-3mm}
\caption{Probability of detection at the ST as a function of the PT   power for active and passive RISs.} \label{fig_Pd_ActiveRIS}\vspace{-5mm}
\end{figure}

Fig.~\ref{fig_Pd_ActiveRIS} compares $\pd$ at the ST for three scenarios: i) no RIS, ii) passive RIS, and iii) active RIS, using the same simulation setup as the previous case study. For the active RIS, the reflecting coefficients are jointly optimized under a per-RIS power constraint, with the maximum transmit power budget of each RIS set to $P_{{\rm RIS},n}^{\max}=1.5$. The active RIS consistently achieves a higher $\pd$ than both the passive RIS and no-RIS cases: by jointly optimizing amplitude and phase, it mitigates the double-fading effect and provides a stronger effective sensing link, particularly in the low- and moderate-power regimes where passive RIS is limited by cascaded channel attenuation. Correspondingly, the detection curve for active RIS shifts leftward relative to passive RIS, indicating that a target $\pd$ can be reached at lower PT transmit power, improving EE and sensing sensitivity, though at the cost of the additional hardware complexity, power consumption, and amplification noise inherent to active elements.



\begin{mybox}[frametitle={Key Insights}]
\begin{enumerate}[leftmargin=*]
\item \textit{Active RIS enhances detection, especially at low power:} Joint amplitude-phase control mitigates the double-fading effect and boosts sensing power at the ST, with the gain over passive RIS most pronounced under low-to-moderate PT transmit power.

\item \textit{Active RIS lowers the transmit power needed for reliable sensing:} The leftward shift in the detection curve shows that a target $\pd$ is achievable at lower PT power than with passive RIS, at the cost of increased hardware complexity, power consumption, and amplification noise.
\end{enumerate}
\end{mybox}

\subsection{STAR RCNs}Conventional RISs are limited to reflecting incident signals over a half-space, requiring both PUs and SUs to reside on the same side of the surface in CRNs \cite{Liu2021}. This constraint is restrictive, particularly in dynamic or mobile scenarios. STAR-RIS overcomes this limitation by enabling simultaneous transmission and reflection of incident signals, thereby supporting full-space coverage \cite{Liu2021}. 

Specifically, a STAR-RIS partitions the incident signal into transmission and reflection components using protocols such as time switching (TS), mode switching (MS), and energy splitting (ES) \cite{Liu2021}. The transmitted and reflected signals are independently controlled via their respective coefficients, allowing flexible manipulation of signal propagation on both sides of the surface. This capability provides additional DoF for beamforming design and significantly enhances system performance in RCNs \cite{Liu2021}.

The use of STAR-RIS in CRNs has been investigated in \cite{Wen2022, Haochen2024}. In \cite{Wen2022}, a STAR-RIS is employed to enhance communication quality, where both the transmission and reflection regions each contain a PU and an SU communicating with a single-antenna BS. Operating under the TS protocol, the objective is to minimize the total SU transmit power by jointly optimizing BS power allocation and STAR-RIS transmission/reflection phase shifts, subject to interference temperature constraints at the PUs and minimum rate requirements for the SUs. As this problem is non-convex, the joint optimization is decomposed into tractable subproblems, where SDR is used to handle the unit-modulus constraints on the STAR-RIS phase shifts, and SCA is employed to convexify the non-convex rate and interference constraints. These subproblems are solved iteratively, leading to a locally optimal solution that balances computational complexity and performance while ensuring feasibility with respect to the system constraints.

In \cite{Haochen2024}, a STAR-RIS is considered for underlay CRNs with multiple PUs and SUs, where both primary and secondary BSs are equipped with multiple antennas. The objective is to maximize the secondary sum rate by jointly optimizing the secondary BS beamforming and the STAR-RIS configuration, subject to interference constraints at the PUs. A BCD algorithm is proposed, considering two STAR-RIS phase-shift models: (i) an independent phase-shift model solved via an SCA-based BCD approach, and (ii) a coupled phase-shift model addressed using a penalty dual decomposition-based BCD method. In the latter case, closed-form updates are derived for the STAR-RIS amplitudes and phase shifts.

\textit{Case Study 5:}
Here, we extend the case study in Section~\ref{sec_RA_pssive} to STAR RCN, where each RIS element simultaneously performs both transmission and reflection. Thus, the transmission and reflection coefficients of the $l$-th element for the $n$-th STAR-RIS are given by  $t_{n,l} = \sqrt{\beta_{n,l}^t} e^{j\theta_{n,l}^t}$ and $r_{n,l} = \sqrt{\beta_{n,l}^r} e^{j\theta_{n,l}^r}$, respectively, where $\beta_{n,l}^t$ and $\beta_{n,l}^r$ denote the transmission and reflection power coefficients, and $\theta_{n,l}^t$ and $\theta_{n,l}^r$ are the corresponding phase shifts. Since the STAR-RIS is passive, each element satisfies the energy conservation constraint $\beta_{n,l}^t + \beta_{n,l}^r = 1$, $\forall n,l$. Accordingly, the transmission and reflection matrices of the $n$-th STAR-RIS are given as $\mathbf{\Phi}_n^t = \operatorname{diag}(t_{n,1}, \dots, t_{n,L_n})$ $\mathbf{\Phi}_n^r = \operatorname{diag}(r_{n,1}, \dots, r_{n,L_n})$.

Following the same system and channel model as in Section~\ref{sec_RA_pssive}, we assume that all PUs are located on the reflection side of the STAR-RISs, while the ST is located on the transmission side. Therefore, the reflection matrices $\{\mathbf{\Phi}_n^r\}$ affect the PT-to-PU channels, whereas the transmission matrices $\{\mathbf{\Phi}_n^t\}$ affect the PT-to-ST sensing channel. Under the above assumption, the effective channel from the PT to PU $k$ becomes $\mathbf{h}_k^{\mathrm{H}}(\boldsymbol{\Phi}^r) = \mathbf{h}_{d,k}^{\mathrm{H}} + \sum_{n=1}^{N} \mathbf{h}_{n,k}^{\mathrm{H}} \mathbf{\Phi}_n^r \mathbf{G}_n$, where $\boldsymbol{\Phi}^r \triangleq \{\mathbf{\Phi}_n^r\}_{n=1}^{N}$. Similarly, the effective PT-to-ST channel during spectrum sensing is given by $\mathbf{H}_s(\boldsymbol{\Phi}^t) = \mathbf{H}_{d,s} + \sum_{n=1}^{N} \mathbf{F}_n \mathbf{\Phi}_n^t \mathbf{G}_n$, where $\boldsymbol{\Phi}^t \triangleq \{\mathbf{\Phi}_n^t\}_{n=1}^{N}$. Accordingly, received SINR at PU $k$ is expressed as\begin{equation}
    \Gamma_k(\mathbf{W},\boldsymbol{\Phi}^r) = \frac{\left|\mathbf{h}_k^{\mathrm{H}}(\boldsymbol{\Phi}^r)\mathbf{w}_k\right|^2} {\sum_{i \neq k}^{K}\left|\mathbf{h}_k^{\mathrm{H}}(\boldsymbol{\Phi}^r)\mathbf{w}_i\right|^2 + \sigma^2}.
\end{equation}
Similarly, the received signal at the ST is given by $\mathbf{y}_s = \mathbf{H}_s(\boldsymbol{\Phi}^t) \sum_{k=1}^{K} \mathbf{w}_k s_k$, and the corresponding received primary signal power is given by
\begin{equation}
    P_{\mathrm{sen}}(\boldsymbol{\Phi}^t) = \sum\nolimits_{k=1}^{K} \left|  \mathbf{H}_s(\boldsymbol{\Phi}^t)\mathbf{w}_k \right|^2.
\end{equation}

Compared to the conventional RIS case, the STAR-RIS introduces two coupled optimization variables, $\boldsymbol{\Phi}^t$ and $\boldsymbol{\Phi}^r$, which are linked through the per-element ES constraint. This coupling significantly increases the problem complexity and requires more advanced optimization techniques. For a given beamforming matrix $\mathbf{W}$, the STAR-RIS optimization problem is formulated as
\begin{subequations}\label{prob_star_phi}
\begin{align}
\mathbf{P}_{\phi}^{\rm STAR}:~
& \max_{\boldsymbol{\Phi}^t,\boldsymbol{\Phi}^r} ~~
\sum\nolimits_{k=1}^{K} \log_2\big(1 + \Gamma_k(\mathbf{W},\boldsymbol{\Phi}^r)\big)
+ P_{\mathrm{sen}}(\boldsymbol{\Phi}^t), \\
\text{s.t.} \quad
& |t_{n,l}|^2 + |r_{n,l}|^2 = 1, \quad \forall n,l.
\end{align}
\end{subequations}

Problem $\mathbf{P}_{\phi}^{\rm STAR}$ is non-convex due to the coupled transmission and reflection coefficients as well as the per-element ES constraints. To efficiently handle this, we adopt an MO-based approach. First, we define $\mathbf{v}_n^t = [t_{n,1}, \ldots, t_{n,L_m}]^{\mathrm T} \in \mathbb{C}^{L_n \times 1}$ and $\mathbf{v}_n^r = [r_{n,1}, \ldots, r_{n,L_n}]^{\mathrm T} \in \mathbb{C}^{L_n \times 1}$, i.e., the transmission and reflection coefficient vectors of the $n$-th STAR-RIS, respectively, yeilding $\mathbf{\Phi}_n^t = \operatorname{diag}(\mathbf{v}_n^t)$ and $\mathbf{\Phi}_n^r = \operatorname{diag}(\mathbf{v}_n^r)$. Then, by stacking all STAR-RIS coefficients, we define $\mathbf{v}^t = \big[ (\mathbf{v}_1^t)^{\mathrm T}, \ldots, (\mathbf{v}_N^t)^{\mathrm T} \big]^{\mathrm T}$ and $\mathbf{v}^r = \big[ (\mathbf{v}_1^r)^{\mathrm T}, \ldots, (\mathbf{v}_N^r)^{\mathrm T} \big]^{\mathrm T}$, where $L = \sum_{n=1}^{N} L_n$. Thereby, for the stacked vectors, the feasible set can be written as $\mathcal{M}_{\rm STAR} = \left\{ (\mathbf{v}^t,\mathbf{v}^r)\in \mathbb{C}^{L}\times\mathbb{C}^{L} \;\middle|\; |v_{l}^t|^2 + |v_{l}^r|^2 = 1,\ \forall l \right\}.$
Note that each pair $(v_{l}^t,v_{l}^r)$ lies on a complex sphere of unit radius in $\mathbb{C}^2$. Thus, $\mathcal{M}_{\rm STAR}$ forms a product manifold of $L$ complex spheres, leading to an oblique complex factory manifold \cite{lee2003introduction, Boumal2023book}. The reformulated problem on $\mathcal{M}_{\rm STAR}$ can be addressed by employing MO techniques \cite{lee2003introduction, Boumal2023book}. By iteratively updating $(\mathbf{v}^t,\mathbf{v}^r)$ using Riemannian conjugate gradient or trust-region methods, the proposed MO algorithm converges to a stationary point of $\mathbf{P}_{\phi}^{\rm STAR}$ \cite{lee2003introduction, Boumal2023book}.

\begin{figure}[!t]\vspace{-2mm}	
\centering
    \includegraphics[width=0.40\textwidth]{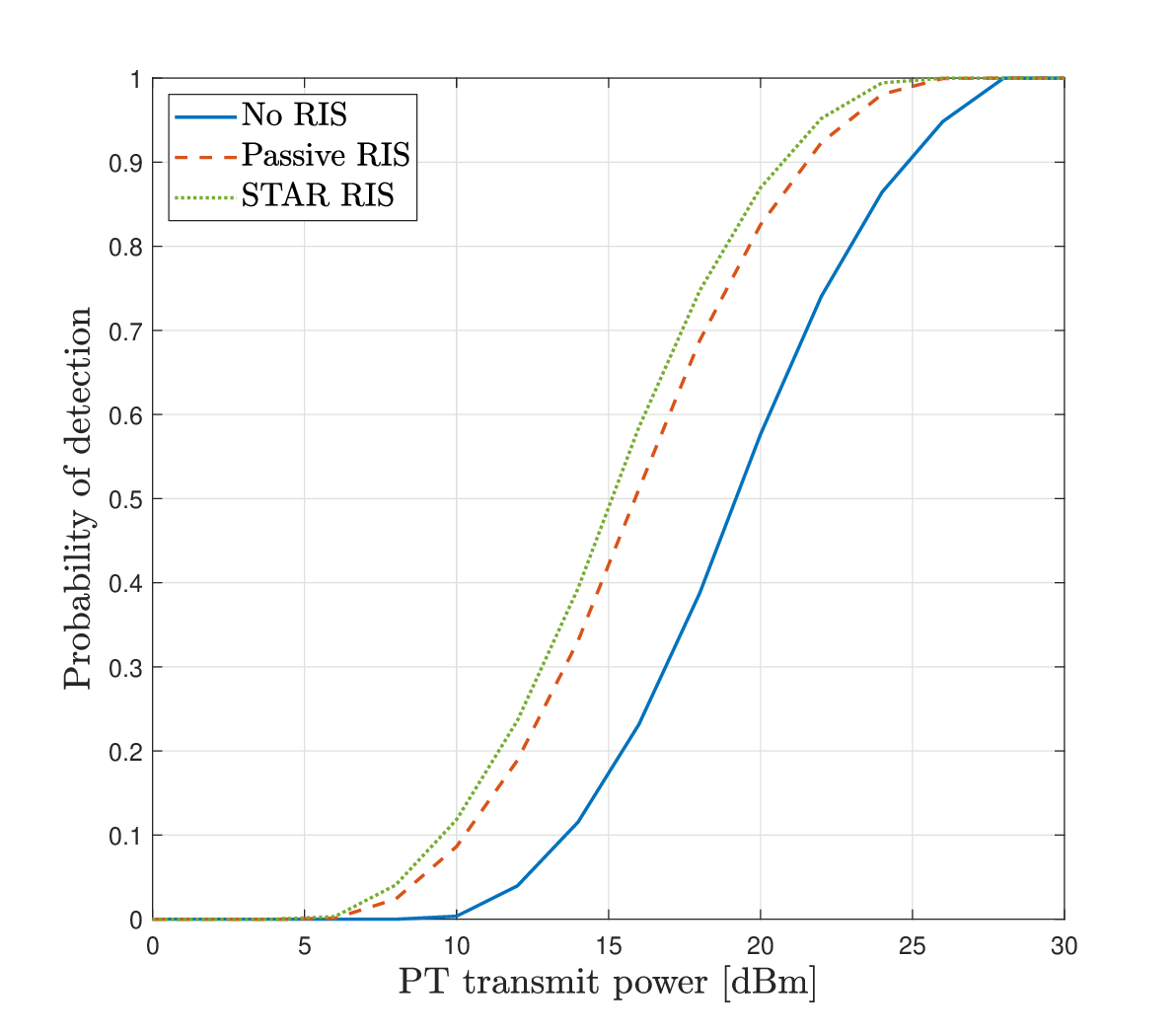}\vspace{-3mm}
\caption{Probability of detection at the ST as a function of the transmit power of the PT with STAR-RIS and passive RIS.} \label{fig_Pd_StarRIS}\vspace{-3mm}
\end{figure}

Fig.~\ref{fig_Pd_StarRIS} compares the $Pd$ of the ST for three scenarios: i) no RIS, ii) passive RIS, and iii) STAR-RIS, using the same simulation setup as the previous case studies. STAR-RIS consistently outperforms both passive RIS and no-RIS configurations across the entire transmit power range, since its ability to simultaneously transmit and reflect via adaptive ES enhances both the PT-to-PU and PT-to-ST links, unlike passive RIS, which reflects toward only one side. The advantage is most pronounced at moderate power: at $P_{\max}=\qty{15}{\dB m}$, $\pd$ improves from approximately \num{0.40} (passive RIS) to \num{0.50} (STAR-RIS), a nearly \qty{25}{\percent} relative gain, and reaching a target $\pd$ of \num{0.9} requires roughly \qtyrange{2}{3}{\dB m} less transmit power than with passive RIS, a gap that remains consistent across the power range, reflecting STAR-RIS's advantage in jointly optimizing transmission and reflection without the single-sided trade-off of conventional RIS.




\begin{mybox}[frametitle={Key Insights}]
\begin{enumerate}[leftmargin=*]
\item \textit{STAR-RIS enables simultaneous dual-link enhancement:} By jointly controlling transmission and reflection through per-element ES, STAR-RIS enhances both the sensing link (PT-to-ST) and communication links without requiring separate RIS deployments, yielding consistent gains over passive RIS across all operating regimes.

\item \textit{STAR-RIS improves sensing efficiency and EE:} The leftward shift of the detection curve relative to passive RIS shows that STAR-RIS reaches target detection performance at lower PT transmit power, at the cost of increased optimization complexity from the coupled transmission-reflection coefficients under the energy-conservation constraint.
\end{enumerate}
\end{mybox}









\subsection{ML for RCNs} 

\begin{table}[t]
\centering
\caption{Summary of ML Paradigms}
\label{tab_ml_summary}\vspace{-2mm}
\begin{tabular}{p{1.8cm}p{2.7cm}p{2.7cm}}
\hline
\textbf{Type} & \textbf{Data Requirement} & \textbf{Typical Tasks} \\ \hline \hline

Supervised learning 
& Labeled input-output pairs 
& Classification, regression \\ \hline

Unsupervised learning 
& Unlabeled data 
& Clustering, dimensionality reduction \\ \hline

Reinforcement learning 
& Agent-environment interaction 
& Sequential decision-making \\ \hline

Semi-supervised learning 
& Limited labeled + abundant unlabeled data 
& Improved generalization \\ \hline

Self-supervised learning 
& Data with automatically generated labels 
& Representation learning \\ \hline

\end{tabular}
\label{tab:ml_summary}\vspace{-5mm}
\end{table}

CRNs use adaptive thresholding, energy detection, matched filtering, and cyclostationary feature extraction for spectrum sensing. However, these approaches exhibit degraded performance at low SNR, rely on prior knowledge of the signal, and are computationally complex \cite{Khalek2024}. Furthermore, classical model-based analysis and convex optimization methods may be inadequate in RCNs due to model mismatches and incomplete system characterization, limiting their ability to fully exploit RIS capabilities \cite{Khalek2024}.

To address these limitations, data-driven ML-based approaches have gained significant attention, as they can learn complex channel and signal characteristics directly from data without relying on accurate analytical models \cite{Cao2022, Pivoto2026, Shi2023, Khalek2024}. In general, ML algorithms can be categorized as summarized in Table~\ref{tab_ml_summary}, based on their data requirements and learning paradigms \cite{Cao2022, Pivoto2026, Shi2023, Khalek2024}. Supervised learning relies on labeled datasets and is commonly used for classification and regression tasks, such as signal detection and modulation recognition in spectrum sensing. Unsupervised learning, which operates on unlabeled data, can be applied to clustering and anomaly detection, enabling the identification of spectrum usage patterns without prior knowledge. Reinforcement learning is particularly well-suited to RCNs, as it enables agents to learn optimal spectrum-access and RIS-configuration policies through interaction with the environment. In addition, semi-supervised and self-supervised learning approaches can improve generalization and reduce the need for large labeled datasets, which is especially beneficial in practical wireless systems where labeled data is limited.

ML thus provides a powerful framework for RCNs, offering improved adaptability, robustness, and computational efficiency \cite{Khalek2024}. In particular, ML enables efficient optimization of spectrum sensing, dynamic resource allocation, beamforming, and security by learning complex channel interactions, predicting interference patterns, and adaptively configuring RIS parameters.

Despite these benefits, ML  remains largely unexplored in RCNs, with the exception of \cite{Kayraklik2024}. This work proposes a practical RIS-assisted spectrum-sensing framework enhanced by DL for CRNs. The key idea is to combine the RIS's signal-enhancement capability with the pattern-recognition power of DL to improve the detection of PU signals. The RIS is used to enhance the received signal at the SU, thereby improving sensing under fading, blockage, or other impairments.

A central innovation of the paper is the formulation of spectrum sensing as an image-based detection problem. Specifically, the received signals (e.g., 4G LTE and 5G NR) are converted into spectrograms, which are then treated as images and processed using state-of-the-art DL object detection models such as Detectron2 and YOLOv7. This allows the system not only to detect the presence of a signal, but also to identify its type and its time-frequency occupancy with high accuracy.

Importantly, the authors of \cite{Kayraklik2024} also validate their approach using a real RIS prototype, demonstrating that RIS consistently improves the performance of DL-based detectors compared to conventional setups. The results show significant gains in detection accuracy and robustness, particularly in scenarios with weak or difficult-to-detect signals. This makes the approach highly relevant for mitigating issues such as hidden-node problems. 

Overall, the paper demonstrates that integrating RIS with DL provides a practical and powerful solution for spectrum sensing, enabling more reliable detection and better spectrum utilization. It also highlights a broader trend toward data-driven and environment-aware sensing, paving the way for intelligent spectrum management in future wireless systems.

\begin{figure*}[!t]\vspace{-2mm}
\centering
    \def\svgwidth{300pt} 
    \fontsize{8}{8}\selectfont 
    \graphicspath{{Figures/}}
    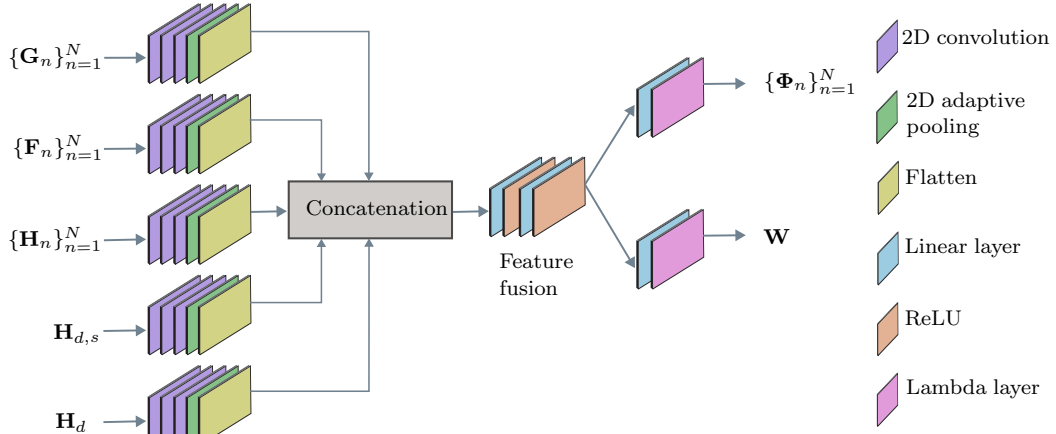 \vspace{-3mm}
    \caption{The CNN architecture.}\vspace{-5mm} \label{fig_CNN_architecture}
\end{figure*}

\textit{Case Study 6:}
Here, we extend the case study in Section~\ref{sec_RA_pssive} from classical optimization to a CNN-based ML solution. We employ unsupervised learning to optimize a custom loss function, thereby eliminating the need for labeled training data. The overall framework is described as follows:

\textit{Inputs and outputs:} 
Inputs are the communication channels among the ST, PT, PU, and RISs, following the case study in Section~\ref{sec_RA_pssive}. Specifically, these channels are given by $\q{G}_n \in \mathbb{C}^{L_n \times M_t}$, $\q{F}_n \in \mathbb{C}^{M_s \times L_n}$, $\q{H}_{d,s} \in \mathbb{C}^{M_s \times M_t}$, $\q{H}_n = [\q{h}_{n,1}, \cdots, \q{h}_{n,K}] \in \mathbb{C}^{L_n \times K}$, and $\q{H}_d = [\q{h}_{d,1}, \cdots, \q{h}_{d,K}] \in \mathbb{C}^{M_t \times K}$. The  outputs are the beamforming matrix $\q{W} \in \mathbb{C}^{M_t \times K}$ and the diagonal phase-shift matrix of each RIS, $\boldsymbol{\Phi}_n \in \mathbb{C}^{L_n \times L_n}$.

\textit{Preprocessing:} 
Since CNNs cannot directly process complex-valued matrices, an I/Q transformation is applied to represent each input by stacking its real and imaginary components, yielding $\bar{\q{J}} \in \mathbb{R}^{2 \times a \times b}$ for $\q{J} \in \mathbb{C}^{a \times b}$, where $\q{J} \in \{\q{G}_n, \q{F}_n, \q{H}_{d,s}, \q{H}_n, \q{H}_d\}$. The first and second channels of $\bar{\q{J}}$ contain the real and imaginary parts, respectively.

\textit{CNN-based architecture:} Fig.~\ref{fig_CNN_architecture} illustrates the proposed multi-branch CNN architecture for jointly optimizing the transmit beamforming matrix and the diagonal phase-shift matrix of each RIS from the input channel matrices. To exploit the heterogeneous characteristics of the different channels, each input is first processed independently before feature fusion. Specifically, each preprocessed channel matrix $\q{J}$ is fed into a dedicated CNN branch composed of three sequential convolutional layers. Each layer includes a 2D convolution, batch normalization, and ReLU activation. The first, second, and third layers produce $d_1$, $d_2$, and $d_3$ feature maps, respectively. After the third convolutional layer, a 2D adaptive average pooling with an output size of $(1, 1)$ is applied. The resulting output is then flattened. The extracted features are then mapped to compact latent representations through fully connected layers. The latent representations from all branches are subsequently concatenated into $d_4 = (3N + 2) d_3$ features and passed to a feature fusion module composed of fully connected layers that capture the joint relationships among the propagation channels. This module consists of two layers, each containing a linear layer followed by a ReLU activation. The outputs of these layers are denoted as $d_5$ and $d_6$, respectively. The fused representation is then divided into $N+1$ output heads: one head predicts the complex transmit beamforming matrix $\q{W}$, while the remaining $N$ heads estimate the RIS phase-shift vector $\boldsymbol{\phi}_n$ for each $n \in \{1,\cdots, N\}$. Each output head begins with a fully connected linear layer. The output size of this layer is $d_7 = 2 K M_t$ for the transmit beamforming head and $d_8 = L_n$ for the $n$th RIS phase-shift head.

\textit{Postprocessing:}
At each output head, constraint-aware Lambda layers are incorporated to ensure the feasibility of the predicted variables. In the beamforming head, the output of the fully connected layer is passed through a custom Lambda function that enforces the transmit power budget constraint in \eqref{eqn_PT_power_const}. Specifically, let the output of the fully connected layer be $\q{z} \in \mathbb{R}^{d_7}$. The Lambda layer first normalizes $\q{z}$ to satisfy the power constraint, after which an I/Q transformation is applied to map the normalized real-valued output into the complex beamforming matrix $\q{W} \in \mathbb{C}^{M_t \times K}$, i.e., from $\mathbb{R}^{2KM_t}$ to $\mathbb{C}^{M_t \times K}$. For each reconfigurable RIS phase-shift head, a separate Lambda function is implemented to enforce the unit modulus constraint in \eqref{eqn_RIS_phi_const}. This function converts the phase shifts into complex phase-shift coefficients, which are then arranged into a diagonal matrix, denoted as $\boldsymbol{\Phi}_n = {\rm diag} (\exp(j \q{y} / \max(|\q{y}|))$. In this context, $\q{y} \in \mathbb{R}^{d_8}$ denotes the output of the fully connected layer of the $n$th RIS phase-shift head.

\textit{Custom loss function:} 
A custom loss function is designed for the proposed unsupervised learning solution to jointly optimize communication and spectrum sensing performance without requiring labeled data. Based on $\q{P}_w$ in \eqref{eqn_P_w} and $\q{P}_{\phi}$ in \eqref{eqn_P_phi}, the loss function is defined as
\begin{eqnarray}
    \mathcal{L} = -\frac{1}{N_s} \sum\nolimits_{i = 1}^{N_s} \left(\sum\nolimits_{k = 1}^{K} \log_2 \left(1 + \Gamma_k^{(i)} \right) + P_{\rm sen}^{(i)} \right),
\end{eqnarray}
where $N_s$ denotes the number of samples, $\Gamma_k^{(i)}$ is the achievable rate of the $k$-th user, and $P_{\rm{ sen}}^{(i)}$ is the received sensing power at the ST for the $i$-th channel realization. The PT transmit power constraint in \eqref{eqn_PT_power_const} and the RIS phase-shift constraint in \eqref{eqn_RIS_phi_const} are enforced through Lambda layers in the post-processing stage.

\textit{Model training and deployment:} 
The proposed CNN-based solution is implemented in PyTorch. Unlabeled training samples are generated via Monte Carlo simulations of the communication and sensing channels described in Section~\ref{sec_RA_pssive}, and the corresponding input matrices are formed according to the specified input–output structure. Mini-batch training is then carried out using the Adam optimizer, with early stopping and ReduceLROnPlateau employed to minimize the custom loss function. Once trained offline, the model can be deployed online to predict the beamforming matrix and RIS phase-shift matrices for a given system configuration.

\begin{figure}[!t]\vspace{-2mm}	
\centering
    \includegraphics[width=0.40\textwidth]{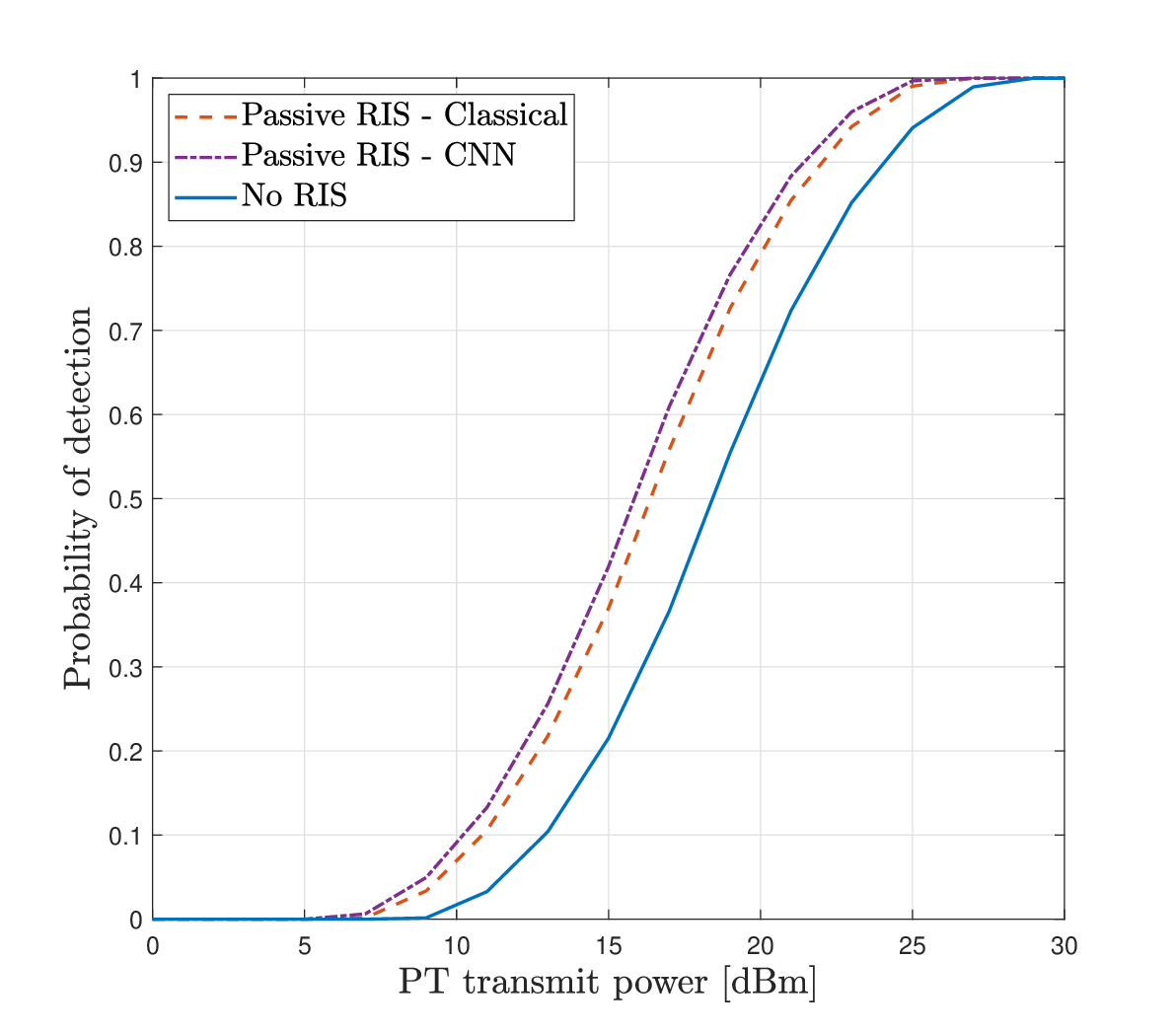}\vspace{-3mm}
\caption{Probability of detection at the ST as a function of the transmit power of the PT with classical optimization and CNN-based optimization for a passive RIS.} \label{fig_DL_case_study}\vspace{-5mm}
\end{figure}

Fig.~\ref{fig_DL_case_study} compares the detection probability of the ST for three scenarios: i) no RIS, ii) passive RIS with classical optimization (Case study 2 in Section~\ref{sec_RA_pssive}), and iii) passive RIS with CNN-based optimization, using the same simulation setup as the previous case studies. Both RIS-assisted schemes substantially outperform the no-RIS baseline, which relies solely on the direct PT-to-ST link and shows a noticeably delayed rise in $\pd$. The CNN-based solution, which jointly learns the beamforming matrix and RIS phase shifts directly from channel realizations without iterative optimization at deployment time, achieves slightly higher $\pd$ than classical optimization in the low-to-moderate power regime. This advantage is most evident in the transition region: at $P_{\max}=\qty{18}{\dB m}$, $\pd$ improves from \num{0.63} (classical) to \num{0.68} (CNN), a \qty{8}{\percent} relative gain, and reaching $\pd=\num{0.9}$ requires about \qty{1}{\dB m} less PT transmit power with the CNN-based design. At high transmit powers, the gap vanishes as all schemes saturate toward $\pd\rightarrow 1$, indicating that the CNN's benefit lies primarily in the challenging, non-saturated operating region.

\begin{mybox}[frametitle={Key Insights}]
\begin{enumerate}[leftmargin=*]

\item \textit{Learning-based optimization preserves and modestly extends RIS sensing gains:} Both classical and CNN-based RIS configurations substantially outperform the no-RIS case, with the CNN achieving a further leftward shift in the detection curve, i.e., a target $\pd$ at lower PT transmit power, particularly in the challenging low-to-moderate SNR regime where the gap with classical optimization is largest.

\item \textit{CNNs offer a low-complexity, generalization-dependent alternative to iterative optimization:} Once trained offline, the CNN directly predicts the beamforming matrix and RIS phase shifts online, reducing computational overhead and enabling fast adaptation, though its effectiveness hinges on representative training data and generalization across channel conditions and network geometries.

\end{enumerate}
\end{mybox}

\section{Open Issues and Future Research Directions}\label{sec_future}
Despite significant advancements in RCNs, several critical challenges remain unexplored and warrant further research. This section discusses these and future research directions.

\subsection{Scalability and Deployment}
The challenges include network management, hardware limitations, and dynamic spectrum coordination \cite{Mehdi2025, Zhao2024, Sangeeta2022}. The presence of numerous RIS elements and multiple RIS deployments significantly increases computational overhead, control complexity, and coordination demands. Additionally, optimal RIS placement is crucial (see Fig.~\ref{fig_RA_RIS_location}), as performance gains depend on factors such as the spatial distribution of PUs and SUs, network topology, and environmental conditions, including fading, obstacles, and multipath propagation \cite{Zhao2024}.

With multiple RIS panels, the number of elements is large, making phase-shift optimization highly complex and necessitating scalable real-time algorithms \cite{Feng2021}. These algorithms can improve performance metrics, including data rate, EE, and coverage.  As we have shown in the case studies, \cite{Feng2021} also highlights that RIS design leads to highly non-convex problems due to the coupling between beamforming and RIS phase shifts, as well as unit-modulus constraints, rendering conventional approaches inadequate.

To address these challenges, \cite{Feng2021} reviews representative optimization methods, including AO, SDR, SCA, FP, and majorization-minimization (MM), as well as low-complexity heuristics for large-scale systems. The optimization problems are systematically classified according to their structure, phase-shift model, and decoupling strategy. The authors highlight several limitations of existing approaches, including high computational complexity, imperfect CSI, hardware constraints, and the lack of convergence guarantees for many AO-based methods. To address these limitations, they develop a unified convex-approximation framework for continuous phase shifts with convergence to a stationary point under mild conditions. The study further identifies scalable and distributed optimization, robust design, and learning-based methods as promising directions toward practical large-scale RIS deployments.

Another promising approach is ML, where multiple RIS controllers collaboratively learn optimal configurations while reducing communication overhead \cite{Puspitasari2023}. Additionally, graph-based optimization techniques can be utilized to model interactions between RIS elements and CR nodes, enabling efficient large-scale optimization \cite{Syed2022}. 

Another key challenge is RIS's adaptability in dynamic wireless environments. While theoretical studies can assume ideal channel conditions, practical deployments must account for hardware impairments, imperfect CSI, and environmental factors such as mobility, obstructions, and multipath fading \cite{Zhao2024}. Robust ML-based algorithms/models for RIS phase optimization could enhance real-time adaptability, but they must be computationally efficient to support large-scale networks \cite{Puspitasari2023}. Additionally, hardware limitations, such as finite-resolution phase shifts, phase noise, and non-ideal reconfigurable elements, pose significant challenges \cite{Mehdi2025}. Open problems include hardware-aware optimization algorithms that accommodate practical imperfections and ensure robust operation under real-world constraints.

\subsection{Integration with Emerging Technologies}
IoT, UAVs, MEC, ISAC, NF communications, advanced multiple access techniques (e.g., RSMA), and FAS present both unique opportunities and challenges. However, they can be spectrum-intensive and can significantly benefit from DSA, interference management, and EE improvements provided by RCNs \cite{Hilal2023}. However, the seamless coordination between RCNs and these systems requires novel architectures, resource allocation strategies, and protocol enhancements. We will next describe such potential integrated systems. 

\subsubsection{IoT}
IoT and ambient IoT networks (e.g., BC and SR) can support a large number of low-power, low-processing-capability devices \cite{Rezaei2023Coding, rezaei2023timespread, Diluka2022}. These networks require efficient spectrum access while maintaining low latency and minimizing energy consumption \cite{Rezaei2023Coding, rezaei2023timespread, Diluka2022}. Traditional CRNs can enable DSA for IoT devices but face challenges, including interference mitigation and spectrum scarcity in densely connected environments. Integrating RCNs with IoT networks can improve SE by intelligently reflecting and directing signals toward IoT nodes, thereby enhancing spectrum-sensing accuracy \cite{Balachander2024}. However, a key challenge is the scalability of spectrum management. Since IoT networks often operate with constrained computational and energy resources, implementing complex spectrum sensing and decision-making algorithms at each node is impractical. Future research should explore strategies for RCNs to dynamically manage spectrum access for IoT devices. 

\subsubsection{UAV}UAVs enable a wide range of applications, including aerial monitoring and disaster response \cite{Sun2024}. However, DSA is essential because UAVs operate in bands already occupied by terrestrial systems, such as sub-\qty{6}{\GHz} cellular bands (e.g., \qtyrange{0.7}{3.5}{\GHz}) and L-/S-band satellite and UAV communications (approximately \qtyrange{1}{4}{\GHz}) \cite{Sun2024}. Their high mobility and dynamic deployment lead to rapidly varying spectrum availability across locations and altitudes, rendering static allocation inefficient. Moreover, the scarcity of dedicated spectrum necessitates opportunistic access to maintain reliable communication while avoiding harmful interference. Consequently, spectrum scarcity and interference remain key challenges for UAV networks \cite{Sun2024}.

RCNs can facilitate more efficient spectrum sharing between UAVs and terrestrial networks. A critical challenge is the UAV's mobility, which requires the rapid reconfiguration of the RIS to maintain optimal signal reflection. Unlike static RIS deployments, UAV-assisted RIS-enabled CRNs must adapt to frequent topological changes, requiring advanced spectrum management techniques. Future research should focus on joint trajectory and spectrum optimization for UAV-assisted RCNs to ensure optimal UAV trajectories and improved spectrum utilization \cite{Pandey2025}. Thus, significant research opportunities remain. 

\subsubsection{MEC}
MEC enables real-time data processing at the network edge, reducing reliance on centralized cloud infrastructure and improving latency for time-sensitive applications \cite{Yang2024}. Integrating RCNs with MEC can enable intelligent, distributed spectrum management, in which spectrum access decisions are made locally at edge nodes rather than at a central controller \cite{Yang2024}. This reduces decision latency and enables dynamic spectrum allocation based on localized traffic demands and interference conditions. However, the computational overhead of real-time RIS optimization at the edge remains a challenge \cite{Yang2024}. MEC nodes have limited computational resources, making it inefficient to implement complex spectrum-sensing and RIS phase-shift optimization algorithms. Future research should explore hybrid edge-cloud architectures in which MEC nodes handle lightweight, short-term spectrum reallocation tasks, while cloud-based controllers manage long-term spectrum planning and RIS configuration.

\subsubsection{ISAC}
ISAC has recently emerged as a key paradigm for 6G systems, enabling the joint utilization of radio resources for both communication and environmental sensing \cite{Diluka2025CFBook, Diluka2025CFISAC, Liu2022ISAC, zargari2025riemannian, zargari2024CFISAC,Vaezi2026}. The integration of ISAC with RCNs introduces new opportunities to enhance spectrum awareness and utilization. In particular, RIS can simultaneously assist communication links and improve sensing performance by shaping the propagation environment to enhance target detection, localization, and environmental mapping \cite{Liu2021RIS}. In RCNs, ISAC can complement conventional spectrum sensing by enabling joint sensing of PU activity and environmental context, improving detection reliability under low-SNR and dynamic conditions \cite{Liu2022ISAC}. However, the coexistence of sensing and communication functionalities introduces new trade-offs between sensing accuracy and communication performance \cite{Liu2022ISAC, Diluka2025CFBook}. RIS configuration must be jointly optimized to balance these objectives, which significantly increases system complexity. Furthermore, ISAC-assisted RCNs require new waveform designs, resource allocation strategies, and protocol adaptations to support simultaneous sensing and communication while ensuring minimal interference to PUs \cite{Liu2022ISAC, Diluka2025CFBook}. Open problems include joint RIS-assisted ISAC and spectrum access frameworks, low-complexity dual-functional waveform design, and robust optimization under imperfect channel and sensing information.

\subsubsection{NF Communication}
NF communication, enabled by extremely large-scale RIS and antenna arrays operating in the recently adopted upper mid-band and mmWave frequency ranges (e.g., \qtyrange{7}{24}{\GHz} and above) \cite{Bjornson2025NF, Baduge2026}, is another promising direction for future RCNs \cite{Azar2024, Liu2023Survey, Diluka2024NF, Galappaththige2025, Diluka2026WB, Diluka2026WBNF, Diluka2026NFSecure}. Unlike conventional far-field models, NF propagation occurs when users are located within the Rayleigh distance, $d_r = \frac{2D^2}{\lambda}$, which can be substantial for large apertures. For example, for a \qty{1}{\meter} RIS operating at \qty{10}{\GHz} (i.e., $\lambda \approx \qty{0.03}{\meter}$), the Rayleigh distance is approximately $d_r \approx \qty{67}{\meter}$. At higher frequencies, this region expands further; for instance, at \qty{28}{\GHz} ($\lambda \approx \qty{0.0107}{\meter}$), the Rayleigh distance increases to approximately $d_r \approx \qty{187}{\meter}$. These values indicate that users in typical urban and indoor deployments are often located within the NF region, where spherical wavefronts and spatially nonstationary channel characteristics must be accounted for. This regime provides additional spatial DoF for precise signal focusing and interference control \cite{Azar2024, Liu2023Survey}.

Integrating NF communication with RCNs can significantly enhance spectrum sharing by enabling highly localized signal transmission, thereby reducing unintended interference to PUs and improving spectrum reuse efficiency \cite{Azar2024, Liu2023Survey}. RIS operating in the NF regime can perform fine-grained wavefront shaping to focus energy on intended SUs while creating spatial nulls toward PUs \cite{Azar2024, Liu2023Survey, Diluka2024NF, Galappaththige2025, Diluka2026WB, Diluka2026WBNF, Diluka2026NFSecure}. However, NF-enabled RCNs introduce several challenges, including accurate NF channel modeling, high-dimensional channel estimation, and increased computational complexity for RIS configuration \cite{Azar2024, Liu2023Survey}. Moreover, conventional optimization techniques based on planar-wave assumptions are no longer applicable and require new models and algorithms tailored to NF propagation \cite{Galappaththige2025, Diluka2026WBNF, Diluka2026NFSecure}. Future research should investigate NF channel estimation techniques for RCNs, scalable optimization methods for extremely large RISs, and hybrid NF/far-field system designs to enable practical deployment.

\subsubsection{Multiple Access}Advanced multiple-access techniques, such as RSMA, offer a promising approach to improving spectrum utilization and interference management in RCNs \cite{Mao2022, Galappaththige2024RSMA, Zakari2026, Diluka2025RSMA}. Multiple access refers to the mechanism by which multiple users share limited wireless resources, such as time, frequency, code, or spatial dimensions. Classical multiple-access schemes include orthogonal multiple access (OMA), where users are separated using orthogonal resources (e.g., TDMA, FDMA, and OFDMA), and NOMA, where users simultaneously share the same resources through power-domain or code-domain multiplexing with successive interference cancellation (SIC). While OMA provides simple interference management, it suffers from limited SE due to strict orthogonalization. In contrast, NOMA improves SE and user connectivity but is highly sensitive to channel disparities, user ordering, and imperfect SIC, particularly under imperfect CSI and heterogeneous network conditions.

RSMA has recently emerged as a generalized and more flexible multiple-access framework that bridges OMA and NOMA by enabling partial interference decoding. To understand this, consider a BS serving $ K > 1 $ users. The BS splits each user’s message into a common part and a private part. The $K$  common parts are then combined into a common stream. The BS then transmits the common stream and $K$  private data streams. The common stream is decoded by the $K$ users, and each private stream is decoded only by its assigned users. Thus, for each user, part of the interference is decoded and canceled, while the remaining interference is treated as noise \cite{Mao2022, Galappaththige2024RSMA, Zakari2026, Diluka2025RSMA}. Unlike conventional NOMA, which requires full decoding of interference, RSMA flexibly controls the extent of interference decoding, thereby improving robustness to imperfect CSI, reducing sensitivity to SIC errors, and enhancing SE and fairness across diverse channel conditions. These advantages make RSMA particularly attractive for RCNs, where DSA, heterogeneous interference levels, and uncertain CSI significantly complicate interference management.

RSMA can also be applied to RCNs, where PNs and SNs must operate under strict interference constraints. By integrating RSMA with RIS, the spatial domain can be exploited to further enhance interference mitigation and improve SE through adaptive beamforming \cite{Zakari2026}. However, the joint design of RSMA parameters, RIS phase shifts, and spectrum-access decisions introduces significant complexity due to the coupled, dynamic nature of CRNs. Moreover, the performance of RSMA relies on accurate CSI, which is challenging to obtain in RIS-assisted environments. Thus, many open problems exist regarding low-complexity and robust RSMA-based resource allocation frameworks, adaptive transmission strategies for DSA, and the integration of RSMA with emerging paradigms such as ISAC and NF communication in RCNs.

\subsubsection{FAS}Fluid antennas, also called movable antennas or position-reconfigurable antennas, have recently emerged as a promising wireless technology, as they can dynamically adjust antenna positions within a confined region \cite{Zhu2026, Zhao2026, Ma2026, Xie2026}, unlike conventional antennas with fixed spatial locations. By adaptively repositioning antenna elements according to channel variations, FAS exploits favorable propagation conditions to improve signal strength and suppress interference \cite{Zhu2026, Zhao2026, Ma2026, Xie2026}. Despite these advantages, the integration of FAS into RCNs remains largely unexplored, creating significant opportunities to enhance spectrum sensing and secondary communications.

In particular, FAS can be utilized to improve sensing reliability at the ST by relocating antennas to positions with stronger received PU signals, thereby mitigating hidden-node problems. Furthermore, when combined with RIS, fluid antennas introduce an additional spatial DoF, enabling the joint optimization of antenna positions and RIS phase shifts to improve interference management and achieve higher SE. However, integrating FAS into RCNs also introduces several challenges, including dynamic channel modeling, increased signaling overhead, and the need for real-time joint optimization of antenna positions, RIS configurations, and spectrum-access decisions. Consequently, important research opportunities arise in developing low-complexity joint FAS-RIS optimization algorithms, robust designs under mobility and imperfect CSI, and practical implementations for dynamic-spectrum environments.

\subsection{Coexistence with Emerging Network Architectures}
New systems, such as CF mMIMO, O-RANs, NTNs, agentic AI, and others, are emerging. The opportunities, challenges, and open research directions for the use of RCNs with them are described below.

\subsubsection{CF Architectures}
CF wireless networks arise from the limitations of conventional cellular architectures, in which users are associated with a single BS, leading to inter-cell interference, uneven service quality, and poor performance at cell edges. Early solutions, such as network MIMO and coordinated multi-point (CoMP) transmission, introduced cooperation among BSs to mitigate these issues, but they remained constrained by cell boundaries, coordination overhead, and scalability limitations. With advances in mMIMO, cloud/centralized processing, and high-capacity fronthaul, the CF paradigm emerged as a user-centric alternative in which many distributed APs jointly serve users, providing more uniform coverage, improved reliability, and enhanced SE \cite{Diluka2025CFBook, Diluka2025CFISAC, Diluka2024CFFD, zargari2024CFISAC}.

However, CF networking is not explicitly standardized in the current specifications of the 3rd Generation Partnership Project (3GPP). Existing standards remain fundamentally cell-based, but incorporate several related concepts that move toward distributed cooperation. These include CoMP, distributed MIMO (or multi-TRP transmission in Releases 16-18), and cloud-RAN architectures that enable centralized coordination and functional splitting. Such features allow users to be served by multiple transmission points, partially capturing the benefits of CF operation while maintaining a cell-centric framework.

Looking ahead, 3GPP developments on 5G-Advanced and early 6G discussions increasingly emphasize distributed, cooperative, and user-centric designs. Although a fully CF architecture has not yet been adopted in standards, its key principles are being gradually incorporated through enhanced coordination, flexible association, and network-wide optimization. As a result, CF networking is widely regarded as a 6G candidate, offering a pathway toward scalable, interference-limited, and globally consistent wireless connectivity.

RCNs can be deployed within a CF architecture \cite{Diluka2025CFBook, Diluka2025CFISAC, Diluka2024CFFD, zargari2024CFISAC} by leveraging underlay, overlay, and interweave spectrum sharing techniques.   Distributed deployment of APs and RIS can enhance spectrum-sensing capabilities, enabling SUs to use the spectrum while minimizing secondary interference on the PN. Moreover, the distributed deployment of RIS can mitigate secondary interference through RIS optimization. For example, with underlay spectrum sharing,  \cite{Kulathunga2025} focuses on RIS phase-shift optimization to minimize interference to the PN. This work reveals that the RIS-assisted CF architecture improves the overall performance of both PN and SN. The performance gains arise from interference-aware communication enabled by eliminating cell boundaries, intelligent reflections, and the ability of distributed APs to collaboratively serve PUs/SUs in a CF mMIMO-based RCN. However, several design challenges remain. These include developing low-complexity, low-overhead channel estimation procedures to enable distributed beamforming, designing efficient spectrum sensing/sharing techniques, and optimizing RIS phase shifts, subject to intertwined design trade-offs between PNs and SNs.     

\subsubsection{O-RAN}
O-RAN is an emerging, disaggregated wireless network architecture that uses interoperable interfaces between the open radio unit (O-RU), the distributed unit (O-DU), and the central unit (O-CU) \cite{Mahmoud2026,Liang2026,Kirana2026}. It promotes physical layer technologies, enabling the use of an open-source software stack and replacing vendor-proprietary hardware modules. The O-RU in an O-RAN-aided RCN architecture can perform RF and low-level physical layer operations, such as dynamic spectrum sensing, access, and sharing, to optimize the overall network efficiency of the primary/secondary systems. Specifically, near-real-time radio access network (RAN) intelligent controllers embedded in the O-RAN architecture can be leveraged to automate these spectrum-related cognitive operations of the RCN. In particular, the O-DU can handle high-level physical layer and medium access control operations of RCN, enabling lower computational complexity and increased flexibility while driving down implementation costs.  Finally, the O-CU in the O-RAN can be leveraged to perform data packet protocols and control radio resources of the RCN, overcoming the closed network limitations of traditional RANs. While the integration of O-RAN into RCN can improve end-to-end performance, several design challenges must be overcome. Among them, security and trust vulnerabilities due to open, multi-vendor environments stand out when an SN is operated within the licensed PN \cite{Kirana2026}. Real-time intelligent radio resource and interference management for RCN within the O-RAN architecture can also be challenging.  

\subsubsection{NTN}
NTNs have recently garnered increasing attention for providing communication services \cite{Ntontin2025,Mahboob2025, Maiolini2024}.  They use spaceborne/airborne communication platforms, such as low Earth orbit (LEO) and geosynchronous Earth orbit (GEO) satellites, high-altitude platform systems (HAPS), and drones to provide direct connectivity, bypassing the need for classical ground-based infrastructure. Specifically, NTNs focus on bridging the connectivity gap in rural areas by enabling direct-to-device communication and broadband internet services, providing reliable and ubiquitous coverage. On the one hand, integrating RIS with spaceborne platforms of NTNs enables intelligent control of EM waves, thereby improving coverage, capacity, and reliability \cite{ Khan2026, Khan2025}. On the other hand, the coexistence of NTNs and traditional terrestrial networks can be established by leveraging  CRN concepts \cite{ Choi2024}. Specifically, efficient spectrum sensing, access, and sharing can enable an interplay between NTNs and RCNs to enhance SE and EE \cite{ Khan2024, Khan2025a}.  Key design challenges of RCN-aided NTNs include accurate channel estimation for fast-moving platforms, joint optimization of active and passive beamformers under strict interference constraints for PUs, and high power consumption for spaceborne wireless platforms.

As elaborated above, the coexistence of RCN with emerging CF, O-RAN, and NTN architectures has significant potential to enhance overall SE and EE. However, the key design challenges of this coexistence, such as faster channel estimation, joint multi-objective optimization of the beamformer and RIS phase shifts, and efficient spectrum sensing and sharing, require extensive investigation in future research. Most of these open research problems may not be amenable to classical signal processing and model-based optimization tools due to model inadequacy, unknown optimal or suboptimal algorithms, and prohibitively complex iterative algorithms for real-time implementation. Towards this end, ML-based solutions are appealing for addressing the key design challenges in signal processing and optimization that must be first established to enable the coexistence of RCN with emerging wireless architectures \cite{Kirana2026,Umer2025}. Hence, developing ML-based designs for RCN is an important direction for future research.            

\subsubsection{Agentic AI-native RCNs}
In traditional ML-driven wireless network architectures, AI is often incorporated as an optimization tool \cite{Saad2025}. Such AI-augmented network architectures typically rely on classical ML models that have centralized control and predefined constraints. This makes them less generalizable and adaptive to real-world network changes and dynamic propagation environments. Consequently, the traditional ML-driven architecture renders itself unamenable to deploying RCNs, which require dynamic spectrum sensing, access, and allocation among PT/PU and ST/SU, as well as joint BS precoder and RIS phase-shift matrix optimizations. To this end, the emerging AI-native wireless network architecture is particularly well-suited for deploying RCNs.

Shifting from the classical AI-augmented network architecture that employs ML optimization tools, the AI-native counterpart aims to deeply embed AI into the design foundations, from the physical layer to the application layer \cite{Saad2025}. This fundamental paradigm shift enables wireless architectures to be continuously adaptive through self-optimizing and autonomous service orchestration features. In this context, the agentic AI paradigm offers a pragmatic approach to the practical realization of AI-native wireless architectures \cite{Dev2025}. Agentic AI generally refers to any autonomous AI system that independently pursues goal-driven proactive reasoning, planning, and operating adaptively within a dynamic environment. It aims to establish intelligent integrated interactions with learning tools, hardware models, and software stacks without continuous human intervention. Specifically, the agentic AI-native wireless architecture sharply contrasts with the classical reactive AI-augmented counterpart, which requires human-aided continuous prompting.  Hence, the agentic AI-native wireless architecture can be leveraged to perform sophisticated tasks of RCNs without constant human oversight.  

In particular, RCNs can be deployed within an agentic AI-native wireless architecture by replacing classical reactive, predefined rule-based spectrum management with proactive, goal-driven alternatives that are more suitable for handling dynamic network topology changes in fast-varying EM propagation conditions. The radio intelligence agents can leverage a combination of foundational large language models, reinforcement learning, and generative AI models to perform intelligent network management for RCNs \cite{Jiang2026}. They enable real-time optimization of BS precoders and RIS phase-shift matrices, interference mitigation via optimal dynamic spectrum sensing and management, and autonomous decision-making among PT/PU and ST/SU deployed within dynamic EM propagation environments.  The agentic AI-native architecture can also leverage an automated spectrum governance framework to monitor the spectrum, ensuring strong compliance with regulations enforced by government agencies, such as the Canadian Radio-television and Telecommunications Commission (CRTC) and the Federal Communications Commission (FCC) in the United States. Embedding this agentic AI-native notion within the O-RAN architecture will enable a pragmatic approach to spectrum sensing and system optimization for RCNs under tight latency constraints. The above discussion raises a myriad of challenging research problems; thus, establishing foundations for agentic AI-native RCNs is a crucial direction for future research.

\subsection{Standardization Challenges}
Integrating RIS into CRNs requires well-defined regulatory frameworks to ensure compatibility with existing spectrum policies \cite{Nasser2021, Sangeeta2022}. Traditional spectrum allocation relies on static licensing models, which do not account for the dynamic spectrum reconfiguration enabled by RIS. Meanwhile, current CR policies primarily govern DSA through sensing and opportunistic transmission \cite{Nasser2021, Sangeeta2022}. However, RIS introduces new forms of spectrum manipulation that necessitate revised interference management policies; e.g.,  active RIS elements emit additional power, increasing interference for users. Standardized control signaling protocols are needed to support dynamic RIS reconfigurability and network architectures.

\subsection{Security and Privacy Concerns}
The use of RCNs introduces new security vulnerabilities that require advanced mitigation strategies \cite{Jacek2022}. A key concern is RIS-assisted eavesdropping, where an adversarial RIS manipulates reflections to intercept confidential communications \cite{Jacek2022}. Another emerging threat is RIS-based PU/SU emulation attacks, in which a malicious RIS alters its reflection pattern to mimic a legitimate PU/SU, thereby deceiving SUs into avoiding certain spectrum bands. The deployment of adversarial RIS poses physical layer security threats to CRNs. For example, an adversarial RIS can aid active eavesdroppers in enhancing pilot contamination attacks during the channel estimation phase of PT/ST. Consequently,  PT/ST implicitly estimates the channels of the active eavesdroppers and inadvertently beamforms confidential data towards them, compromising the achievable secrecy rates \cite{Dassanayake2026}.  To address these challenges, secure beamforming techniques can be developed that leverage RIS reconfigurability to generate artificial-noise-based jamming, thereby preventing unauthorized signal interception. Additionally, secure spectrum sensing methods that exploit multipath diversity and cooperative sensing should be explored. Furthermore, research should investigate secure RIS optimization, where multiple RISs collaborate to enhance security without directly sharing user data.

\section{Conclusion} Although cognitive radio networks (CRNs) have been extensively investigated, important challenges remain in improving spectrum utilization, enhancing spectral and energy efficiency (SE and EE), and maintaining reliable operation under adverse propagation and low SNR conditions. RISs offer a promising means of addressing these limitations by enabling programmable control of the wireless propagation environment. By combining the spectrum awareness and dynamic access capabilities of CRNs with the propagation control capabilities of RISs, RCNs have emerged as a promising architecture for future 6G wireless systems. RISs can strengthen desired links, suppress interference, and improve spectrum sensing and access, thereby enhancing the overall efficiency and reliability of CRNs. Despite the rapidly growing interest in RCNs, a comprehensive survey that systematically consolidates their architectures, enabling techniques, recent advances, and outstanding challenges has been lacking.

This paper provides a comprehensive survey of RCNs and their potential for next-generation wireless networks. It first introduces the fundamentals of CRNs and RISs, including DSA models, spectrum sensing techniques, RIS operating principles, and RIS architectures. It then examines the key design aspects of RCNs, including channel estimation, RIS-assisted spectrum access, deployment strategies, multi-RIS configurations, system architectures and communication protocols, and joint optimization of RIS and CRN parameters. The existing literature is systematically organized into six major research directions: performance analysis, resource allocation and optimization, secure RCNs, active RISs, simultaneously transmitting and reflecting RISs (STAR-RISs), and machine-learning-enabled RCNs. Finally, the paper identifies key challenges, open problems, and future research directions, including scalability and practical deployment, integration with emerging 6G technologies, coexistence with evolving network architectures, standardization, and security and privacy. Overall, this survey provides a unified perspective on the state of the art and establishes a roadmap for developing practical, intelligent RCNs.

\balance

\bibliographystyle{IEEEtran}
\bibliography{IEEEabrv,ref}

\end{document}